\documentclass[twocolumn]{aastex63} 
\usepackage[normalem]{ulem}
\usepackage[version=4]{mhchem}

\newcommand{\rev}[1]{\textcolor{black}{#1}}

\shorttitle{Water excitation atlas, diagnostic diagrams, and Doppler mapping}
\shortauthors{JDISCS collaboration}

\begin{document}

\title{Water in protoplanetary disks with JWST-MIRI: spectral excitation atlas and radial distribution from temperature diagnostic diagrams and Doppler mapping}

\correspondingauthor{Andrea Banzatti}
\email{banzatti@txstate.edu}

\author[0000-0003-4335-0900]{Andrea Banzatti}
\affil{Department of Physics, Texas State University, 749 N Comanche Street, San Marcos, TX 78666, USA}

\author[0000-0003-3682-6632]{Colette Salyk}
\affil{Department of Physics and Astronomy, Vassar College, 124 Raymond Avenue, Poughkeepsie, NY 12604, USA}

\author[0000-0001-7552-1562]{Klaus M. Pontoppidan}
\affiliation{Jet Propulsion Laboratory, California Institute of Technology, 4800 Oak Grove Drive, Pasadena, CA 91109, USA}

\author[0000-0002-6695-3977]{John Carr}
\affiliation{Department of Astronomy, University of Maryland, College Park, MD 20742, USA}

\author[0000-0002-0661-7517]{Ke Zhang}
\affil{Department of Astronomy, University of Wisconsin-Madison, Madison, WI 53706, USA}

\author[0000-0003-2631-5265]{Nicole Arulanantham}
\affiliation{Space Telescope Science Institute 3700 San Martin Drive Baltimore, MD 21218, USA}

\author[0000-0002-3291-6887]{Sebastiaan Krijt}
\affiliation{School of Physics and Astronomy, University of Exeter, Stocker Road, Exeter EX4 4QL, UK}

\author[0000-0001-8798-1347]{Karin I. \"Oberg}
\affiliation{Center for Astrophysics, Harvard \& Smithsonian, 60 Garden St., Cambridge, MA 02138, USA}

\author[0000-0003-2076-8001]{L. Ilsedore Cleeves}
\affil{Astronomy Department, University of Virginia, Charlottesville, VA 22904, USA}

\author[0000-0002-5758-150X]{Joan Najita}
\affiliation{NSF’s NOIRLab, 950 N. Cherry Avenue, Tucson, AZ 85719, USA}

\author[0000-0001-7962-1683]{Ilaria Pascucci}
\affil{Department of Planetary Sciences, University of Arizona, 1629 East University Boulevard, Tucson, AZ 85721, USA}

\author[0000-0003-0787-1610]{Geoffrey A. Blake}
\affiliation{Division of Geological \& Planetary Sciences, MC 150-21, California Institute of Technology, Pasadena, CA 91125, USA}

\author[0000-0001-7152-9794]{Carlos E. Romero-Mirza}
\affiliation{Center for Astrophysics, Harvard \& Smithsonian, 60 Garden St., Cambridge, MA 02138, USA}

\author[0000-0003-4179-6394]{Edwin A. Bergin}
\affil{Department of Astronomy, University of Michigan, 1085 S. University Ave, Ann Arbor, MI 48109}

\author[0000-0002-2828-1153]{Lucas A. Cieza}
\affil{N\'ucleo de Astronom\'ia, Facultad de Ingenier\'ia y Ciencias, Universidad Diego Portales, Av Ej\'ercito 441, Santiago, Chile}

\author[0000-0001-8764-1780]{Paola Pinilla}
\affiliation{Mullard Space Science Laboratory, University College London, Holmbury St Mary, Dorking, Surrey RH5 6NT, UK.}

\author[0000-0002-7607-719X]{Feng Long}
\altaffiliation{NASA Hubble Fellowship Program Sagan Fellow}
\affil{Department of Planetary Sciences, University of Arizona, 1629 East University Boulevard, Tucson, AZ 85721, USA}

\author{Patrick Mallaney}
\affil{Department of Physics, Texas State University, 749 N Comanche Street, San Marcos, TX 78666, USA}

\author[0000-0001-8184-5547]{Chengyan Xie}
\affil{Department of Planetary Sciences, University of Arizona, 1629 East University Boulevard, Tucson, AZ 85721, USA}

\author[0000-0002-1566-389X]{Abygail R. Waggoner}
\affil{Department of Astronomy, University of Wisconsin-Madison, Madison, WI 53706, USA}

\author[0000-0001-8240-978X]{Till Kaeufer}
\affiliation{School of Physics and Astronomy, University of Exeter, Stocker Road, Exeter EX4 4QL, UK}

\author{the JDISCS collaboration}

\begin{abstract}
This work aims at providing fundamental general tools for the analysis of water spectra as observed in protoplanetary disks with JWST-MIRI. We analyze 25 high-quality spectra from the JDISC Survey reduced with asteroid calibrators as presented in \cite{pontoppidan24}. First, we present a spectral atlas to illustrate the clustering of \ce{H2O} transitions from different upper level energies ($E_u$) and identify single (un-blended) transitions that provide the most reliable measurements. With that, we demonstrate two important excitation effects: the opacity saturation of ortho-para line pairs that overlap, and the non-LTE excitation of $v=1-1$ lines scattered across the $v=0-0$ rotational band. 
Second, we define a shorter list of fundamental lines spanning $E_u=$~1500--6000 K to develop simple line-ratio diagnostic diagrams for the radial temperature distribution of water in inner disks, \rev{which can be interpreted using discrete temperature components or a radial gradient}.
Third, we report the detection of disk-rotation Doppler broadening of molecular lines, which confirms the radial distribution of water emission including, for the first time, the radially-extended $\approx 170$--220~K reservoir close to the snowline. \rev{The combination of measured line ratios and broadening suggests that drift-dominated disks have shallower temperature gradients with an extended cooler disk surface enriched by ice sublimation.}
We also report the first detection of a \ce{H2O}-rich inner disk wind from narrow blue-shifted absorption in the ro-vibrational lines. We summarize these findings and tools into a general recipe to make the study of water in planet-forming regions reliable, effective, and sustainable for samples of $> 100$ disks.
\end{abstract}

\section{Introduction}  \label{sec: intro}
The infrared spectrum of water vapor has attracted increasing interest in the community since the discovery of its forest of lines in protoplanetary disks with the Spitzer-IRS \citep{irs} and ground-based spectrographs almost two decades ago \citep{cn08,salyk08}. With $> 1000$ transitions significantly contributing to spectra at 10--37~$\mu$m, but only $\approx 150$ blended features of multiple transitions that could be distinctly observed with IRS \citep[e.g.][]{pont10}, water spectra are at the same time fascinating and challenging. Most of the attraction comes from the fact that they trace inner disks at radii where rocky and super-Earth planet formation is expected to happen \citep[e.g.][]{bitsch19_rocky,lambrechts19,Izidoro21}, as shown by water line profiles when observed at high resolving power from the ground \citep{pont10b,najita18,salyk19,banz23}. 
Water is expected to trace and significantly contribute to a number of processes that are fundamental in star formation all the way to late phases of disk evolution and planet formation \citep[e.g.][]{vandishoeck14}, from the dynamics and accretion of solids \citep[e.g.][]{cieslacuzzi06,ros13} to the development of habitable conditions \citep[e.g.][]{krijt22}. Determining the water abundance in disks as a function of orbital distance and time remains one of the major goals for understanding these fundamental processes \citep[e.g.][]{meijerink09,pont14}.

On the other hand, challenges come from the wide range of excitation conditions that water spectra trace in inner disks, including radial and vertical temperature and density gradients and non-local thermodynamic equilibrium (LTE) excitation \citep[e.g.][]{glassgold09,najita11,meijerink09,bosman22,banz23}. The key to distinguishing different processes and correctly interpreting the observed spectra to obtain a global view of water properties as a function of disk radius lies largely in the possibility to measure the flux and kinematics of transitions from a wide range of Einstein-$A$ coefficients ($A_{ul}$) and a large range in upper level energy ($E_u$). 
With Spitzer-IRS spectra, the most critical limitation to the analysis of water emission properties has come from the low resolving power (R~$\sim$~600--700, or 450~km/s), blending multiple transitions and leaving fundamental degeneracies between temperature and column density in slab model fits \citep{meijerink09,cn11,salyk11_spitz}. Despite the identification of some important trends with stellar temperature \citep{pont10,pascucci13}, with stellar and accretion luminosity \citep{salyk11_spitz,banz17,banz20}, with the formation of an inner disk dust cavity \citep{salyk11_spitz,banz17,banz20}, and with the dust disk mass or radius as observed at millimeter wavelengths \citep{najita13,banz20}, the need for higher-resolution spectroscopy has clearly emerged as a priority for making progress in the field after Spitzer.

In spite of the paucity of high-resolution spectra due to the difficulty of observations through the Earth's atmosphere, recent work using spectrally-resolved data from ground-based spectrographs has made some progress after the Spitzer surveys. Infrared water spectra have been found to trace a temperature gradient in the inner regions of protoplanetary disks, when velocity-resolved line widths from a large range in energy levels have been obtained by combining data from different instruments \citep{banz23}. A gradient is naturally expected by disk models \citep[e.g.][]{glassgold09,meijerink09} but evidence from the data has been slow to emerge. Early fits to Spitzer-IRS spectra initially assumed a single temperature that could well reproduce at least part of the data \citep{cn11,salyk11_spitz}, while pointing out that spectral lines at longer wavelengths seemed to come from colder temperatures \citep{salyk11_spitz,banz12}. Later, some works explored using temperature gradients that could better reproduce a broader spectral range, in some cases including also far-infrared spectra \citep{kzhang13,blevins16,liu19}. This is an aspect in which the sharper spectral view provided by JWST-MIRI can now advance the community understanding of water in inner disks \citep{MINDS23,MINDS23k,MINDS24}. Since the first MIRI observations, water emission from multiple temperatures has been confirmed by single-temperature fits to different parts of the emission \citep{banz23b,gasman23,schwarz24} as well as two- or three-temperature simultaneous fits \citep{pontoppidan24,temmink24b} and radial gradient fits to parts of the rotational spectrum at MIRI wavelengths \citep{munozromero24b,kaeufer24}. 

The identification of multiple temperature regions and characterization of their properties is a significant step towards the goal of determining the water abundance as a function of orbital distance and disk age/evolution. However, the detailed analysis and interpretation of water spectra remains a challenge after two years of JWST observations. Slab model fits to the observed spectra across MIRI wavelengths give different temperature and column density estimates even when similarly simple slab model tools are used \citep[e.g.][and Salyk et al. in prep.]{banz23b,gasman23,pontoppidan24,schwarz24,temmink24b}, and it is currently unclear how much of that depends on the spectral ranges and specific lines being included (or excluded) from the fits rather than from species contamination, spectra quality (including residual fringes), and non-LTE excitation effects that have long been known \citep{meijerink09} but have not yet been accounted for in fitting MIRI water spectra.

At the increased resolving power of MIRI, line blending is still an issue for carefully measuring the emission from different energy levels, due to the overlap of multiple water transitions as well as the contamination from a number of atomic and molecular species. A common approach to correctly analyze water spectra and compare results from different studies has not yet emerged, and the identification of reliable lines for breaking degeneracies and isolating or controlling different effects (temperature from opacity gradients, LTE from non-LTE excitation) is still a priority.

It is desirable as a community to agree on a reliable line list and some simple diagnostics that will provide comparisons across samples without biases from different implementations of slab modeling tools and their (often untested) dependence on fitting different line ranges that do not properly account for contamination or non-LTE effects.
As a contribution towards defining a common ground for community comparisons, in this work we provide a curated list of reliable single (un-blended) lines and demonstrate their use to describe the excitation and kinematic properties of water spectra as observed with MIRI, to provide the community with useful tools for their analysis and interpretation in the growing number of disks being collected in current and future observing cycles. In particular, we develop a general definition of water temperature and column density diagnostic diagrams that will enable empirical, line-flux-ratio comparisons across samples and datasets, to support a broad understanding of water in different conditions and its study across multiple processes that shape inner disk evolution and planet formation.

This paper is organized as follows. In Section \ref{sec: atlas} we present a spectral atlas of water emission as a general reference to support line identification and analysis, with specific focus on aspects of line blending, contamination, opacity overlap, and non-LTE excitation that are important for a correct analysis of observed spectra. With the atlas, we demonstrate a curated list of single, un-blended lines that provides the most reliable flux and broadening measurements to support global analyses of water in inner disks.
Using this line list, in Section \ref{sec: cool excess} we define four line ratios that provide a simple general definition of diagnostic diagrams for the temperatures and column density of water emission in individual sources as well as large samples.
In Section \ref{sec: kinematics} we present, for the first time, evidence for Doppler broadening of water lines in MIRI-MRS spectra, as well as the detection of unresolved wind blue-shifted absorption on top of disk-rotation broadened emission in a disk observed at high inclination. 
In Section \ref{sec: doppler_mapping} we will use this information to perform a simple Doppler mapping of water lines, which can provide an independent estimate of the emitting radii of lines from different upper level energies.
In Section \ref{sec: discussion} we combining all these findings into the discussion of the radial distribution of water in inner disks in terms of radial gradients and of the processes that regulate them, and we summarize the tools into a general recipe to support the analysis of water MIRI-MRS spectra in future samples.

\section{Sample \& data reduction} 
The data analyzed in this work were taken with the James Webb Space Telescope \citep[JWST,][]{Gardner23} between February 2023 and March 2024. The disks were observed with the Medium Resolution Spectrometer \citep[MRS,][]{jwst-mrs,Argyriou23} mode on the Mid-Infrared Instrument \citep[MIRI,][]{miri,miri2}. \rev{MIRI covers the full wavelength range of 4.9--28\,$\mu$m with resolving power of 1800-4000 \citep[][see also Section \ref{sec: MIRI R} in this work]{Argyriou23,pontoppidan24} with spatial resolution down to 0.2 arcsec at the shortest wavelengths.}
The sample included in this work comes from three GO programs in Cycle 1 that defined the JWST Disk Infrared Spectral Chemistry Survey \citep[JDISCS,][ Arulanantham et al. (in prep.)]{pontoppidan24}: 14 disks from GO-1584 (PI: C. Salyk; co-PI: K. Pontoppidan), 8 disks from GO-1640 (PI: A. Banzatti), and 3 disks from GO-1549 (PI: K. Pontoppidan) for a total of 25 disks. Sample properties are reported in Appendix \ref{app: sample measurements}. The full sample is described in the overview paper by Arulanantham et al. (in prep.); here we focus on disk spectra of T~Tauri stars, i.e. we exclude 3 disks around intermediate-mass stars (HD~163296, HD~142666, and HD~143006), and we also exclude one disk in a binary system (AS~205~S) that has strong fringe residuals and binary contamination at $> 18 \mu$m in MIRI. Some spectra have been analyzed and presented before in papers from the JDISCS collaboration: CI~Tau, GK~Tau, IQ~Tau, and HP~Tau in \cite{banz23b}, Sz~114 in \cite{cy23_Sz114}, FZ~Tau in \cite{pontoppidan24}, AS~209 and GQ~Lup in \cite{munozromero24a,munozromero24b}, DoAr33 in \citet{colmenares24}.

All spectra were extracted with a wavelength-dependent aperture radius of $1.4 \times 1.22 \lambda/D$ and wavelength-calibrated with the JDISCS pipeline as described in \cite{pontoppidan24}, which adopts the standard MRS pipeline \citep{MIRI_pip} up to stage 2b and then uses asteroid spectra observed as calibrators to provide high-quality fringe removal and characterization of the spectral response function to maximize S/N in channels 2--4 (while a standard star is used in channel 1). Target acquisition with the MIRI imager was adopted to place all targets and asteroids on exactly the same spot on the detector with sub-spaxel precision, to ensure similar fringes and maximize the quality of their removal. For the data included in this work, we used the latest available JDISCS reduction (version 8.2) that uses the MRS pipeline version 11.17.19 and Calibration Reference Data System context jwst\_1253.pmap. Before the gas emission line analysis presented in this work, all spectra were continuum-subtracted using a median smoothing and a 2nd-order Savitzky-Golay filter with the procedure presented in \cite{pontoppidan24} updated to apply a final offset based on line-free regions as demonstrated in Appendix \ref{app: cont_sub}. The continuum-subtracted spectra are included in Appendix \ref{app: atlas_compact}.

\begin{figure*}
\centering
\includegraphics[width=1\textwidth]{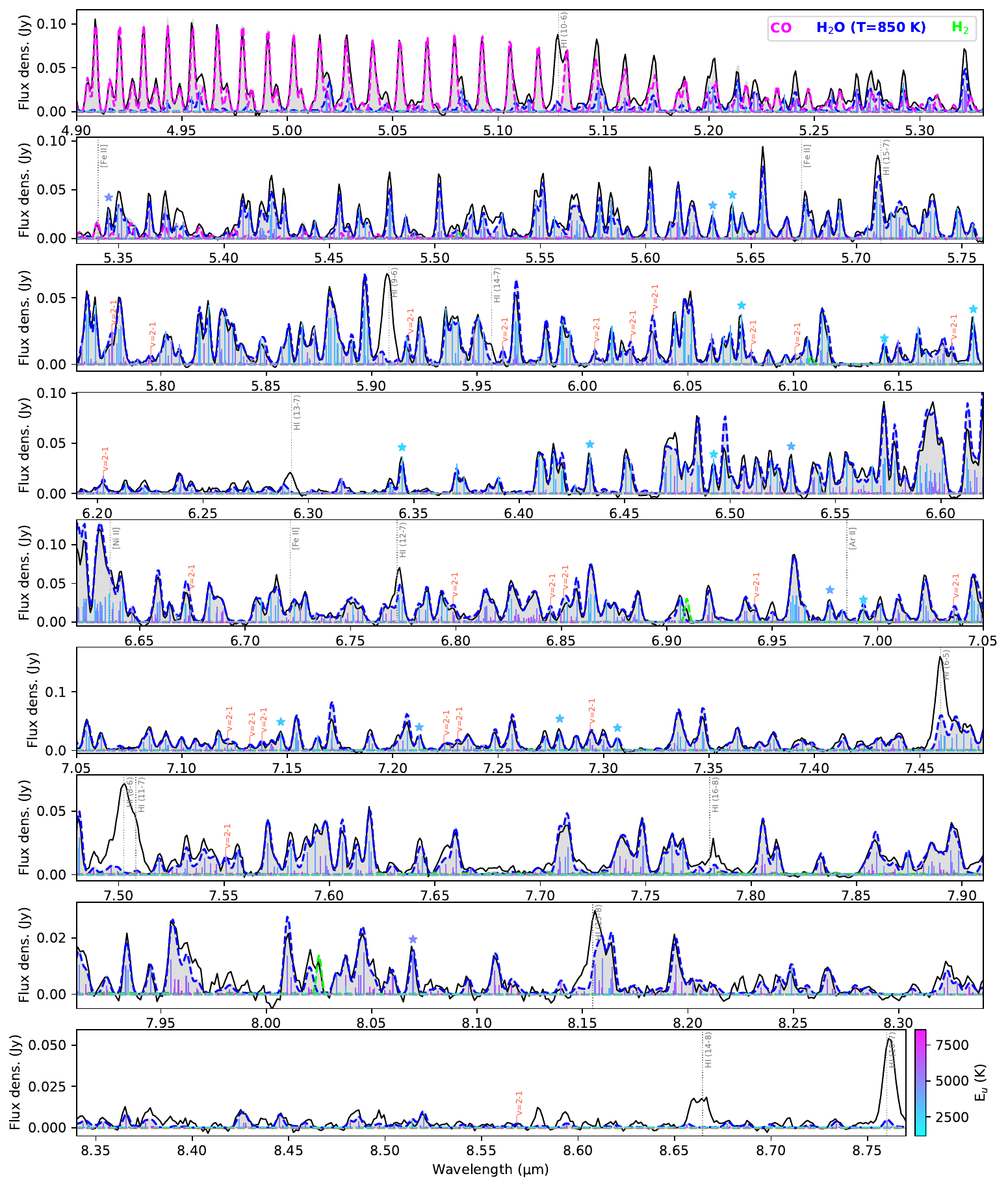}
\caption{Atlas of water emission in MIRI-MRS spectra (see Section \ref{sec: atlas}), using CI~Tau as an example (shown in black). Vertical lines mark transition wavelengths, color-coded according to the upper level energy $E_u$ as shown in the color bar. Lines marked with a star (color-coded in the same way) are from the list of single (un-blended) transitions used in this work (Appendix \ref{app: line list}). Slab models (top-right inset, Appendix \ref{app: slab models params}) are added with dashed lines on top of the data, their sum is shown as a grey shaded region. The ro-vibrational band at 5--8~$\mu$m is dominated by $v=1-0$ lines (all lines unless labelled), but there are some prominent $v=2-1$ lines that are detected (see Section \ref{sec: atlas}). }
\label{fig: WATER_atlas_ROVIB}
\end{figure*}

\begin{figure*}
\centering
\includegraphics[width=1\textwidth]{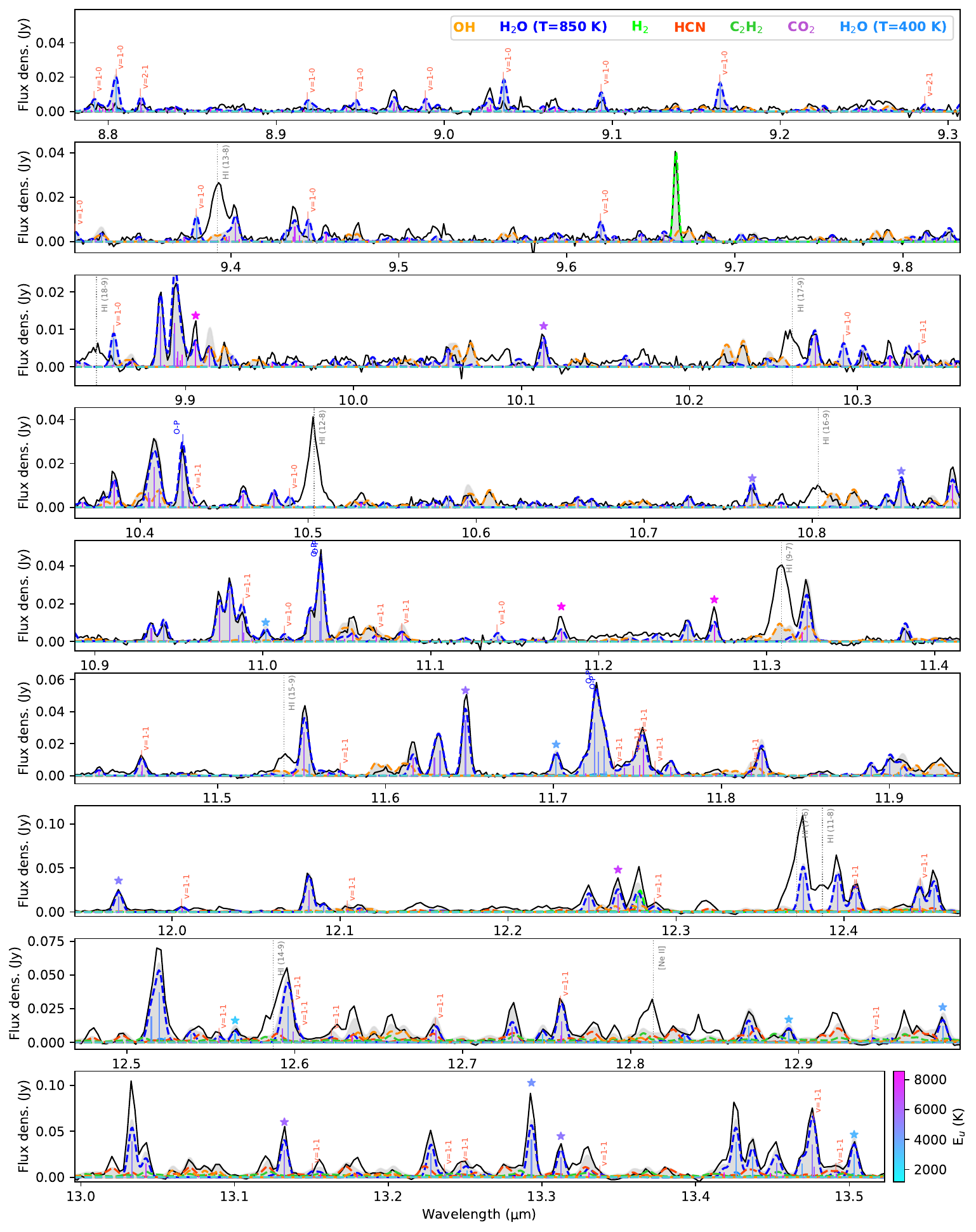}
\caption{Atlas of water emission in MIRI-MRS spectra (continued from Figure \ref{fig: WATER_atlas_ROVIB}): the rotational spectrum at intermediate wavelengths. The emission is dominated by $v=0-0$ lines (all lines unless labelled), but there are lines from the first vibrational state including several prominent $v=1-1$ lines (Section \ref{sec: v1-1lines}). Ortho-para line pairs are labelled ``O-P" (see Section \ref{sec: saturation}).}
\label{fig: WATER_atlas_ROTshort1}
\end{figure*}

\begin{figure*}
\centering
\includegraphics[width=1\textwidth]{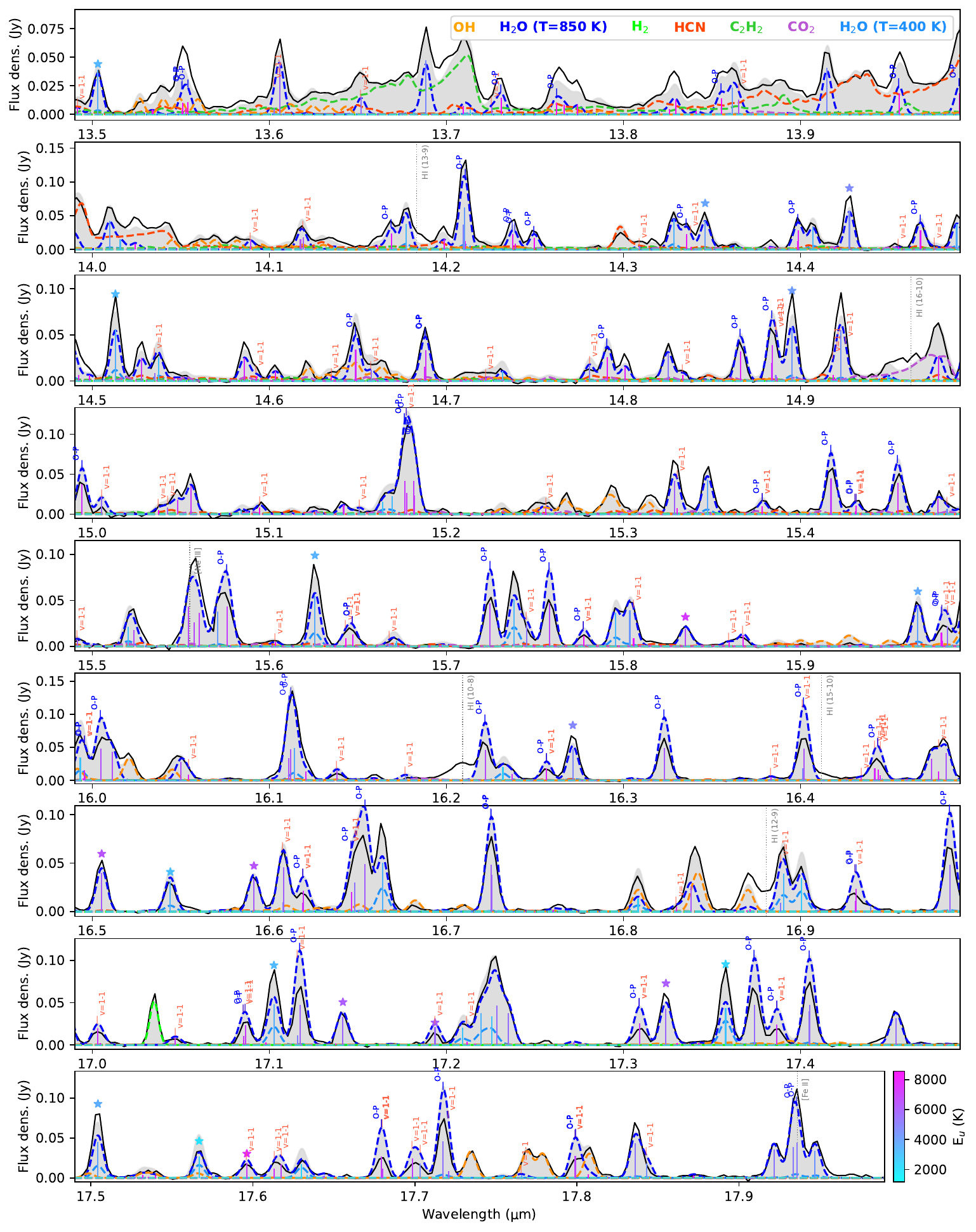}
\caption{Atlas of water emission in MIRI-MRS spectra (continued from Figure \ref{fig: WATER_atlas_ROTshort1}): the rotational spectrum at intermediate wavelengths. The emission is dominated by $v=0-0$ lines (unlabelled), but there are several prominent $v=1-1$ lines (Section \ref{sec: v1-1lines}). Ortho-para line pairs are labelled ``O-P" (see Section \ref{sec: saturation}).}
\label{fig: WATER_atlas_ROTshort2}
\end{figure*}

\begin{figure*}
\centering
\includegraphics[width=1\textwidth]{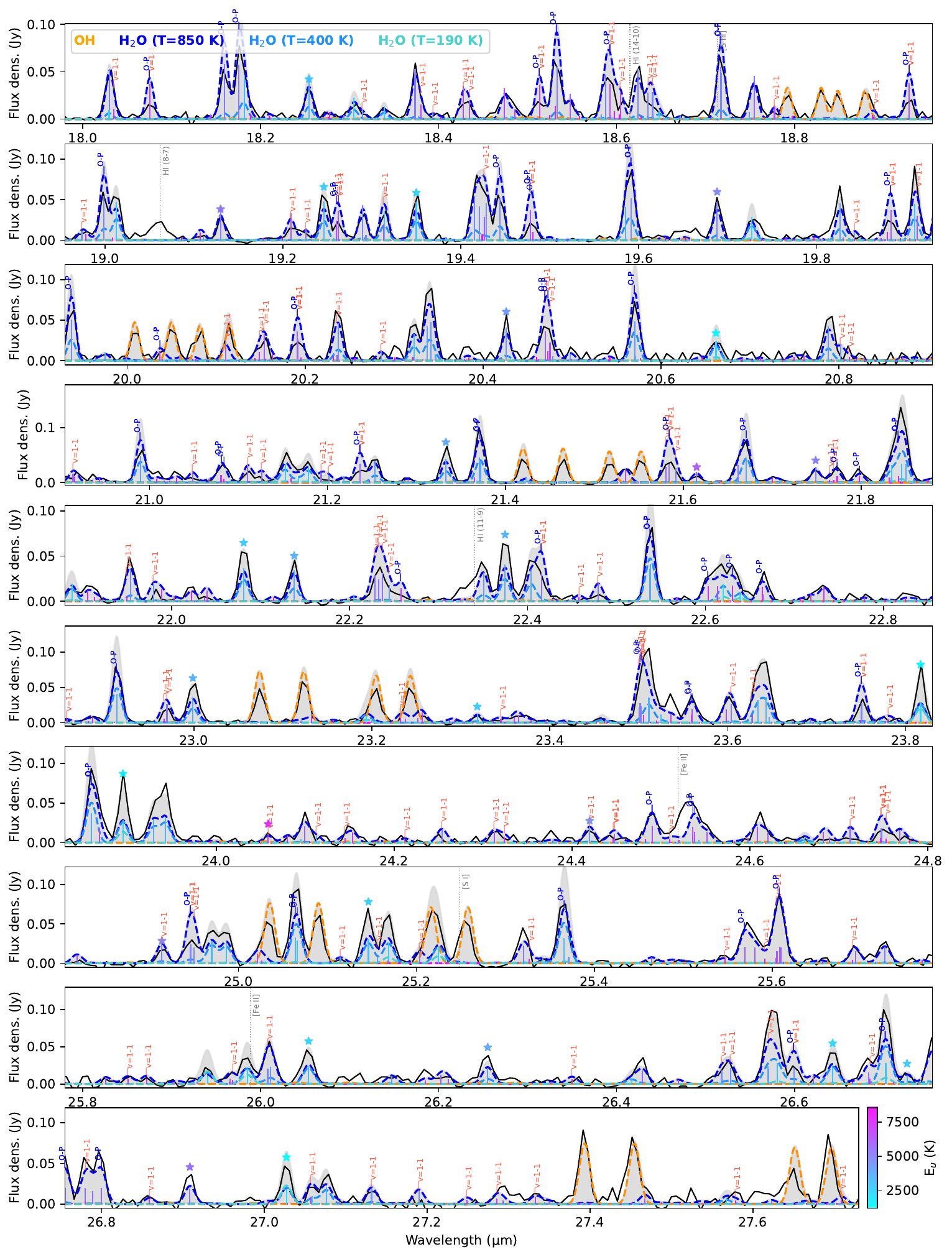}
\caption{Atlas of water emission in MIRI-MRS spectra (continued from Figure \ref{fig: WATER_atlas_ROTshort2}): the long MIRI wavelengths. A colder component of water emission ($\sim 190$~K) is prominent in two low-energy ($E_u \sim 1500$~K) lines near 23.85~$\mu$m (Section \ref{sec: cool excess}).}
\label{fig: WATER_atlas_ROTlong}
\end{figure*}

\section{A spectral atlas of water emission} \label{sec: atlas}
Figures \ref{fig: WATER_atlas_ROVIB} to \ref{fig: WATER_atlas_ROTlong} present an atlas of water emission as observed across MIRI spectra, using as an example the spectrum of CI~Tau from \cite{banz23b}. This spectrum is chosen for multiple reasons: its high S/N ($\approx 450$ around 17~$\mu$m), the detection of all molecules commonly found in disks (CO, \ce{H2O}, \ce{OH}, \ce{HCN}, \ce{C2H2}, \ce{CO2}, \ce{H2}), and the detection of a large number of atomic lines (dominated by H~I with $\sim30$ lines), which provide a useful template to illustrate contamination to water lines from other species. Moreover, CI~Tau has been used in previous work as a base reference to identify the presence of cooler water at larger disk radii from excess emission in low-energy lines, for being a spectrum that can be described almost entirely by a single hot component \citep{banz23b,munozromero24b}. This property also makes it a good example to isolate other excitation effects related to line opacity and non-LTE excitation, which will be demonstrated below in this section.

The atlas is made using the functionalities of iSLAT \citep{iSLAT, iSLAT_code} to compile and export simulated spectra for single-temperature slabs of gas in LTE \citep[defined by an excitation temperature $T$ in K, column density $N$ in cm$^{-2}$ and a slab emitting area $A_{slab} = \pi R_{slab}^{2}$ in au$^{2}$, e.g.][]{cn11,salyk11_spitz,banz12} for multiple molecules and visualize the individual transitions blended at the resolving power of MIRI, using line properties from HITRAN \citep{hitran20}. Each figure shows model spectra for emission at different temperatures: a ``hot" water component (850~K) and a ``warm" water component (400~K) as representative of typical temperatures found from slab fits to ground-based or MIRI spectra for a total sample of ten disks in previous work \citep{banz23,banz23b,munozromero24b,temmink24b}. 
While these temperature components have been found to describe well the rotational water emission across MIRI wavelengths (12--27~$\mu$m), with hotter emission generally dominating shorter wavelengths, an additional ``cold" $\sim$~170--200~K component, where detected, becomes prominent at longer MIRI wavelengths, in particular in two lines near 23.85~$\mu$m with upper level energies of $\approx 1500$~K \citep{kzhang13,banz23b,munozromero24b,temmink24b}. This component is of particular interest because it traces a temperature consistent with the ice sublimation/condensation front (the ``snowline") in inner disks \citep[120--180~K, e.g.][]{pollack94,sasselov00,lodders03}. The different spectral line flux distribution of these three components across infrared wavelengths provides a useful general tool for their identification in MIRI spectra, as will be demonstrated in Section \ref{sec: cool excess}.

Other molecules are plotted in different colors to illustrate where water emission is contaminated by other species (and vice versa), including atomic line (which are marked with dotted lines and labeled for identification). The adopted slab model parameters for each molecular model are reported in Appendix \ref{app: slab models params}; these are just representative models, not fits to the data, and include different ro-vibrational components for CO \citep{banz22} and two components of OH to approximately reproduce their non-LTE excitation, but not the OH asymmetry from prompt emission, though very visible in the data \citep[for detailed analyses on OH excitation in disks, see e.g.][]{carr14,tabone21,tabone24,zannese24}. In the case of CO, the highest $J$ level detected in CI~Tau is P~69 with $E_u \sim 16000$~K, at 5.544~$\mu$m.

\subsection{Using the atlas}
Figures \ref{fig: WATER_atlas_ROVIB} to \ref{fig: WATER_atlas_ROTlong} provide a general reference for several aspects that are helpful for the analysis of water spectra as observed with MIRI-MRS, which will be elaborated on below:
\begin{enumerate}
    \item the level of blending and confusion between different energy levels in each observed water line;
    \item water lines that are single, i.e. not blended (at the MRS resolution) with other water transitions from different levels at typical disk temperatures;
    \item the relative emission in higher- versus lower-$E_u$ across MIRI wavelengths, which is connected to the emission from different temperatures;
    \item contamination of water emission from other molecules and atoms;
    \item saturation from line opacity overlap and non-LTE excitation of higher vibrational levels ($v>0$).
\end{enumerate}

\subsubsection{Water transitions from different $E_u$} 
The first three points in the list can be visualized as follows. Under the water models in Figures \ref{fig: WATER_atlas_ROVIB} to \ref{fig: WATER_atlas_ROTlong}, the individual transitions that make the observed blended emission are plotted with vertical dashed lines, color-coded with a gradient from magenta (higher-$E_u$) to cyan (lower-$E_u$) to reflect the upper level energy. As in iSLAT, the height of these lines in the plot is proportional to their intensity, to visualize their relative contribution in each blend. To provide a maximum case of line blending, we visualize lines from the 850~K model; at lower temperatures, the higher-energy transitions are less excited and therefore more lines may be dominated by a single, lower-energy transition. However, in this sample we find that disk spectra generally have the high-energy lines significantly excited and detected aside from a few exceptions (see Appendix \ref{app: atlas_compact} and Salyk et al. submitted), therefore the spectral atlas figures presented here should provide a useful general guidance. 

From these figures, it is possible to obtain a quick idea of how many transitions contribute significantly to each observed line/blend, and whether higher- rather than lower-energy transitions are blended in a given observed line. 
In cases where line clustering is too dense and/or the coloring system is not sufficient to identify individual lines in Figures \ref{fig: WATER_atlas_ROVIB} to \ref{fig: WATER_atlas_ROTlong}, we recommend using iSLAT\footnote{Available at \url{https://github.com/spexod/iSLAT}.} to inspect interactively any line blends of choice for different temperature and column density values. With this procedure, we have identified a list of single (un-blended) water lines whose properties reflect the emission from individual upper levels (Section \ref{sec: line selection}); this list provides reliable measurements for a number of analysis geals, which will be demonstrated below.

\subsubsection{Contamination from different species} 
For the fourth point, the atlas provides quick identification of contamination from other molecules and atoms. For lines with lower contrast in the figure, we suggest the use of iSLAT for interactive visualization and inspection. Depending on their relative strength, most water lines between 12 and 16~$\mu$m are typically contaminated by lines from \ce{OH}, \ce{HCN}, \ce{C2H2}, and \ce{CO2}. For this reason, in this work we only use a few lines in this region, lines that in conditions found to be typical in this sample are dominated by water emission (Figures  \ref{fig: WATER_atlas_ROTshort1} and \ref{fig: WATER_atlas_ROTshort2}). Atomic lines, in particular from H~I, are scattered across MIRI wavelengths and also need to be carefully checked to avoid contamination. 

Vice versa, the atlas can be used to identify atomic emission lines that are the least contaminated by water or other molecules. In the case of HI, the most clean, strong lines are (upper-lower levels): 9-6 (5.91~$\mu$m), 14-8 (8.66~$\mu$m), 10-7 (8.76~$\mu$m), 12-8 (10.50~$\mu$m), 10-8 (16.21~$\mu$m), 8-7 (19.06~$\mu$m). These lines, when detected, can generally be measured from MIRI spectra even when water emission is present; measuring other HI lines requires subtracting a model for the contaminants first, usually water but sometimes OH or the organics. Other atomic species of common interest, including [NeII] and [NeIII], are all contaminated by water or other molecules and should generally be measured in water-subtracted spectra, unless water emission is absent or very weak. Examples of the water-subtraction process using Spitzer spectra can be found in \cite{salyk11_spitz,rigliaco15}, and with MIRI spectra in \cite{grant23}.

\begin{figure*}
\centering
\includegraphics[width=1\textwidth]{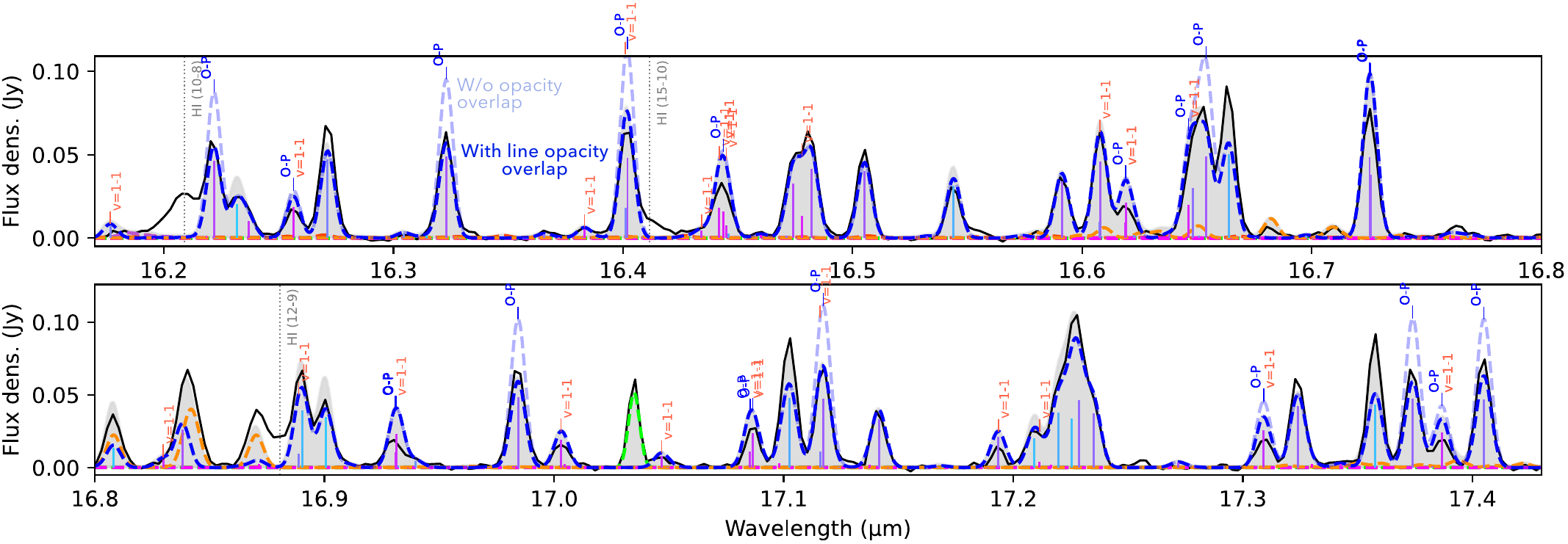}
\caption{Portion of Figure \ref{fig: WATER_atlas_ROTshort2} illustrating the effect of mutual line opacity saturation. The iSLAT hot water model from Figure \ref{fig: WATER_atlas_ROTshort2} is now shown in light blue, and a new model with the same parameters but including line opacity overlap is shown in bold blue \citep[using \texttt{spectools\_ir},][]{spectools_ir}. Accounting for line saturation correctly reproduces the reduced emission from the ortho-para line pairs, but the $v=1-1$ lines are still over-predicted because both models assume LTE (compare to Figure \ref{fig: WATER_atlas_ROTshort_ROVsuppr}).}
\label{fig: WATER_atlas_ROTshort_COLcomp}
\end{figure*}

\begin{figure*}
\centering
\includegraphics[width=1\textwidth]{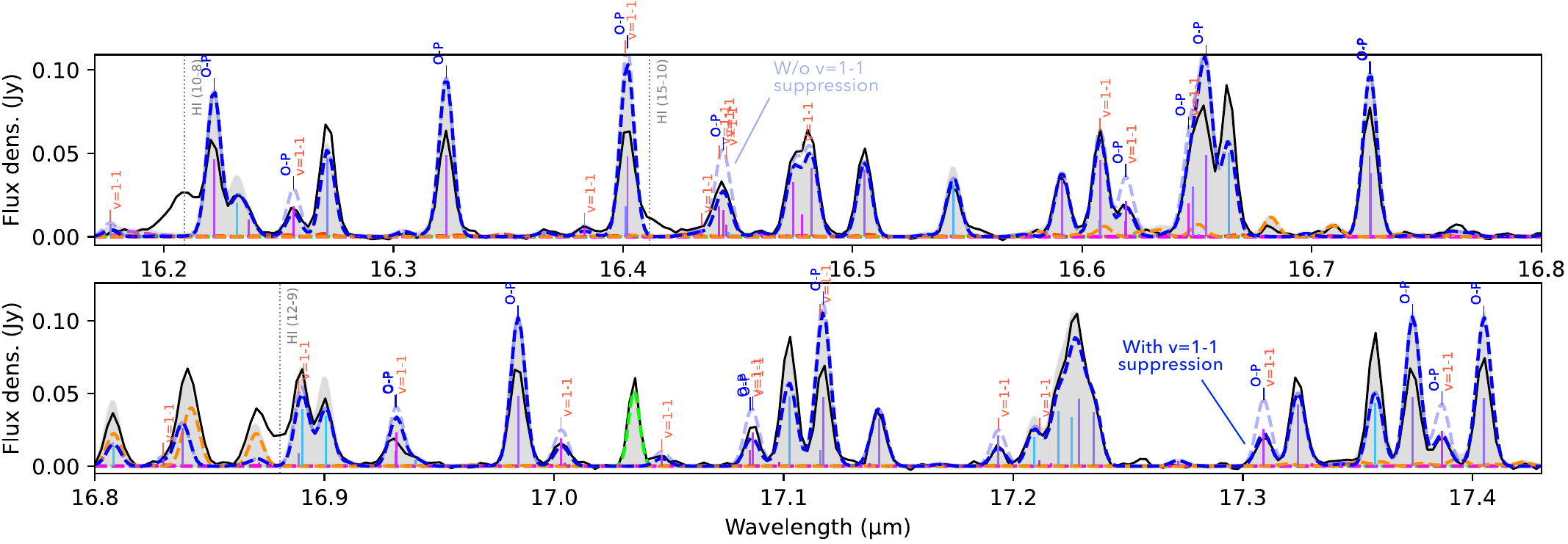}
\caption{Same as Figure \ref{fig: WATER_atlas_ROTshort_COLcomp}, but here illustrating the effect of non-LTE excitation. The iSLAT model is shown in light blue, and the same model with the $v=1-1$ lines divided by a factor of 2 is shown in bold blue.}
\label{fig: WATER_atlas_ROTshort_ROVsuppr}
\end{figure*}

\subsubsection{Mutual line opacity saturation} \label{sec: saturation}
For the fifth point in the list above, the atlas identifies transitions that are of special interest for understanding the excitation and observed flux of specific water lines. 
Previous work showed that parts of the infrared organic emission features at 13--14~$\mu$m may become highly optically thick and saturate due to the dense line clustering. This was observed in extreme high-column density conditions by \cite{tabone23} in the emission feature of \ce{C2H2} as observed with MIRI in the protoplanetary disk of a low-mass star. It has not been demonstrated yet where this effect may matter for infrared water spectra, instead, in spite of the fact that line opacity overlap has previously been included in modeling Spitzer spectra \citep{cn11,salyk11_spitz}. With the increased resolving power of MIRI-MRS, we can now better observe where line opacity overlap matters the most.
In the case of water, the mutual opacity saturation becomes relevant in several ortho-para (O-P) line pairs that share the same $E_u$ but have different statistical weight and overlap exactly or very closely in wavelength; in these cases, each line contributes to the opacity of the other line and the observed blended line saturates at lower values of the column density. This effect can be easily identified with iSLAT, which over-predicts the flux of all these line pairs by ignoring this mutual saturation effect. 

To demonstrate the opacity saturation effect in the case of water lines, we add in Figure \ref{fig: WATER_atlas_ROTshort_COLcomp} a new model that uses the same parameters as the iSLAT hot water model, but made with a code that includes mutual line opacity saturation \citep[\texttt{spectools-ir},][]{spectools_ir}. The simulated spectra from the two codes perfectly match as they should (since they are equivalent, otherwise) except for the orto-para line pairs, where \texttt{spectools-ir} provides a very good match to the data in contrast to the iSLAT model.
Therefore, line overlap should be accounted for when fitting water spectra, because many O-P lines pairs are close or coincident in wavelength, such that their fluxes do not simply add when the lines optical depth increases towards thick conditions. For details on existing code that implements this effect for water emission, see \texttt{spectools-ir}\footnote{Available at \url{https://github.com/csalyk/spectools_ir}.} \citep{spectools_ir} and \texttt{iris}\footnote{Available at \url{https://github.com/munozcar/IRIS}.} \citep{iris}.

\subsubsection{Non-LTE excitation of lines in $v=1$ and $v=2$} \label{sec: v1-1lines}
Another important effect included in the fifth point in the list above is related to non-LTE excitation.
Non-LTE excitation of infrared water emission has been discussed in detail in previous work \citep{meijerink09}, and recently proposed to explain the observed suppression of the ro-vibrational bands in comparison to the excitation of pure rotational lines \citep[see][]{bosman22,banz23}. The origin of this effect comes from the very different critical densities of water lines from different bands, of the order of $10^{13}$~cm$^{-3}$ for the ro-vibrational lines in comparison to $10^{8}-10^{11}$~cm$^{-3}$ for the $v=0-0$ rotational lines \citep[see summary in Table 1 and Figure 13 of][]{banz23}. This difference implies that rotational lines thermalize at lower gas densities, while ro-vibrational lines may be sub-thermally excited if the gas density of the emitting layers in inner disk does not reach $\sim 10^{13}$~cm$^{-3}$.

To simulate the sub-thermal excitation of ro-vibrational lines, in Figure \ref{fig: WATER_atlas_ROVIB} we used a factor 4 decrease in emission in comparison to the model of the rotational lines at longer wavelengths (see model parameters in Appendix \ref{app: slab models params}), which is enough to approximately reproduce the ro-vibrational band in CI~Tau. A similar factor was found to approximately reproduce the ro-vibrational band also in the case of FZ~Tau in \cite{pontoppidan24}. 
Even with this global reduction factor, $v=2-1$ lines are generally over-predicted in comparison to the $v=1-0$ lines (Figure \ref{fig: WATER_atlas_ROVIB}), pointing again to non-LTE excitation \citep[for this effect in the case of CO lines, see e.g.][]{banz22,ramireztannus23}. 
We remark that fits to higher-resolution (but much smaller spectral range) spectra from the ground suggested slightly higher temperatures of $\approx 1000$~K for the ro-vibrational water band near 5~$\mu$m \citep{banz23}, which is dominated by higher energy levels (4500--9500~K) and therefore it is more sensitive to gas at higher temperature. However, these higher-energy lines are much weaker and generally blended at the resolution of MIRI, where strong un-blended lines are dominated by lower-energy levels (Figure \ref{fig: WATER_atlas_ROVIB}). We also remark that the suppression factor needed by the $v=1-0$ lines in comparison to the rotational lines cannot be explained simply by a smaller emitting area \citep{munozromero24b}, which is nonetheless demonstrated by velocity-resolved line widths from ground-based surveys \citep{banz23} and now also from MIRI spectra (Section \ref{sec: broadening}).
 
The excitation mismatch between different vibrational levels is also very visible in the $v=1-1$ lines, as resolved by MIRI-MRS and shown in this work for the first time. There is a large number of strong $v=1-1$ lines intermixed to the $v=0-0$ lines in the main rotational emission region at longer wavelengths ($> 10$~$\mu$m, Figures \ref{fig: WATER_atlas_ROTshort2} and \ref{fig: WATER_atlas_ROTlong}), and these are over-predicted by fits to the $v=0-0$ lines as expected in case of sub-thermal excitation of higher vibrational bands \citep{meijerink09}.  
It should be noted that the $v=1-1$ lines are still over-predicted by the \texttt{spectools-ir} model in Figure \ref{fig: WATER_atlas_ROTshort_COLcomp}, since it still assumes LTE. This is not an issue of the single-temperature approximation: the over-prediction of $v=1-1$ lines still happens even when modeling the water spectrum as the sum of two temperatures or as a temperature gradient \citep[by inspecting figures reporting the best-fit models, the over-prediction of some of these lines can in fact be recognized in][]{pontoppidan24,munozromero24b,temmink24b}.

\begin{figure}
\centering
\includegraphics[width=0.4\textwidth]{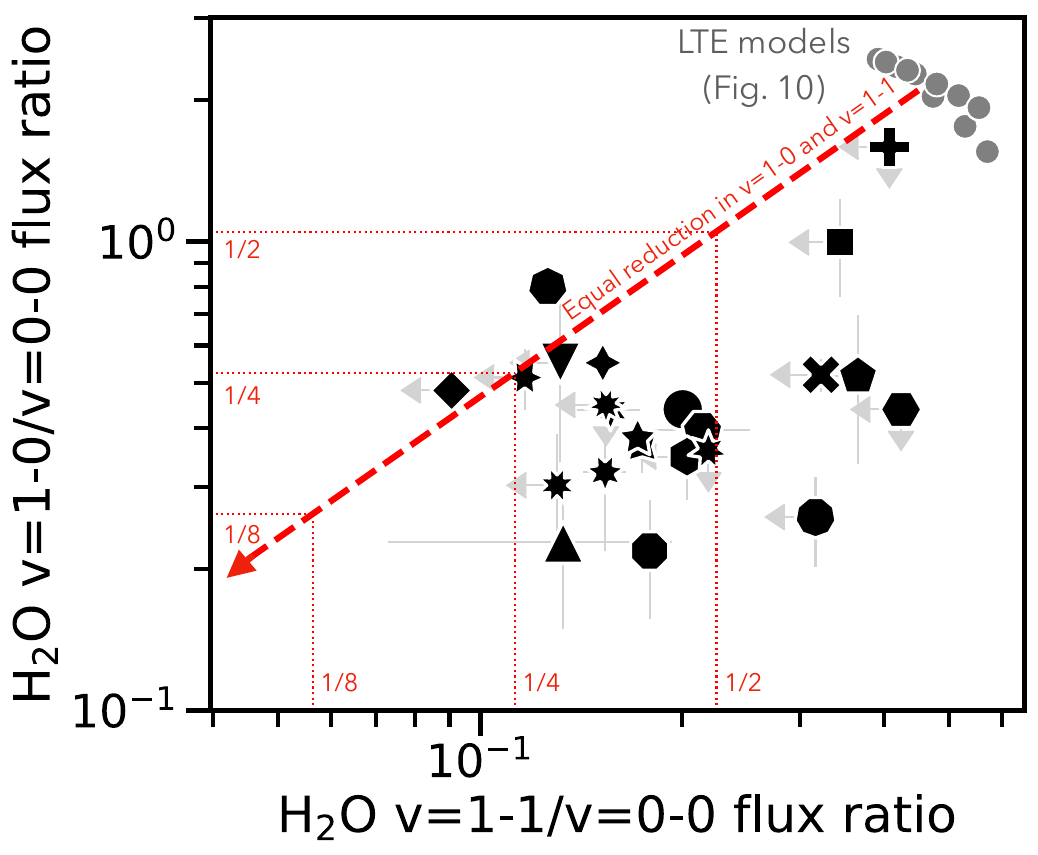}
\caption{\rev{Measured line flux ratios of a $v=1-0$ and a $v=1-1$ line with a $v=0-0$ line (see Section \ref{sec: v1-1lines}). All lines from the $v=1$ level are significantly weaker than what they should be in LTE, as found previously from ground-based spectra \citep{banz23}. The red lines with labeled fractions mark suppression factors in reference to a median LTE prediction (models and sample symbols are the same as used in Section \ref{sec: cool excess}). In case of non detections, 2-$\sigma$ limits are marked with arrows.} }
\label{fig: nonLTE_diagram}
\end{figure}

As a simple test, we show in Figure \ref{fig: WATER_atlas_ROTshort_ROVsuppr} that these lines, similarly to the $v=1-0$ band at shorter wavelengths, match the observed data reasonably well with the simple suppression of their flux by a single factor common to all lines. In the case of CI~Tau, for the $v=1-1$ lines this factor is $\sim 2$. \rev{For reference, Figure \ref{fig: nonLTE_diagram} shows the measured flux ratios in a $v=1-0$ line (8.0696~$\mu$m) and a $v=1-1$ line (24.91403~$\mu$m) with a $v=0-0$ line (16.27136~$\mu$m), all with similar upper level energy of $\approx 4800$~K and $A_{ul} \approx 10$~s$^{-1}$. In comparison to LTE models used in this work (Section \ref{sec: cool excess}), the measured line fluxes cluster around a suppression of $\sim 1/6$ for the $v=1-0$ lines and $\sim 1/3$ for the $v=1-1$ lines. The greater suppression in $v=1-0$ lines is consistent with them being further away from LTE due to their higher critical densities ($\approx 10^{13}$~cm$^{-3}$), requiring denser gas for thermalization. Based on this argument, and if the $v=0-0$ lines at 12--28~$\mu$m with lower critical density ($\approx 10^{8}-10^{10}$~cm$^{-3}$) are instead in LTE, the density of the water-emitting region should be $> 10^{10}$~cm$^{-3}$ and $<< 10^{13}$~cm$^{-3}$ but probably different across the different disk radii that emit the different water lines (see Sections \ref{sec: cool excess} and \ref{sec: kinematics}). }
A detailed analysis of the relative excitation of $v=1-0$ and $v=1-1$ lines to derive more specific gas density estimates is deferred to future work.

\subsection{List of single un-blended lines used in this work} \label{sec: line selection}
By applying all the criteria described above, we selected a list of un-blended water lines for the analysis presented in this work. Each line in the list is from a single upper level and avoids contamination from other species as observed with MIRI-MRS. We stress that, depending on the specific spectrum and the analysis goals and methods, other lines can certainly be measured and used in MIRI spectra; here we aim at providing a reliable line list that can be measured directly from the spectra for the typical conditions found in this sample, without subtracting other molecular models or attempting to de-blend lines first. 
To obtain more reliable data for line broadening measurements, in comparison to the lists of single lines originally provided in iSLAT \citep{iSLAT} we have excluded lines that are significantly blended on their wings (Section \ref{sec: broadening}).
The full line list used in this work is marked with a star in Figures \ref{fig: WATER_atlas_ROVIB} to \ref{fig: WATER_atlas_ROTlong} and reported in Appendix \ref{app: line list}.

\begin{figure*}
\centering
\includegraphics[width=1\textwidth]{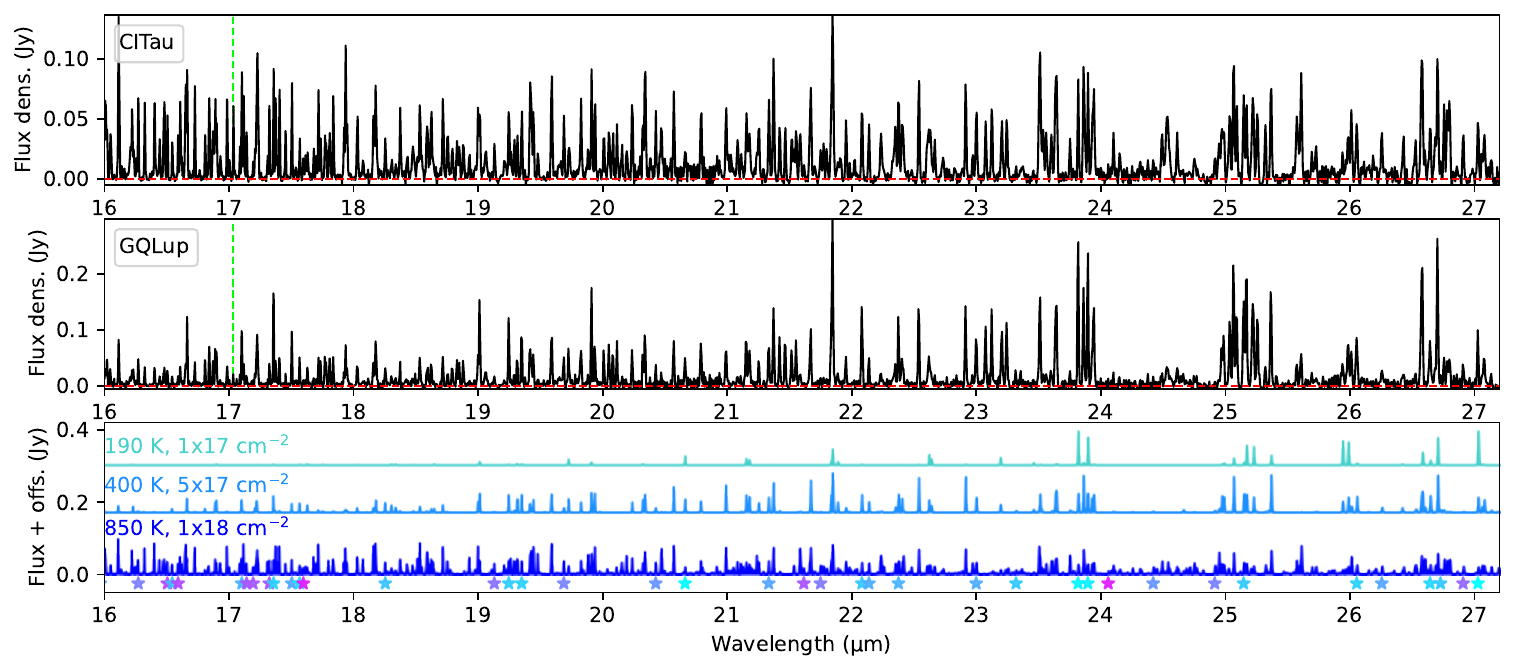}
\caption{Representative water spectra for a hot-dominated (CI~Tau) and cool-dominated (GQ~Lup) disk are shown in comparison to temperature-component models (bottom). The different spectral line flux distribution is very visible, with the strongest lines having similar flux across MIRI wavelengths in the case of a hot-dominated spectrum, versus increasingly stronger lines at longer wavelengths when emission from cooler water is present. The full sample is shown in Appendix \ref{app: atlas_compact}. The \ce{H2} transition near 17~$\mu$m is marked with a green dashed line.}
\label{fig: MIRI_spectra_atlas_RED}
\end{figure*}

A subset of this larger line list that plays a central role in our analysis is between 14.4 and 17.6~$\mu$m (Figure \ref{fig: WATER_atlas_ROTshort2}). These lines will be used for excitation and broadening diagnostics as explained in Sections \ref{sec: cool excess} and \ref{sec: kinematics}. The advantages of using lines in this spectral range are multiple: they are in one of the highest-S/N and highest resolving power parts of the MIRI spectrum \citep{pontoppidan24}, \rev{they cover the entire range in upper level energy provided by the global list of single lines (and therefore are sensitive to the entire range of emitting temperatures except for the cold component, see below Section \ref{sec: line diagnostic})}, they are stronger than lines at shorter wavelengths and are free from contamination from organics, and they are close to the \ce{H2} S1 line at 17.04~$\mu$m, which provides a useful anchor point for the MIRI resolving power (see Section \ref{sec: broadening}). In cases where organic emission is particularly strong relative to water emission, water lines in this list at 12--16~$\mu$m may be significantly contaminated by HCN, \ce{C2H2}, or \ce{CO2}; in this sample, this is the case for DoAr~25 (with HCN), GO~Tau (with \ce{C2H2}), and MY~Lup (with \ce{CO2}).

\begin{figure*}
\centering
\includegraphics[width=0.8\textwidth]{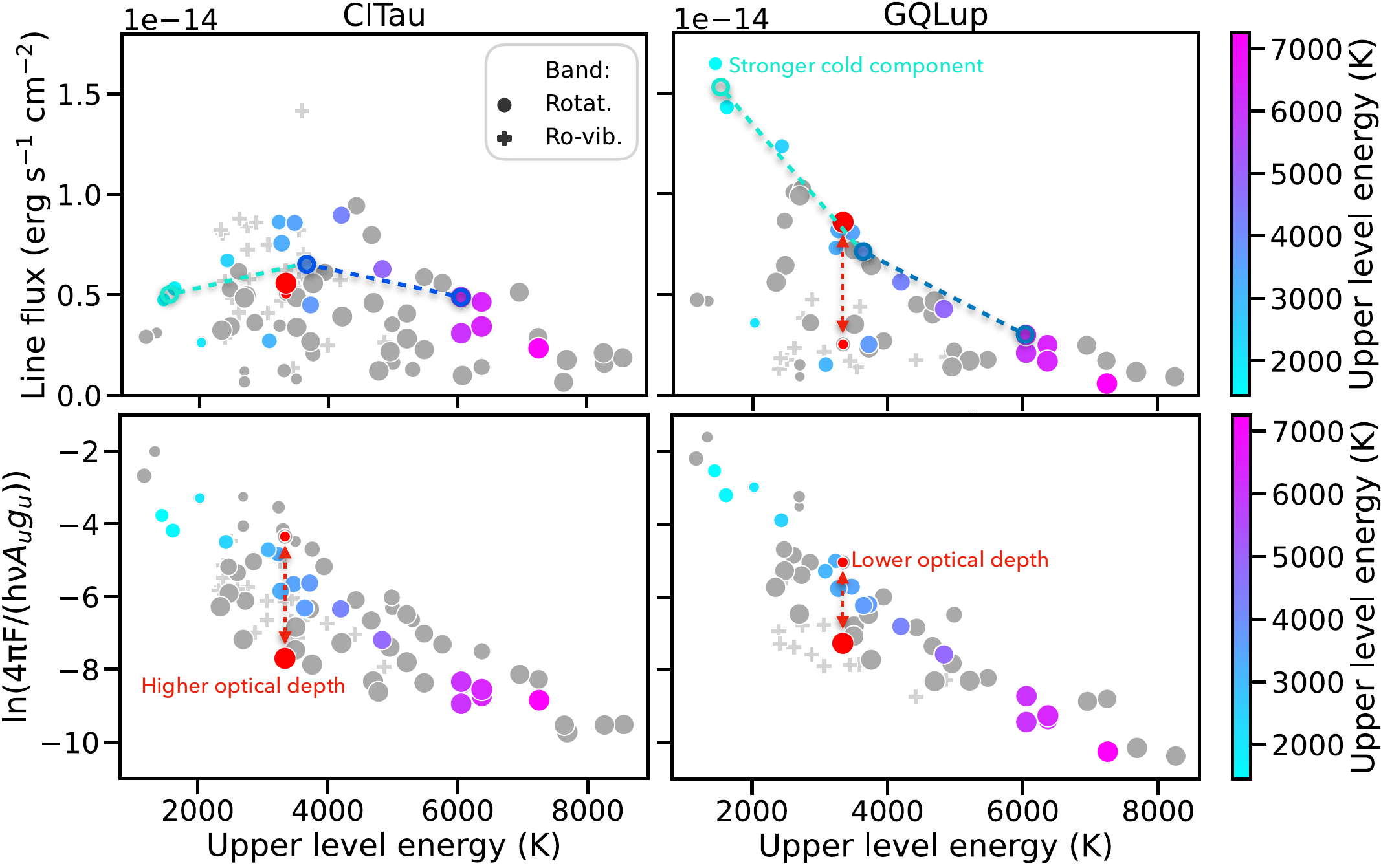}
\caption{Demonstration of line selection for definitions of line ratios used in Figure \ref{fig: cool_excess}, showing line fluxes (top) and the rotation diagram (bottom) as a function of the upper level energy of individual water transitions. The very different spectral line flux distribution of the hot-dominated CI~Tau and cool-dominated GQ~Lup shown in Figure \ref{fig: MIRI_spectra_atlas_RED}, here shown as a function of $E_u$, is once again very visible. Grey datapoints are all lines used in this work (Appendix \ref{app: line list}), colored datapoints are the lines at 14--18~$\mu$m explained in Section \ref{sec: line selection} plus the two lower-energy lines near 23.85~$\mu$m. The line ratios at 1500, 3600, and 6000 K reflect the relative flux at different temperatures. The line ratio at 3340 K between two lines with very different $A_{ul}$ (shown proportional to the symbol size) reflects the optical depth. Rotation diagrams for the whole sample are shown in Figure \ref{fig: RotDiagr_sample} in Appendix \ref{app: sample_plot_grids}.} 
\label{fig: Line_selection_demo}
\end{figure*}

\section{Definition of line-ratio diagnostic diagrams for MIRI-MRS water spectra} \label{sec: cool excess}
With the list of single lines selected in Section \ref{sec: atlas}, we can now proceed to an update and generalization of the distribution of water emission from different temperatures introduced in previous work. \rev{This section demonstrates that by using a few, carefully selected line ratios it is possible to obtain a simple, model-independent view of the relative emission from water at different temperatures in inner disks before performing time-consuming fits with sophisticated models.}
For measuring emission line properties from the spectra we use iSLAT, which implements the least-square minimization code \texttt{lmfit} \citep{lmfit} to perform single-Gaussian fits and measure the line centroid, full width at half maximum (FWHM), flux, and their uncertainties. 

When fitting for two temperatures as well as a temperature gradient, previous work found CI~Tau to be the disk most dominated by a hot water component, while GQ~Lup to be one with strong emission from additional cooler water at larger radii \citep{munozromero24b}. We show these two extreme cases in Figure \ref{fig: MIRI_spectra_atlas_RED} to illustrate how the temperature distribution in a MIRI spectrum can be visible by eye from the spectral line flux distribution, i.e. the distribution of line flux as a function of wavelength. This distribution is rather flat with wavelength (i.e. the strongest lines have similar flux across MIRI wavelengths) in the case of a hot-dominated spectrum (CI~Tau), while it shows increasingly stronger lines at longer wavelengths when emission from cooler water is present (GQ~Lup).
This different spectral line flux distribution is exemplified at the bottom of the figure using slab models at three temperatures representing properties found in previous work as explained in Section \ref{sec: atlas} \citep[assuming a cold component at 190~K, to represent results from fits to MIRI spectra,][]{banz23b,munozromero24b,temmink24b}, for reference. 

Leveraging the different spectral line flux distribution of different temperature components, previous work defined an empirical ``cool water excess" by using the hot-dominated CI~Tau spectrum as a comparative template for other disks, and by measuring excess emission in the lower-energy transitions from a large number of single lines measured across MIRI wavelengths \citep{banz23b}. In this work, we develop a simpler diagnostic that can be applied broadly to future analyses of disk samples observed with MIRI. We use: i) a hot water model as the base to define the minimum flux in low-energy lines in absence of emission from cooler water, and ii) a curated, short list of emission lines from single transitions spanning $E_{u}$ between 1400 and 6000 K that maximize data quality in terms of S/N and resolving power, absence of residual fringes, and absence of contamination from other species in typical emission conditions as observed in this work's sample.

\subsection{Diagnostic lines for temperature and density} \label{sec: line diagnostic}
\begin{figure*}
\centering
\includegraphics[width=1\textwidth]{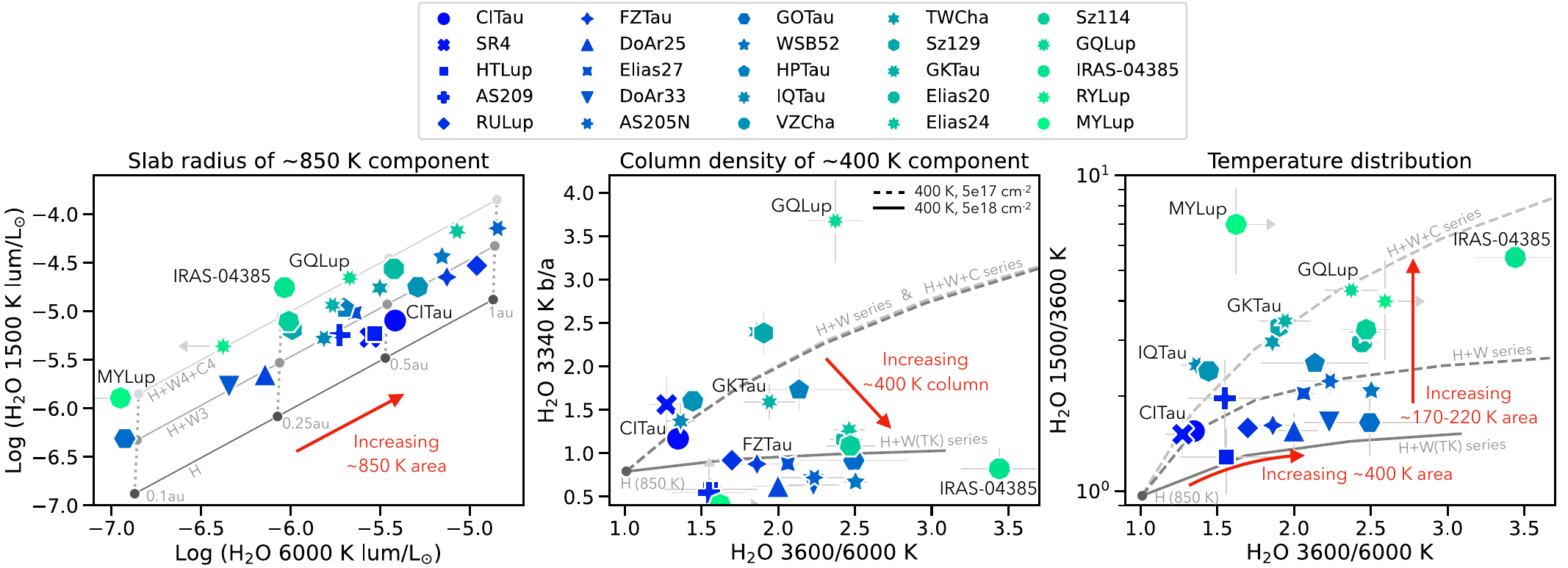}
\caption{\rev{Definition of water diagnostic diagrams using the lines selected in Table \ref{tab: cool excess line list}. In case of non detections, 2-$\sigma$ limits are marked with arrows. Some targets mentioned in the text are labeled for easier visualization. The distribution of the sample in these plots is shown in reference to multiple series of slab models that are fully explained in Appendix \ref{app: slab models params} and Figure \ref{fig: diagnostic_diagram_models}. \textit{Left:} The 6000~K line luminosity reflects the emitting area of a hot component (H model), while the vertical spread in the 1500~K line luminosity reflects a range of emitting areas for colder water (W and C models). \textit{Middle:} The 3340~K line ratio is sensitive to the column density of the warm component, as shown with a dashed line (lower column, W models) and solid line (higher column, W(TK) models), demonstrating that it is typically optically thick in this sample (line ratio $\approx 1$). \textit{Right:} The 3600/6000~K and 1500/3600~K line ratios reflect a range of emitting areas for the warm and cold components respectively as shown by the red arrows. The temperature distribution in terms of radial gradients is discussed in Section \ref{sec: discussion_radial}, Figure \ref{fig: diagnostic_diagram_gradients}.}}
\label{fig: cool_excess}
\end{figure*}

In Figures \ref{fig: Line_selection_demo} and \ref{fig: cool_excess}, we present the selected lines and demonstrate their use as proxies for temperature distribution and column density of the water spectra. These lines are reported in Table \ref{tab: cool excess line list}.
From the main list of single lines defined above in Section \ref{sec: atlas}, three are selected to span the entire range of $E_u$ covered by MIRI at the two extremes (near 1500~K and 6000~K, avoiding lines at higher $E_u$ that are typically weaker and more likely to be affected by non-LTE excitation), and at intermediate $E_u$ (near 3600~K). These transitions provide three flux ratios that capture the relative emission from different temperature components: the line flux ratio between transitions from 3600 and 6000~K (capturing the relative strength of a ``warm" water component, here taken at $\approx 400$~K), and the line flux ratio between transitions from 1500 and 3600~K or 6000~K (for the relative strength of a ``cold" water component, here taken at $\approx 190$~K).

While the low-energy lines near 23.85~$\mu$m were already identified as good tracers of sublimation temperatures ($\sim 150$--180~K) water in previous work \citep{kzhang13,banz23b}, the two lines we use here to trace water at higher temperatures are selected from the 17--17.6~$\mu$m range, which achieves the highest S/N and a high resolving power at MIRI wavelengths (Section \ref{sec: broadening}).
The line selection from different options with similar $E_u$ was made such that the 850~K reference model adopted in this work, labeled with ``H" in Figure \ref{fig: cool_excess}, has all line ratios with a value of $\sim 1$, making it a convenient, easy reference. For this reason, the 1500~K line flux used in the diagnostic diagrams in this work is taken as the sum of the two transitions in Table \ref{tab: cool excess line list}.

\begin{deluxetable}{l l c c}
\tabletypesize{\small}
\tablewidth{0pt}
\tablecaption{\label{tab: cool excess line list} List of \ce{H2O} transitions used as diagnostics.}
\tablehead{\colhead{Wavelength} & \colhead{Transitions (upper-lower)} & \colhead{$A_{ul}$} & \colhead{$E_{u}$} \\
\colhead{($\mu$m)} & \colhead{(format: $v_1 v_2 v_3~~J_{\:K_a \:K_c}$)} & \colhead{(s$^{-1}$)} & \colhead{(K)} }
\tablecolumns{4}
\startdata
\multicolumn{4}{l}{\textbf{Label:  1500~K (temper. diagnostic - Figs. \ref{fig: cool_excess}, \ref{fig: 23um_diagram})}} \\
23.81676 & 000-000 \: $8_{\:3\:6} - 7_{\:0\:7}$ & 0.61 & 1448\\
23.89518 & 000-000 \: $8_{\:4\:5} - 7_{\:1\:6}$ & 1.04 & 1615\\
\hline
\multicolumn{4}{l}{\textbf{Label: 3600~K (temperature diagnostic - Fig. \ref{fig: cool_excess})}}  \\
17.50436 & 000-000 \: $13_{\:4\:9} - 12_{\:3\:10}$ & 4.94 & 3646 \\
\hline
\multicolumn{4}{l}{\textbf{Label: 6000~K (temperature diagnostic - Fig. \ref{fig: cool_excess})}}  \\
17.32395 & 000-000 \: $16_{\:8\:9} - 15_{\:7\:8}$ & 41.5 & 6052 \\
\hline
\multicolumn{4}{l}{\textbf{Label: 3340~K (density diagnostic - Fig. \ref{fig: cool_excess})}}  \\
13.50312 (a) & 000-000 \: $11_{\:7\:4} - 10_{\:4\:7}$ & 0.49 & 3341 \\
22.37473 (b) & 000-000 \: $11_{\:7\:4} - 10_{\:6\:5}$ & 25.3 & 3341 \\
\hline
\multicolumn{4}{l}{\textbf{Non-LTE diagnostics - Fig. \ref{fig: nonLTE_diagram}}}  \\
8.0696 & 010-000 \: $10_{\:5\:6} - 11_{\:6\:5}$ & 5.95 & 4868 \\
16.27136 & 000-000 \: $15_{\:5\:10} - 14_{\:4\:11}$ & 9.23 & 4835 \\
24.91403 & 010-010 \: $9_{\:6\:3} - 8_{\:5\:4}$ & 18.6 & 4778 \\
\hline
\enddata
\tablecomments{For the 1500~K lines, in Figure \ref{fig: cool_excess} we take the combined flux of the two lines listed in this table, as explained in Section \ref{sec: cool excess}. Lines are identified and inspected using iSLAT \citep{iSLAT,iSLAT_code}. Line properties are from HITRAN \citep{hitran20}. The full line list used for other parts of the analysis in this work is reported in Appendix \ref{app: line list}.}
\end{deluxetable}

Additionally, we identify two lines from the same upper level with $E_u$ = 3341~K and statistical weight of 69 that have very different Einstein-$A$ coefficients ($A_{ul}$): 0.49 and 25.3 s$^{-1}$. This line pair traces the spectral line flux distribution in MIRI spectra near the peak of maximum spread in the rotation diagram (Figure \ref{fig: Line_selection_demo}) and it is one of the few transitions with largely different $A_{ul}$ from the same upper level we could find among single un-blended lines that are strong enough to be typically detected in disks. Such transitions can in principle provide useful measurements to help estimating the column density, as recently discussed in \citet{gasman23} where a selection of blended lines was used as an approximation. In fact, in optically thin conditions the lower-$A_{ul}$ transition from the same level will be weaker, while in optically thick conditions the two transitions will have similar flux, and their flux ratio (defined in this work as the higher-$A_{ul}$ line, labeled ``b", divided by the lower-$A_{ul}$ line, labeled ``a") will reflect these two regimes. The hot model in Figure \ref{fig: cool_excess} is optically thick and this line ratio is in fact close to unity. As described in detail in \cite{banz23}, a lower column density in the rotation diagram will be observed as a lower spread in the data, while the opposite will be seen with a larger column density. These different regimes are well illustrated by the two examples included in Figure \ref{fig: Line_selection_demo}. A caveat to keep in mind for the more optically thin of these lines, that at 13.50~$\mu$m, is that it is contaminated in case of strong organic emission relative to water, as note above for DoAr~25, GO~Tau, and MY~Lup.

\subsection{Definition of general water diagnostic diagrams}
\rev{The diagnostic line flux ratios are shown in Figure \ref{fig: cool_excess} as measured across the entire disk sample included in this work. The spectra of CI~Tau and SR~4 lie closest to the hot model (specifically with a slab radius of $\sim 0.5$~au, left panel in Figure \ref{fig: cool_excess}) as expected from visually inspecting their spectra, since both have a flat spectral line flux distribution (Figure \ref{fig: MIRI_spectra_atlas_RED} and Appendix \ref{app: atlas_compact}). 
The rest of the sample shows a wide range of diagnostic line luminosities (a factor of $\approx 100$) and ratios (factors of $\approx$~1--10) that can be interpreted in the framework of sequential combinations of discrete LTE components (the model series are fully explained and reported in Appendix \ref{app: slab models params} and Figure \ref{fig: diagnostic_diagram_models}) as an approximation of radial gradients (Section \ref{sec: discussion_radial}). Each model track in Figure \ref{fig: cool_excess} corresponds to the 850~K model (H) plus a sequentially increasing emitting area of the 400~K component (H+W model series) and of a 190~K component (H+W+C model series). All disks in this sample require some emission from the warm component as indicated by the 3600/6000~K line ratio (middle and right panels), from about $2\times$ the slab radius of the hot component (e.g. CI~Tau) to about $6\times$ that radius (the end of the model track in the figure). }

\rev{The 3340~K line ratio, sensitive to the column density of the 400~K component, indicates that water emission in this sample is moderately to highly optically thick (the datapoints cover the region between the dashed and solid lines in the middle panel, showing column densities from $5 \times 10^{17}$~cm$^{-2}$ to $5 \times 10^{18}$~cm$^{-2}$). As noted above, in disks with strong organic emission relative to water this line ratio is contaminated and shows values as low as $\sim 0.5$, as measured in DoAr~25 and MY~Lup.}

\rev{Emission at colder temperatures becomes prominent in the 1500~K lines and their ratio with the 3600~K and 6000~K lines. The 1500/3600~K line ratio in the right panel in Figure \ref{fig: cool_excess} shows that while about 50\% of the sample lies between the H+W model tracks, consistent with a hot component plus a warm water component, the other 50\% requires additional emission from colder water. The H+W+C model track shown in the figure to approximately align with the datapoints with the highest measured 1500/3600~K ratios shows areas of the cold component between $2\times$ the slab radius of the warm component to about $6\times$ that radius (the end of the model track in the figure), as with the H+W model series above.}

\begin{figure*}
\centering
\includegraphics[width=1\textwidth]{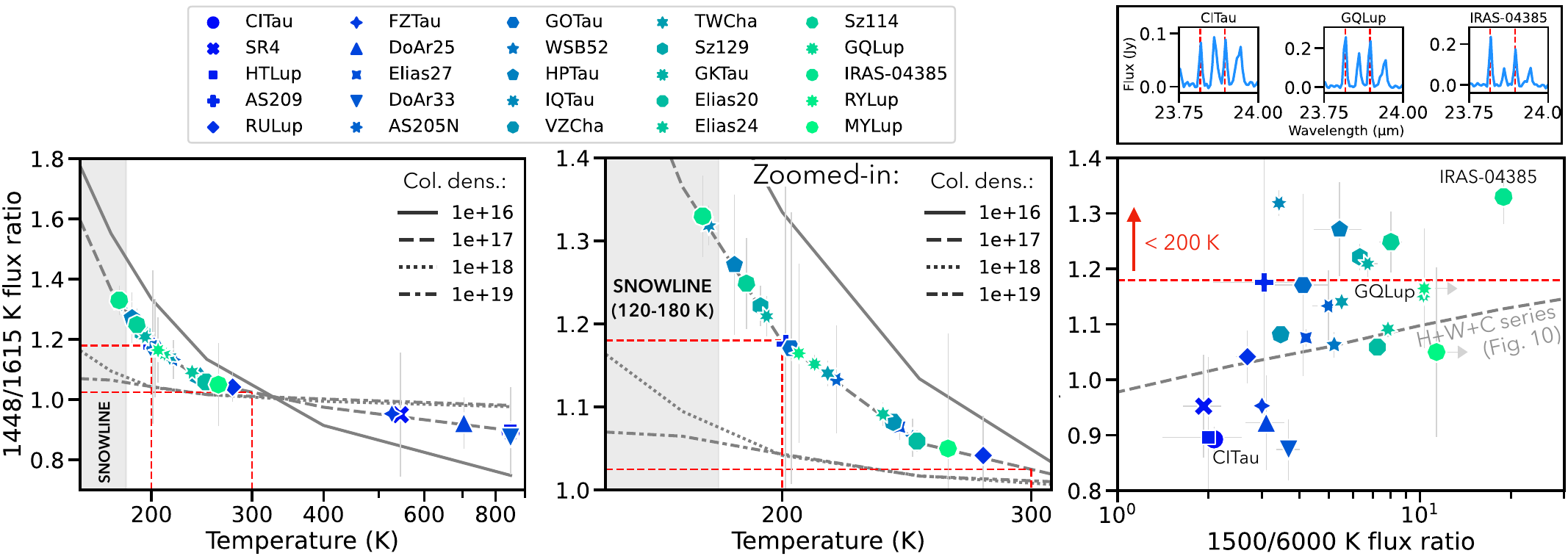}
\caption{\rev{Diagnostic diagram for the coldest water detected at MIRI wavelengths, using the temperature-dependent flux ratio between the two low-energy lines near 23.85~$\mu$m illustrated in three examples in the top-right inset (for their properties, see Table \ref{tab: cool excess line list}). Slab models in LTE are shown with grey lines over a range of column densities between $10^{16}$~cm$^{-2}$ (giving optically thin emission) and $10^{19}$~cm$^{-2}$ (giving optically thick emission), as show in the legend. The sample analyzed in this work is distributed using to the measured line flux ratios along the $10^{17}$~cm$^{-2}$ model, assuming it as an average column for the cold reservoir. The range of snowline temperatures from \citet{lodders03} is shown for reference. The plot to the right shows that disks with larger 1500/6000~K ratio also have larger asymmetry in the 23.85~$\mu$m lines, indicative of prominent colder water.}}
\label{fig: 23um_diagram}
\end{figure*}

\rev{The position of a given disk in the diagnostic diagrams can therefore be used to obtain an estimate of the slab radii, column density, and relative emitting areas of different temperature components that are found from slab fits to their spectra, as well as for comparative analyses of different disks. For instance, both CI~Tau and FZ~Tau are dominated by a hot component with some contribution from warm water, as found in previous fits \citep{pontoppidan24,munozromero24b}. Relative to CI~Tau, the position of FZ~Tau in Figure \ref{fig: cool_excess} indicates a more extended and more optically thick warm component, consistent with the results from radial-gradient fits to the line fluxes \citep{munozromero24b}.
As for disks requiring emission from colder water, three disks that align with the H+W+C track in the diagram (GK~Tau, GQ~Lup, and IQ~Tau), have indeed been found to have water emission down to sublimation temperatures near the snowline \citep[$\sim 170$--200~K, ][]{banz23b,munozromero24b}. The line ratios in GQ~Lup show more extended 400~K and 190~K components in comparison to GK~Tau, consistent with the more extended cold water area found by fits \citep{munozromero24b}. In the larger sample included in this work, we now find a disk with an even stronger cold water component: IRAS~04385+2550, a more embedded disk in Taurus associated with the Herbig-Haro object HH~408 \citep{stapelfeldt99,schaefer09,bally12} that might be in an earlier phase of strong pebble drift.
Cases where the hot water emission is strongly reduced or absent and the observed spectrum can be reproduced mostly by a warm and/or cold water component may provide only lower limits in some line flux ratios, such as is found in MY~Lup that is well reproduced by a single $\sim 300$--400~K component (Salyk et al. submitted). }

The coldest emission observed with MIRI has been modeled in a few disks in previous work, finding temperatures of 170--200~K \citep{banz23b,munozromero24b,temmink24b}. \rev{Here, we introduce a simpler diagnostic for the coldest temperature that is detected with MIRI: the flux asymmetry between the two 1500~K transitions, where cooler temperatures produce stronger emission in the 1448~K line at 23.867$\mu$m. In Figure \ref{fig: 23um_diagram}, we demonstrate this new diagnostic by taking a grid of slab models between a temperature of 150~K and 850~K and a column density between $10^{16}$~cm$^{-2}$ (giving optically thin emission) and $10^{19}$~cm$^{-2}$ (giving optically thick emission), spanning the entire range of conditions adopted above in this work. We simulate their spectra as for the models shown in Figure \ref{fig: WATER_atlas_ROTlong} and measure the line flux of these two transitions over the same range as done for the data. We then plot the model line ratio as a function of the slab temperature for the different column density curves in Figure \ref{fig: 23um_diagram}. This model grid illustrates that temperatures above 320~K produce a stronger 1615~K line, or an equal flux in the two lines (a ratio of $\sim 1$) in case of optically thick emission. At temperatures $< 320$~K, the 1448~K line becomes increasingly stronger, especially for lower values of column density (optically thin emission).} 

\begin{figure*}
\centering
\includegraphics[width=1\textwidth]{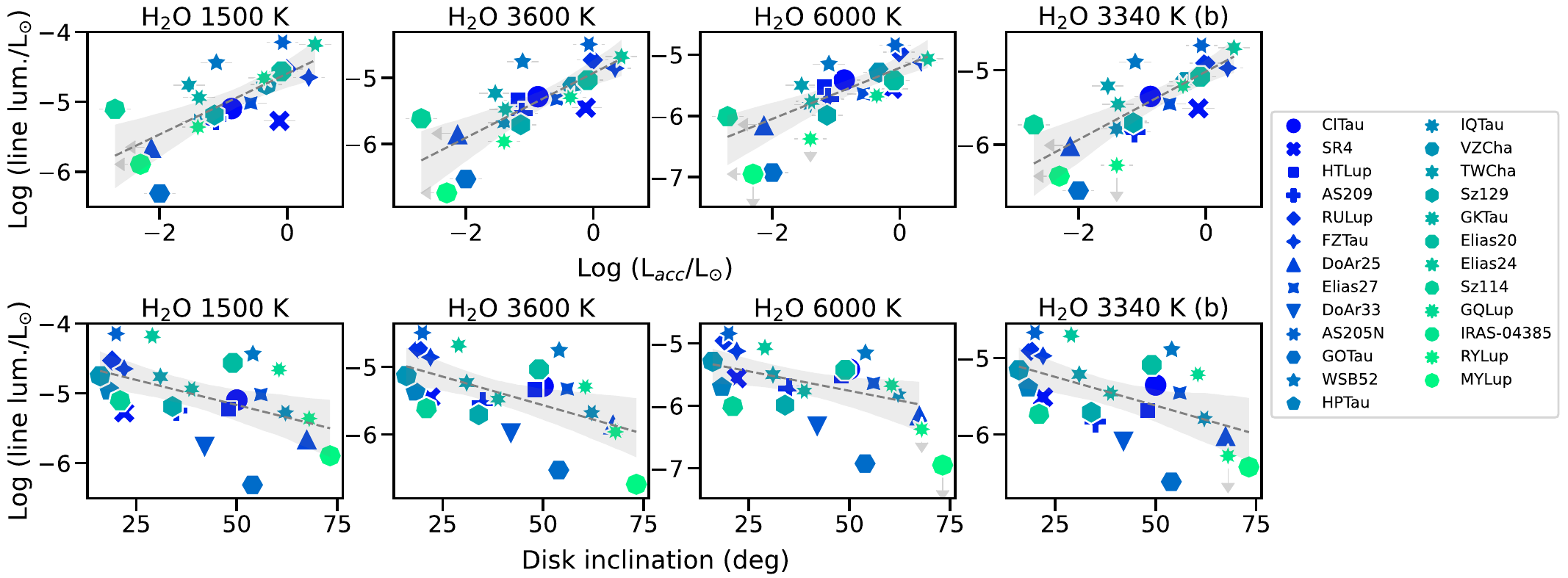}
\caption{Trends between the water lines from Table \ref{tab: cool excess line list} and the accretion luminosity (top) or disk inclination (bottom). Linear fits and their 95\% confidence intervals are shown as dashed lines and shaded regions, and their parameters are reported in Appendix \ref{app: lin_reg_par}. The color-coding follows that of Figure \ref{fig: cool_excess}. }
\label{fig: Lum_corr}
\end{figure*}

\rev{We distribute the line flux ratios measured in the sample along the $10^{17}$~cm$^{-2}$ model, assuming it represents an average column density for this component \citep[as found for the coldest water detected in radial temperature profiles in][]{munozromero24b}. With this column density, the line asymmetry gives evidence for a temperature of 170--200~K in five disks (IRAS-04385, GK~Tau, IQ~Tau, Sz~114, Sz~129)\footnote{In the case of GK~Tau, a cold component of $\sim 170$~K was indeed found from a combined fit to MIRI-MRS and IRS low-energy lines in \citet{banz23b}.}, five more for a temperature up to 220~K (HP~Tau, GQ~Lup, TW~Cha, RY~Lup, AS~205~N) and five more a temperature of 220--250~K (VZ~Cha, Elias~24, Elias~20, WSB~52, Elias~27). In general, these models show that a line ratio $> 1$ requires temperatures $< 350$~K and a ratio of $> 1.2$ needs temperatures $< 220$~K for any column density explored here. In total, therefore, more than 50\% of this sample has evidence for water emission at temperatures 170--250~K, close to and consistent with that expected for the snowline region \citep{lodders03}. These disks also have larger 1500/3600~K and 1500/6000~K ratios (the latter one shown in the right panel in Figure \ref{fig: 23um_diagram}) indicative of having a greater reservoir of cold water, as demonstrated in Figure \ref{fig: cool_excess}.
To confirm the exact temperature and determine the column density of water near and across the snowline, lower-energy lines at longer wavelengths need to be included as previously provided from Spitzer or Herschel \citep{kzhang13,banz23b} or from a future far-infrared observatory \citep{pont18,pontoppidan_prima,kamp21}.}

\subsection{Trends with accretion, inclination, and disk size} \label{sec: trends}
In reference to previous work that found correlations in water emission as observed with Spitzer or ground-based instruments \citep{salyk11_spitz,banz20,banz23}, it is important to test for correlations between the line fluxes and ratios used in Figure \ref{fig: cool_excess} and the accretion luminosity, one of the strongest correlations found before, and the disk inclination, for comparison to the inclination effects that will be shown later in Section \ref{sec: kinematics}. In Figure \ref{fig: Lum_corr} we confirm with the new MIRI spectra that rotational water emission correlates with accretion and anti-correlates with disk inclination. The slope steepens with $E_u$ in both cases, as found in \cite{banz23b} in the case of the correlation with accretion (linear correlation parameters are reported in Appendix \ref{app: lin_reg_par}). \rev{These correlations support the idea that inner disk irradiation and heating determine the size of the emitting areas for water at different temperatures, as suggested in previous work \citep{salyk11_spitz,banz23,banz23b}. The anti-correlations with disk viewing angle are less trivial to interpret as they could depend on the geometry and obscuration of different regions between a vertical inner dust rim and the disk surface at larger radii; these trends will be further analyzed in future work.}
The water line ratios, instead, do not correlate with these properties, suggesting that the relative areas of emitting regions for water reservoirs at different temperatures are not set primarily by accretion and are independent of the viewing angle.

\begin{figure*}
\centering
\includegraphics[width=1\textwidth]{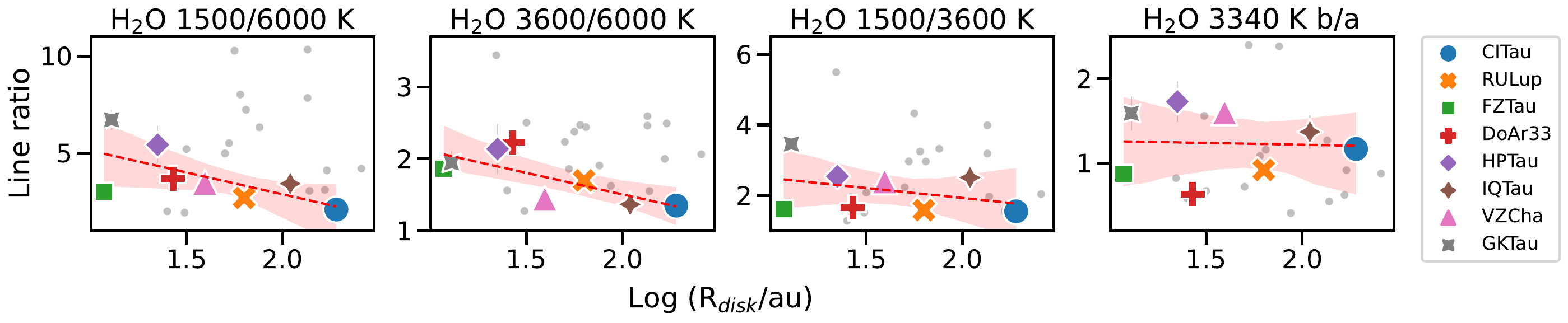}
\caption{Correlations between the line ratios defined in Section \ref{sec: cool excess} and the dust disk size as measured from ALMA continuum images. Colored larger datapoints (used for the linear fit) show disks with single stars, no millimeter cavity, and no cloud contamination (see Section \ref{sec: trends}). The rest of the sample is shown with small grey dots and is excluded from the fit. The first two anti-correlations to the left expand what found in \cite{banz23b} in four disks. Linear fits and their 95\% confidence intervals are shown as dashed lines and shaded regions, their parameters are reported in Appendix \ref{app: lin_reg_par}. }
\label{fig: cool_excess_ALMA}
\end{figure*}

Another known correlation to test in the context of water emission from inner disks is with disk sizes as measured at high angular resolution with ALMA. In Figure \ref{fig: cool_excess_ALMA} we show the trends between the three line ratios and the millimeter dust disk size following previous work \citep[][see Appendix \ref{app: sample measurements} for the millimeter data references]{banz20,banz23b}. In this figure we include in the linear regression only the sub-sample of full disks around single stars and without millimeter inner dust cavities or signs of being in younger embedded phases (cloud/envelope contamination, including the case of IRAS-04385 associated with an Herbig-Haro outflow). The reason for this selection is to isolate the effects of gas and dust transport through the disk from other effects that may regulate the observed inner water vapor due to age and environment, inner disk depletion, and binary interactions \citep[e.g.][]{salyk24,ramireztannus23,perotti23,schwarz24,grant24}. The measured line diagnostics in the context of these effects are being analyzed in upcoming papers.

The anti-correlations detected in line ratios in the sub-sample of eight disks in Figure \ref{fig: cool_excess_ALMA} (with regression parameters reported in Table \ref{tab: lin_fit_params} in Appendix \ref{app: lin_reg_par}) correspond to what previously found in four disks in \cite{banz23b}, which used a larger line list but similar energy levels. The correlation is detected only in the 1500/6000~K and 3600/6000~K line ratios, suggesting that the underlying physical process regulates the flux (here interpreted as emitting area) observed in the 170--200~K and 400~K water components relative to the hot inner water reservoir, which is instead mostly regulated by accretion \citep{banz23b}. The 3340~K line ratio, instead, does not correlate with disk size, suggesting that the column density of the 400~K component is independent from the underlying process driving this correlation. The discussion of these correlations in the context of pebble drift delivering water ice to the snowline, following previous work, is provided in Section \ref{sec: discussion}.

The figures in this section are provided as future reference for fundamental dependencies of the observed line luminosities to system parameters, which will help interpreting water spectra of specific disks. Multiple factors play a role in driving the physical or chemical properties of the water emitting regions in inner disks, and distinguishing their relative role in specific cases will not be trivial. However, based on previous work and Figure \ref{fig: Lum_corr} we can expect that the water luminosity generally correlates with accretion, and with stronger correlation for higher-energy levels. We can also generally expect that the viewing angle plays a role in how we observe water spectra, with a luminosity that decreases at higher inclinations.

\begin{figure*}
\centering
\includegraphics[width=0.95\textwidth]{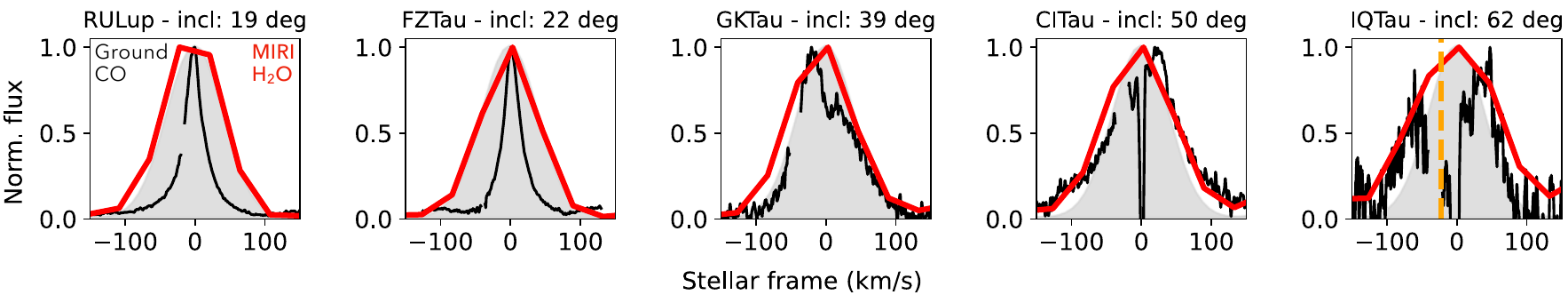}
\caption{Evidence for disk-rotation Doppler broadening of molecular lines observed with MIRI-MRS. High-resolution (R~$>50,000$) ro-vibrational CO line profiles are shown in black \citep[from][]{brown13,banz22}. The single (un-blended) MIRI water line at 17.35766~$\mu$m ($E_u \sim 2400$~K, Appendix \ref{app: line list}) is shown in red. A gaussian broadening that assumes a resolving power equivalent to FWHM = 95~km/s is shown for comparison as a grey shaded area. The orange dashed line in IQ~Tau shows the centroid of the blue-shifted absorption component found in the ro-vibrational lines in its MIRI spectrum (Figure \ref{fig: IQTau_abs}).}
\label{fig: ROTAT_broadening_LINES}
\end{figure*}

\begin{figure*}
\centering
\includegraphics[width=1\textwidth]{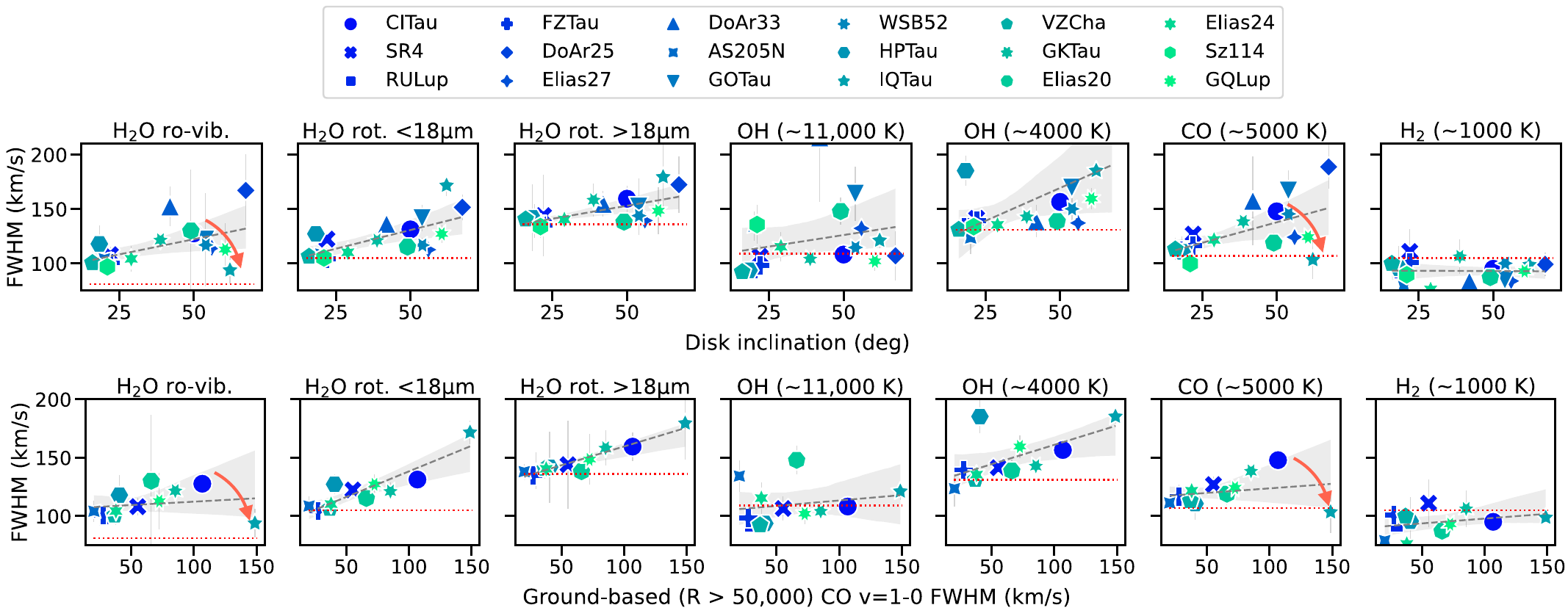}
\caption{Evidence for disk-rotation Doppler broadening of molecular lines observed with MIRI-MRS. In this figure we exclude disks with inner dust cavities or lower S/N (see Appendices \ref{app: sample measurements} and \ref{app: atlas_compact}). The observed line FWHM of \ce{H2O}, OH, and CO correlates with disk inclination (top) and with the high-resolution CO FWHM from ground-based observations where available (bottom). The nominal MIRI resolving power is reported as a horizontal dotted line for reference in each panel from \cite{pontoppidan24}. The direction indicated with an arrow shows the effect of blue-shifted wind absorption observed in IQ~Tau (Sect. \ref{sec: absorption}). Linear fits and their 95\% confidence intervals are shown as dashed lines and shaded regions, and their parameters are reported in Appendix \ref{app: lin_reg_par}. The color-coding follows that of Figure \ref{fig: cool_excess}.}
\label{fig: ROTAT_broadening}
\end{figure*}

\section{Gas kinematics in MIRI spectra} \label{sec: kinematics}
Building upon previous work, the diagnostic diagrams defined above in Section \ref{sec: cool excess} and Figure \ref{fig: cool_excess} demonstrate that different disks have different fractions of temperature components in their inner disk, showing up as a different spectral line flux distribution where hotter emission populates higher-$E_u$ and colder emission emerges at lower-$E_u$ to the extent to which it is present in a given disk. This is consistent with results from ground-based higher-resolving-power spectra, which additionally measured a gradient in line widths with broader higher-energy lines and narrower lower-energy lines directly demonstrating a temperature gradient across disk radii \citep[see Figure 13 in][]{banz23}.
The recent characterizations of the resolving power of MIRI \citep{Argyriou23,pontoppidan24} now open the way to test whether some water lines may be partially spectrally resolved. If this is true, we could determine their orbital region (not just the equivalent emitting area provided by the slab models used so far to fit MIRI spectra) from the observed kinematics and improve estimates of the radial distribution of water indicated by the line flux ratios in Figure \ref{fig: cool_excess}. 

In this section we present the detection and a first general analysis of disk-rotation Doppler line broadening measured in MIRI spectra. Even in this case, the list of single un-blended lines presented in Section \ref{sec: atlas} turns out to be very important to provide the most reliable measurements of line widths across MIRI wavelengths.

\subsection{Doppler broadening of emission lines} \label{sec: broadening}

Potential inclination effects on the observed MIRI spectra were pointed out in \cite{banz23b} in the case of the high-inclination (62~deg) disk of IQ~Tau, which showed broader lines than the nominal resolving power and a more compact slab emitting area than in three other disks. In reference to this finding, Figure \ref{fig: ROTAT_broadening_LINES} shows the single water line at 17.35766~$\mu$m ($E_u \sim 2400$~K) for selected targets where high-quality ro-vibrational CO lines are available at high resolving power from ground-based spectrographs. The figure demonstrates that, just like the high-resolution CO lines, the MIRI water lines also become broader at higher disk inclinations, suggesting a similar effect of rotational Doppler broadening from gas in the disk. By extending the sample and systematically measuring the line widths in all disks we can now conclusively test Doppler broadening as a general effect in MIRI spectra.

\begin{figure*}
\centering
\includegraphics[width=0.95\textwidth]{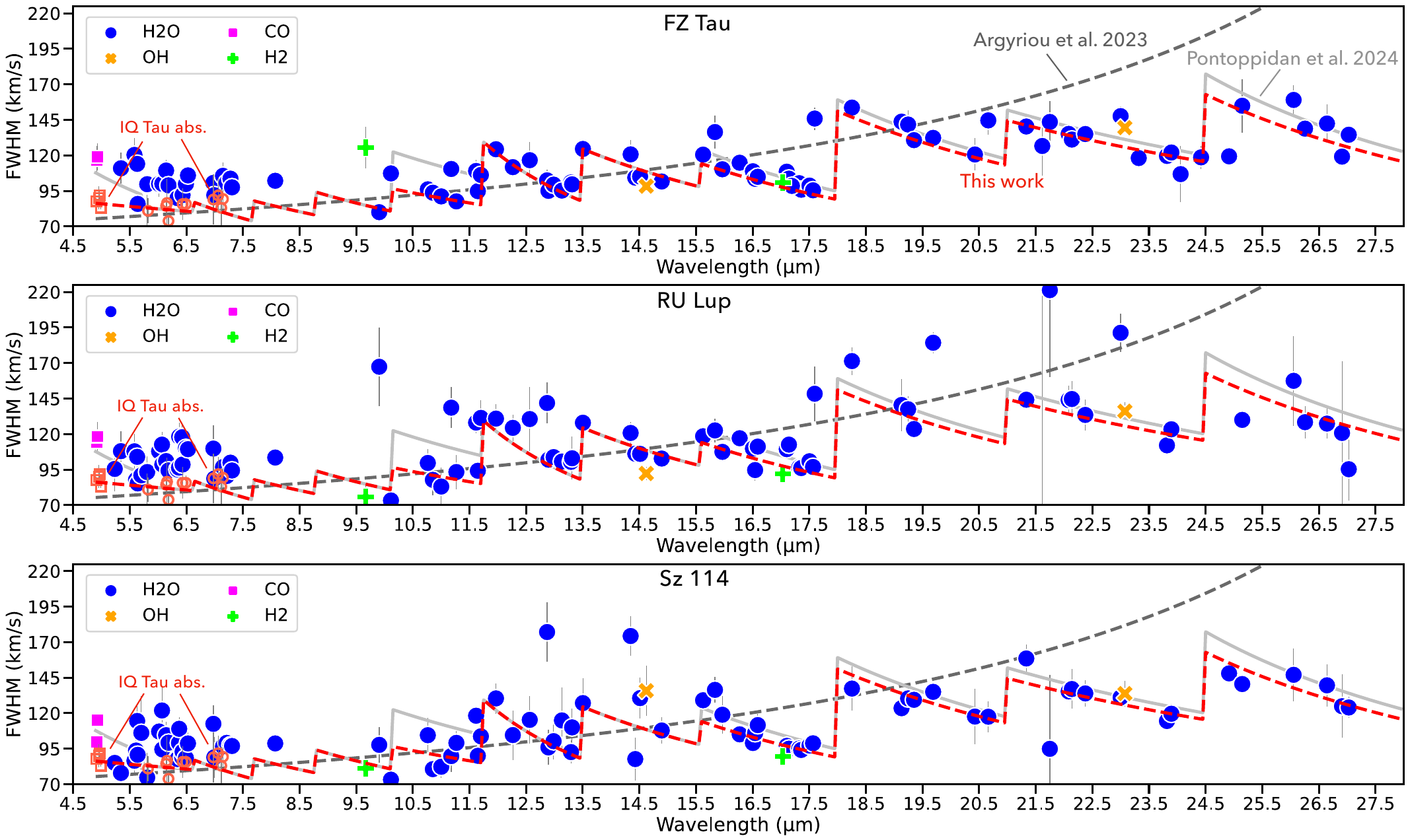}
\caption{MIRI-MRS resolving power from line width measurements in this work, as compared to recent works \citep[][gray dashed and solid lines, respectively]{Argyriou23,pontoppidan24}. The unresolved \ce{H2} lines (0-0 S(1) and S(3), light green crosses) provide useful anchor points suggesting a slightly better resolving power near 17~$\mu$m (see also Figure \ref{fig: ROTAT_broadening}) than what was estimated in \cite{pontoppidan24}. The absorption lines in IQ~Tau (light red open datapoints) demonstrate that the resolving power at the shortest wavelengths is consistent with what was estimated in \cite{Argyriou23}. The resolving power we adopt in this work is shown with a red dashed line (see Table \ref{tab: resolving_power} in Appendix \ref{app: res power} and \cite{pontoppidan24}).}
\label{fig: Res_Pow_MIRI}
\end{figure*}

Figure \ref{fig: ROTAT_broadening} presents clear evidence for disk-rotation Doppler broadening of line widths measured with MIRI. For water, we split the measured line widths (using the same line list presented above and reported in Appendix \ref{app: line list}) into three panels: one for ro-vibrational lines at $<9 \mu$m, one for rotational lines at $<18 \mu$m, one for rotational lines at $> 18 \mu$m. In each case, we show the median FWHM value as measured in single lines from the main line list (Section \ref{sec: line selection}) in each disk to capture a representative line broadening in each wavelength range. \rev{The measured FWHM will be de-convolved to estimate physical emitting radii in Section \ref{sec: doppler_mapping}.}
The top panel shows the line FWHM against disk inclination, which should show a positive trend when lines are broadened by Keplerian rotation in a similar inner disk region as observed from a range of viewing angles. The bottom panel shows the FWHM of MIRI lines against the FWHM of $v=1-0$ CO lines measured at much higher resolving power (R~$> 50,000$) with ground-based spectrographs in previous surveys (RU~Lup, FZ~Tau, GK~Tau, and CI~Tau with iSHELL from \citet{banz22}, IQ~Tau with CRIRES from \citet{brown13}) as available on SpExoDisks\footnote{Accessible at \url{https://spexodisks.com/}} \citep{SpExoDisks}, which should correlate if they are both broadened by Keplerian rotation in the inner disk. High-resolution spectra from the ground have already shown that CO and \ce{H2O} lines share a similar shape and broadening at 4.6--12.4~$\mu$m \citep[for an overview, see][]{banz23}, and the correlations now found with line widths from MIRI spectra extends this finding to wavelengths of $>13$~$\mu$m for the first time. 
At the lowest disk inclinations, the narrowest lines are consistent with the nominal MIRI resolving power (except for the ro-vibrational lines) as measured in previous work \citep{Argyriou23,pontoppidan24}, which is reported in each panel of the figure for comparison. In the case of the three water emission ranges illustrated in the figure, we show median values of the nominal resolving power of MIRI over the relevant wavelength ranges.

\begin{figure*}
\centering
\includegraphics[width=0.8\textwidth]{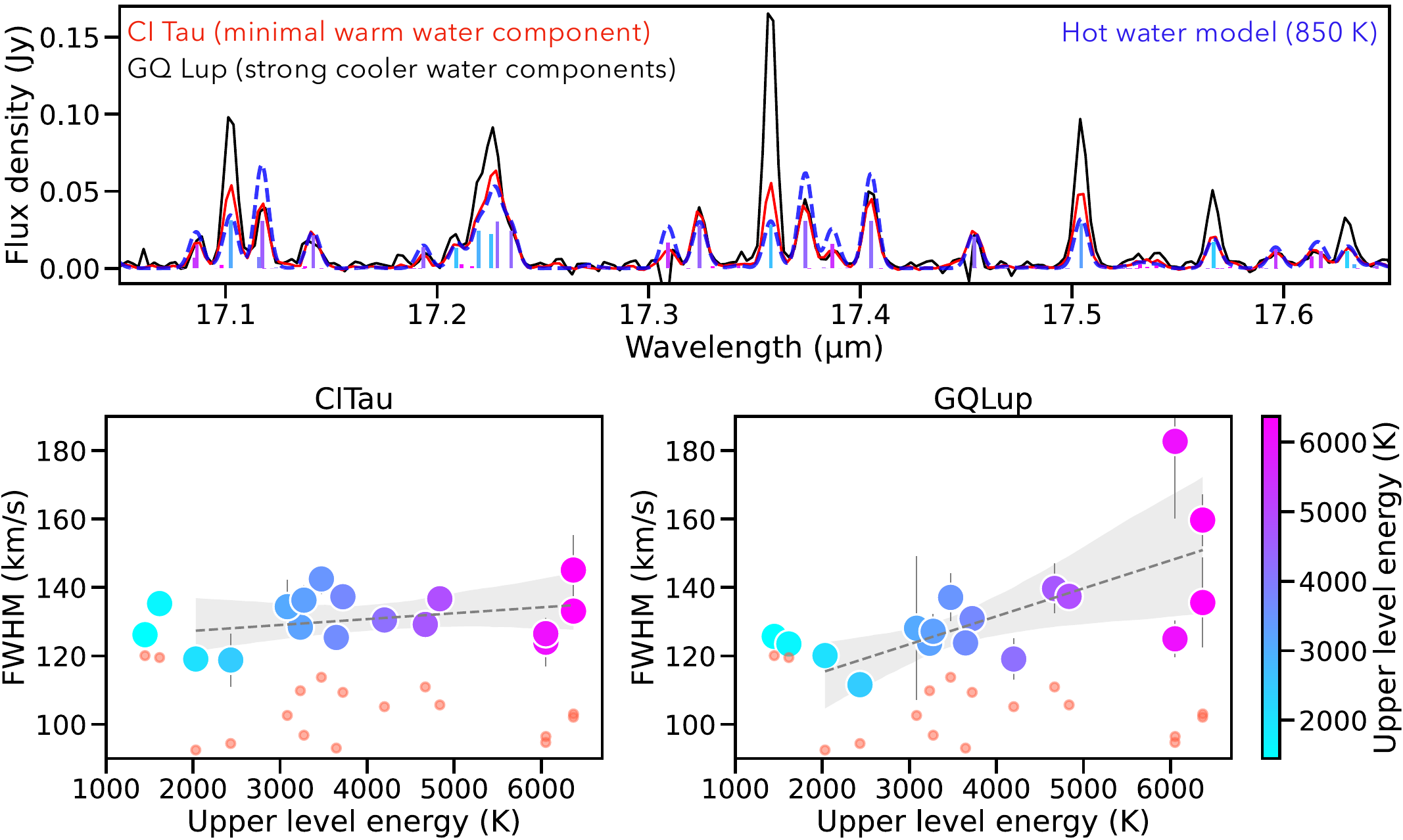}
\caption{Evidence for Doppler broadening of MIRI water lines as a function of upper level energy $E_u$. The figure includes the same two disks that have minimal warm water component (CI~Tau) vs a strong cold water component (GQ~Lup) as defined in Section \ref{sec: cool excess}; the rest of the sample is included in Figure \ref{fig: Eu_broad_sample} in Appendix \ref{app: sample_plot_grids}. In the upper panel, their spectra are scaled as in \cite{banz23b} and show individual transitions as in Figure \ref{fig: WATER_atlas_ROTshort2}. The water lines in GQ~Lup become narrower at lower $E_u$, as expected if the cold water reservoir is in a disk region at larger radii \citep{banz23b,munozromero24b}. Light red dots show the updated MIRI resolving power at each line wavelength from Figure \ref{fig: Res_Pow_MIRI}. The two lines near 1500~K from Table \ref{tab: cool excess line list} are excluded from the fit, due to the lower resolving power at 23~$\mu$m. }
\label{fig: Eu_broadening}
\end{figure*}

For comparison to the \ce{H2O} lines, Figure \ref{fig: ROTAT_broadening} reports the trends observed in lines from other species \rev{(measured with average significance of 12--17 $\sigma$ in this sample)}: a single hot OH line at 14.62~$\mu$m ($E_u = 10,754$~K, the only single OH line we can find at MIRI wavelengths, which may include some contamination from organics depending on the relative strength, see Figure \ref{fig: WATER_atlas_ROTshort2}), a lower-energy OH line pair that overlaps in wavelength at 23.07~$\mu$m (with $E_u \sim 4100$~K), the P26 line of $v=1-0$ CO emission (one of the only three un-blended CO lines in MIRI spectra with minimal contamination from water and other higher-energy CO vibrational lines, together with P25 and P27), and the \ce{H2} 0-0 S(1) line near 17~$\mu$m, which is well separated from emission from other molecules (all other \ce{H2} lines except for S(3) are too weak or significantly blended, see Figures \ref{fig: WATER_atlas_ROVIB}--\ref{fig: WATER_atlas_ROTshort2}). While water, CO, and partly OH show similar evidence for Doppler broadening, supporting their disk origin, the \ce{H2} line width does not increase with inclination, suggesting that the line is typically unresolved. Another \ce{H2} line that can be measured is the S(3) line at 9.665~$\mu$m (Figure \ref{fig: WATER_atlas_ROTshort1}); also this line provides evidence for being uresolved in MRS spectra (Figure \ref{fig: Res_Pow_MIRI}), but it is blended in its wings with \ce{H2O} and OH lines which can contaminate its measured FWHM in case of relative weak emission (e.g. FZ~Tau). The other \ce{H2} lines covered by MRS are all blended with \ce{H2O} but still consistent with the local resolving power at each wavelength. That \ce{H2} is unresolved in MRS spectra is consistent with a non-disk origin for \ce{H2} emission, which is indeed observed to trace outflows and winds in young disks \citep[e.g.][]{yang22,arulanantham24,Delabrosse24}. Another option could be extended emission from larger disk radii (therefore narrower lines) than the emission from other molecules detected in MIRI spectra.

\subsubsection{Updates on the MIRI resolving power} \label{sec: MIRI R}
The data in Figure \ref{fig: ROTAT_broadening} show that the narrowest lines are found at low inclinations, as expected from Keplerian broadening. We can therefore use MIRI spectra of low-inclination disks to measure the MIRI resolving power, as done recently in \cite{pontoppidan24}. This is done in Figure \ref{fig: Res_Pow_MIRI}, where three targets are selected for their low inclination (FZ~Tau, 22 deg, and RU~Lup, 19 deg) and low stellar mass (Sz~114, 21 deg, 0.2~$M_{\odot}$), which provide the best conditions to have unresolved lines across the MIRI spectrum. Their line measurements are shown in reference to the MIRI resolving power from \cite{Argyriou23}, which used HI and forbidden lines from a planetary nebula, and \cite{pontoppidan24}, which used water and CO lines from the protoplanetary disk of FZ~Tau. 

The characterization from \cite{pontoppidan24} is confirmed overall, including the much higher resolving power in channel 4 in comparison to that estimated in \cite{Argyriou23}. In this updated analysis we identify a few sub-bands where the resolving power is slightly better than what was previously found in \cite{pontoppidan24}, and we provide the updated profile in Figure \ref{fig: Res_Pow_MIRI} (the red dashed line) and tabulated in Table \ref{tab: resolving_power} of Appendix \ref{app: res power}.
In particular, we identify the \ce{H2} line at 17.035~$\mu$m as providing a useful anchor point showing a higher resolving power than what the nearby water lines show, suggesting that these might be partially resolved in disks.
By taking the median value of measurements for this \ce{H2} line in this sample, we estimate FWHM$_{\rm{MRS}}$~$\sim 93$~km/s, i.e. about $\sim 10$~km/s better than what was estimated in \cite{pontoppidan24} at this wavelength.

\begin{figure*}
\centering
\includegraphics[width=1\textwidth]{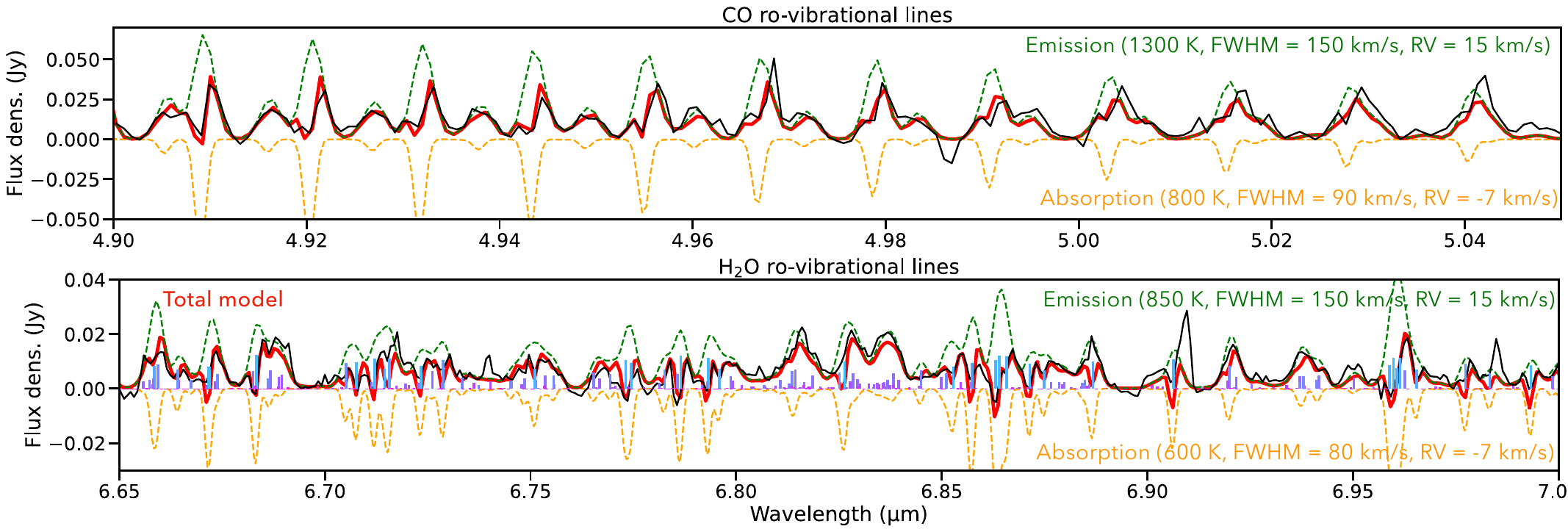}
\caption{Absorption spectra of ro-vibrational CO (top) and \ce{H2O} lines (bottom) in the high-inclination disk of IQ~Tau. Two models with different RV, temperature, and FWHM are shown in the figure to approximately reproduce the data. The MRS data is shown in black, the difference of the two models in red. Individual water transitions are color-coded as in Figure \ref{fig: WATER_atlas_ROVIB}, to illustrate that absorption is prevalent in lower-energy lines (cyan).}
\label{fig: IQTau_abs}
\end{figure*}

\subsubsection{Doppler broadening as a function of $E_u$} \label{sec: broadening-Eu}
Another very interesting trend emerges from considering line broadening as a function of upper level energy $E_u$. This can be easily analyzed over small spectral ranges where the resolving power is approximately constant, to avoid confusing the signal with a changing resolving power with wavelength. This was one of the reasons why we selected lines between 14.4 and 17.6~$\mu$m in Section \ref{sec: line selection}: the resolving power is approximately the same (see Figure \ref{fig: Res_Pow_MIRI}) and there is a good number of single (un-blended) lines that can be used for FWHM measurements.
If water emission comes from a radially narrow region that can be well approximated with a single temperature component, we should not expect an observable trend between the measured line FWHM and $E_u$. Instead, if water emission comes from a more extended disk region that has a larger temperature gradient as suggested by ground-based work \citep{banz23}, we may be able to measure a trend between FWHM and $E_u$, where the lower-$E_u$ lines should become increasingly narrower. 

This is exactly what is observed in the data when comparing two disks that have previously been proposed to be in these two different situations: Figure \ref{fig: Eu_broadening} shows that the spectrum of GQ Lup, with one of the strongest and most extended cool water components found so far, has a clear trend between FWHM and $E_u$, while the same lines in CI~Tau are flat with $E_u$. The rest of the sample is shown in Figure \ref{fig: Eu_broad_sample} in Appendix \ref{app: sample_plot_grids}. This finding provides new, independent confirmation of previous work that proposed water to trace an extended disk region in disks with increased emission from low-energy lines \citep{banz23b,munozromero24b}. Line broadening will be applied in Section \ref{sec: doppler_mapping} to extract radial profiles from the observed water emission.

\subsection{Water and CO absorption at high inclinations} \label{sec: absorption}
If lines observed with MIRI-MRS are rotationally broadened at high inclinations as demonstrated in the previous section, we could also expect to observe absorption on top of emission lines with different Doppler broadening, since this is observed at high-resolution from the ground \citep[e.g.][]{pont11,brown13,banz22,banz23}. The best test case should be a high-inclination disk with deep blue-shifted absorption observed in ro-vibrational CO emission from the ground, since absorption is observed to deepen at higher disk inclinations possibly due to a larger portion of an inner disk wind intercepted along the line of sight \citep{pont11,banz22}. In our sample this is the case of IQ~Tau, whose CO ro-vibrational line shape shows broad double-peaked emission (due to its high inclination of 62 deg) with at least one blue-shifted absorption component (Figure \ref{fig: ROTAT_broadening_LINES}, previously shown in \cite{brown13}). 

Figure \ref{fig: IQTau_abs} shows portions of the ro-vibrational CO and \ce{H2O} spectra from the MIRI spectrum of IQ~Tau, confirming the scenario proposed above. The spectral lines of both molecules show a broader and more complex shape than in other disks in this work, a shape that can be excellently matched with the simple difference of two spectra (red total model in Figure \ref{fig: IQTau_abs}): a hotter spectrum in emission at the RV of the star (with FWHM = 150 km/s, much larger than the nominal resolving power at these wavelengths, as already shown in Figure \ref{fig: ROTAT_broadening}), and a colder spectrum blue-shifted by -7 km/s (with FWHM = 80--90 km/s); the models used for absorption adopt a similar column density of 1--5~$\times 10^{17}$~cm$^{-2}$. While the models in this figure are just for quick demonstration, we remark that the data seem to clearly show two aspects: that the absorption spectrum is blue-shifted, and that it is cooler than the emission spectrum. The different temperature is most evident from the $v=2-1$ CO lines, which are less absorbed than the $v=1-0$, and in the water lines, where it is the low-energy lines to have the most significant absorption (see color-coded transitions in Figure \ref{fig: IQTau_abs}). 

This finding opens up the possibility to distinguish and study blue-shifted absorption from inner disk winds in molecular disk spectra observed with MIRI. At the same time, it highlights the importance of having high-resolution molecular spectra from ground-based instruments, that easily distinguish absorption and blue-shifted winds in the velocity-resolved line profiles \citep[for a recent overview, see][]{banz22}, to support the analysis of MIRI spectra. The detailed analysis of the molecular wind absorption spectrum in IQ~Tau and other disks is left for future work. For a detailed discussion of the analysis of water absorption spectra as observed with MIRI-MRS, see \cite{li24}.

One immediate practical application in this work is that we can use measurements of absorption lines in IQ~Tau, which have FWHM of $\approx 7$~km/s \citep{banz22} and are therefore surely unresolved with MRS, to improve the characterization of the MIRI resolving power at the shortest wavelengths. This is shown in Figure \ref{fig: Res_Pow_MIRI}, where we add with empty red datapoints the measured FWHM of absorption lines from IQ~Tau. These confirm that the resolving power is close to that estimated in \cite{Argyriou23}, and demonstrate that ro-vibrational CO and \ce{H2O} emission lines are generally partially resolved by the MRS in protoplanetary disks, consistent with their small emitting radius obtained from the large line broadening observed at high resolution from the ground \citep{banz23}.

A decrease in ro-vibrational line broadening was in fact visible even in Figure \ref{fig: ROTAT_broadening} in disks at inclinations $> 50$~deg (region marked with a red arrow), with IQ~Tau providing the most extreme case where CO and \ce{H2O} lines decrease down to the MIRI resolving power. This suggests that ro-vibrational CO and \ce{H2O} absorption may be present and observable in MIRI spectra in disks at inclinations $> 50$~deg, which should be checked when the ro-vibrational lines are to be analyzed. In the sample analyzed here, in addition to IQ~Tau absorption seems to be visible in GQ~Lup and possibly Elias~20 and GO~Tau.
In case high-resolution spectra from the ground are available, they can provide useful guidance on whether to expect absorption to be present in a specific disk; however, ground-based observations are not always available nor possible for MIRI targets. Cases where absorption is not as significantly blue-shifted as in IQ~Tau (Figure \ref{fig: ROTAT_broadening_LINES}), e.g. CI~Tau, may only result in weaker lines and not be easily detected from the MIRI spectra alone \citep[see previous discussion in][]{banz23}.

\begin{figure}
\centering
\includegraphics[width=0.45\textwidth]{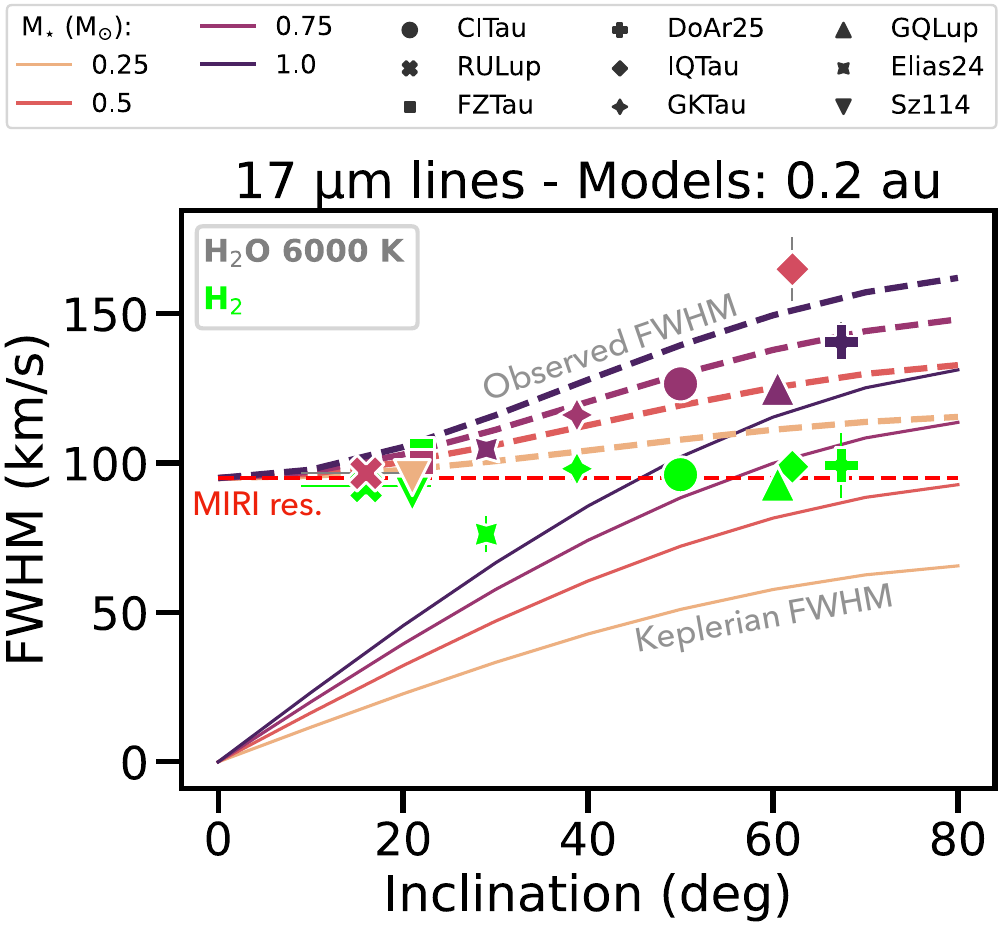}
\caption{Keplerian interpretation for the Doppler broadening of emission lines in MIRI spectra. Line widths from Keplerian models at 0.2~au (with FWHM = 2$V_{\rm{Kepl}}$) are shown with solid lines for a range of stellar masses between 0.25 and 1~$M_{\odot}$, their convolution to the MIRI resolving power is shown with dashed lines. For illustration, we show a selection of disks spanning the whole range in disk inclinations included in this work, and we use the 6000~K line from Table \ref{tab: cool excess line list}. The \ce{H2} line at 17.035~$\mu$m is included as a reference for the MIRI resolving power (see Figure \ref{fig: Res_Pow_MIRI}).}
\label{fig: Doppler_mapping}
\end{figure}

\begin{figure*}
\centering
\includegraphics[width=1\textwidth]{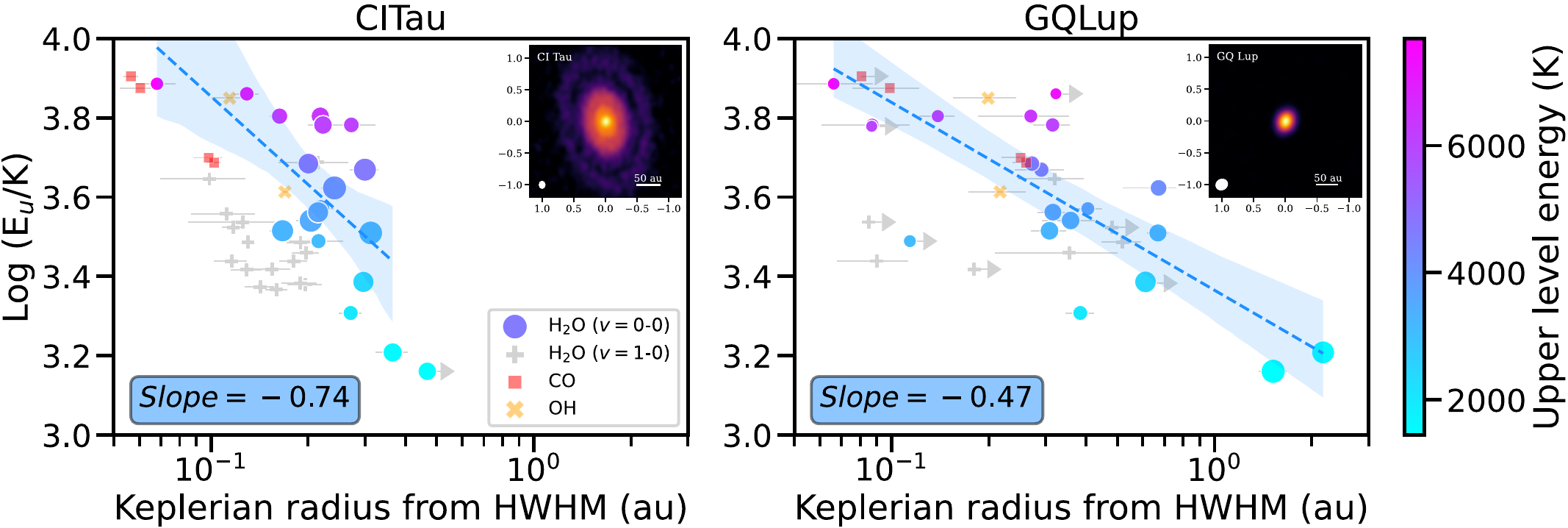}
\caption{Radial profiles of water lines observed in CI~Tau (large, multi-gapped disk dominated by a single hot water component) and GQ~Lup (compact disk with strong warm and cold water components), by applying the simple Doppler mapping technique described in Section \ref{sec: doppler_mapping} and Figure \ref{fig: Doppler_mapping}. ALMA continuum images are shown for reference in each panel \citep[][and in prep.]{long19}. All lines are partially resolved in both disks, enabling estimates of emitting radii from the de-convolved line broadening. Water rotational transitions are shown as filled circles with size proportional to the line flux (to illustrate the different spectral line flux distribution as in the top panel of Figure \ref{fig: Line_selection_demo}), ro-vibrational lines with grey crosses. \rev{A power-law fit to water transitions where Doppler broadening is detected is shown with a dashed line and shaded area (95\% confidence region), the best-fit slope is reported in each plot. Keplerian radii for CO and OH lines measured in this work (Figure \ref{fig: Res_Pow_MIRI} and Appendix \ref{app: line list}) are included for comparison to the water lines.}}
\label{fig: Radial_profiles}
\end{figure*}

\subsection{Doppler mapping of MIRI water lines} \label{sec: doppler_mapping}
The data shown in Section \ref{sec: kinematics} and Figures \ref{fig: ROTAT_broadening_LINES}, \ref{fig: ROTAT_broadening}, and \ref{fig: Eu_broadening} demonstrated that individual molecular lines in MIRI-MRS spectra may be rotationally broadened beyond the MIRI resolving power (except for \ce{H2}), depending on the disk inclination and the upper level energy of the emitting lines. This finding, in addition to the strong correlations with the FWHM of ro-vibrational CO lines as observed from the ground at high resolving power, supports the idea that MIRI lines are rotationally-broadened by Doppler effect in an inner disk with an inside-out temperature gradient. In this section, for simplicity we assume that the Doppler broadening can be simply described by Keplerian rotation in the disk; for an overview and discussion of the possible contamination by a slow disk wind to the narrow CO components see \cite{pont11,banz22}. 

In Figure \ref{fig: Doppler_mapping}, we illustrate one example of how Keplerian rotation in the disk broadens MIRI lines beyond the resolving power. With solid lines we show the FWHM produced by gas in Keplerian rotation at 0.2~au \citep[which should be appropriate for high-energy lines, see overview in][]{banz23} for a range of stellar masses between 0.25 and 1 M$_{\odot}$. With dashed lines, we show how the Keplerian models would be observed after convolution with the MIRI-MRS resolving power, by assuming $\rm{FWHM}_{obs.} = \sqrt{(FWHM_{\rm{MRS}})^2+(FWHM_{Kepl.})^2}$. We do not include any thermal broadening component, as it is negligible ($< 5$~km/s) at the temperatures relevant for this work \citep[$< 1000$~K, see e.g.][]{meijerink09,pontoppidan24,munozromero24b}. As can be seen from the models, Doppler broadening should become detectable in high-S/N spectral lines starting at inclinations as low as $\sim20$~deg, especially in disks around stars with larger mass.

For comparison to the models, we include the 17.32~$\mu$m ($E_u \sim 6000$~K) line from Table \ref{tab: cool excess line list} tracing hot water emission for a representative sample of targets that spans the entire range in disk inclinations included in this work, with masses between 0.2 and 1~$M_{\odot}$ (color-coded in the same way as the models, for comparison). This line is confirmed to be consistent with tracing disk radii at $\sim0.2$~au in a few disks that overlap with their stellar-mass model curve (e.g. CI~Tau); instead, disks falling above their stellar-mass curve indicate emission from smaller radii (IQ~Tau) and those falling below indicate emission at larger radii (GQ~Lup). For reference, we also show the \ce{H2} line that reflects the MIRI resolving power near 17~$\mu$m (Section \ref{sec: MIRI R}).   

The simple Doppler mapping technique illustrated in Figure \ref{fig: Doppler_mapping} is generally applied to multiple MIRI lines in Figure \ref{fig: Radial_profiles}, providing a radial excitation profile with line upper level energies as a function of their Keplerian emitting radius from the line broadening. In this figure, as examples we include two disks that were previously identified to have a hotter and supposedly more compact versus a colder and supposedly more extended emission, the same disks shown in Figure \ref{fig: Eu_broadening}. Here we use the line FWHM/2 = HWHM (for the Keplerian velocity $V_{\rm{Kepl}}$ from one side of the disk) as de-convolved with the local MIRI resolving power (Section \ref{sec: MIRI R}), for the same sample of lines used for Figure \ref{fig: Eu_broadening}. The radial excitation profiles in Figure \ref{fig: Radial_profiles} provide a new demonstration, completely independent from the temperature fits made in previous work, for the different distribution of water in these disks \citep{banz23b,munozromero24b}: in CI~Tau (large, multi-gapped disk dominated by a single hot water component, proposed to have a reduced icy pebble drift) confined within $\lesssim 0.35$~au, and in GQ~Lup (compact disk with strong warm and cold water components, proposed to have a stronger water enrichment from pebble drift) extending out to $> 1.5$~au. A fit to the radial profiles in Figure \ref{fig: Radial_profiles} confirms the relative difference and slopes in radial gradients estimated from power-law fits to the line fluxes in \cite{munozromero24b}, highlighting the potential in future work to improve model fits to MIRI spectra by including both line excitation and line broadening.

\begin{figure*}
\centering
\includegraphics[width=1\textwidth]{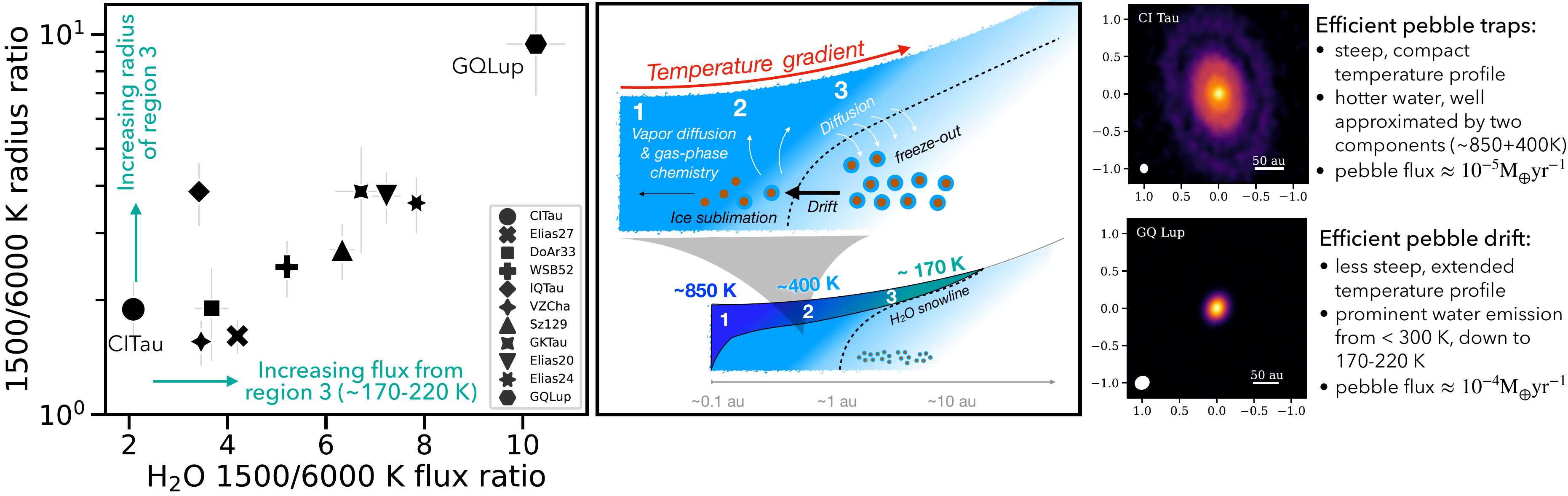}
\caption{Left: Ratios of Keplerian radii estimated from the deconvolved line widths (Figure \ref{fig: Radial_profiles}) as a function of their line flux ratio, using the 1500 and 6000~K lines from Table \ref{tab: cool excess line list}. This figure includes only disks where Doppler broadening is detected (Appendix \ref{app: sample_plot_grids}). With the exception of IQ~Tau (the disk where wind absorption is detected, see Figure \ref{fig: IQTau_abs}), the trend in this figure shows that a larger flux in the 1500~K lines corresponds to their emission coming from larger radii. Middle: Illustration of the temperature gradient approximated by the three temperature regions at increasing disk radii, in reference to processes of icy pebble drift, sublimation, and gas diffusion \citep[modified from][]{banz23b}. \rev{Right: Summary of differences between disks that have efficient pebble traps vs drift (Section \ref{sec: discussion}), based on the results from this work and \citet{munozromero24b}.}}
\label{fig: Kepl_ratios+cartoon}
\end{figure*}

An important caveat to consider is that these regions are derived from the line broadening (which can be resolved in MIRI spectra) and only provide a characteristic emitting radius that does not reflect the full radial extent of the emission, which would be shown from the velocity of the typical double peaks from Keplerian rotation (which instead cannot be resolved in MIRI spectra). However, the Keplerian radii estimated from MIRI spectra can be used for relative comparisons between different disks and for comparing the relative emitting region of different lines in a given disk. For instance, CI~Tau shows that ro-vibrational lines are emitted from smaller disk radii than the rotational lines, as found before in higher-resolution spectra from the ground \citep{banz23}. The ro-vibrational lines in GQ~Lup (incl = 60~deg) are instead very weak and possibly include absorption (see discussion in Section \ref{sec: absorption}), providing only a few uncertain estimates in Figure \ref{fig: Radial_profiles}. 

\rev{For comparison to the water radial profiles, we include in Figure \ref{fig: Radial_profiles} the Keplerian radii estimate in the same way for the CO and OH lines that have been measured in this work (Figure \ref{fig: Res_Pow_MIRI} and Appendix \ref{app: line list}). The radial gradient in CO lines shows a very similar slope to the water gradient in both disks. The comparison to water profiles can be useful, but it should be kept in mind that these are only simple approximations of the CO distribution in the inner disk, because MIRI-MRS spectra cannot resolve the multiple CO components shown by ground-based spectra that fully resolve the line profiles \citep[see e.g. discussions in][]{banz22,banz23}.}

\section{Discussion} \label{sec: discussion}
In this section we discuss some important applications of the tools and findings presented above to study MIRI water spectra in protoplanetary disks, with the intent to provide a helpful framework for community efforts and a common ground for comparisons across different samples.

\subsection{The radial distribution of water in inner disks} \label{sec: discussion_radial}

The diagrams in Figure \ref{fig: cool_excess} have been introduced for providing a simple, general view of the relative emission from water at different temperatures. These simple diagnostics can provide a helpful starting point before performing detailed fits with slab or more sophisticated models, and provide an empirical framework for comparisons across datasets and samples independently from different modeling tools. As described in Section \ref{sec: cool excess}, the position of a given disk in the diagnostic diagrams informs on whether a $\sim 400$~K and $\sim 170$--200~K components significantly contribute to its water spectrum in addition to a hot $\sim850$~K component that is commonly (but not always) present, and on the column density of a $\sim 400$~K component. These discrete components are only a convenient approximation of the radial gradient previously found in inner disks \citep{banz23}, as shown in recent work from fits to the MIRI line fluxes \citep{munozromero24b,temmink24b,grant24} and demonstrated for the first time from their Doppler broadening in this work (Figure \ref{fig: Radial_profiles}).

\begin{figure*}
\centering
\includegraphics[width=1\textwidth]{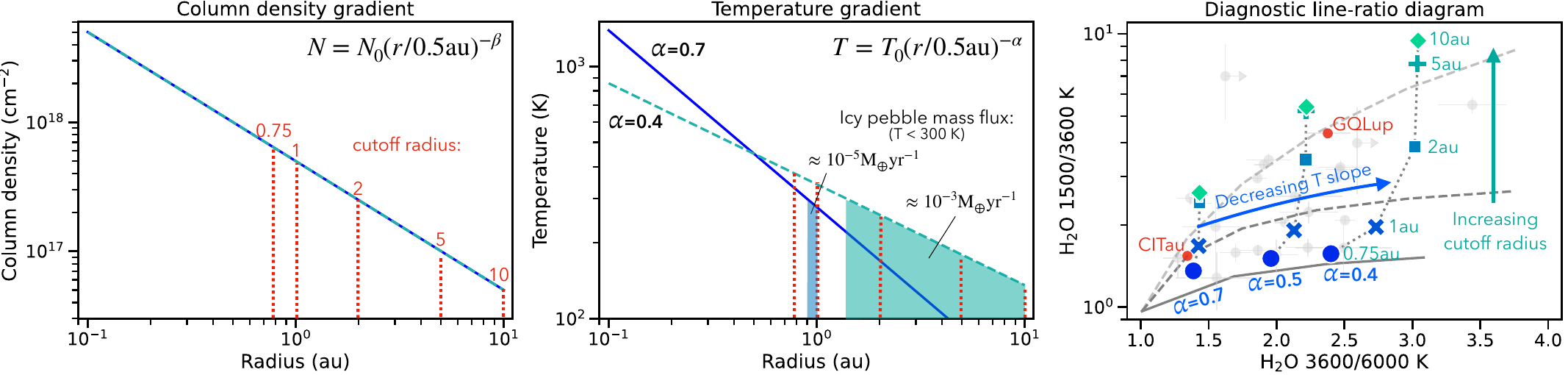}
\caption{\rev{Power-law radial gradients profiles with a cutoff radius based on the slab model fit approach presented in \cite{munozromero24b}. A range of temperature gradient slopes between 0.4 and 0.7 and cutoff radii between 0.75~au and 10~au are sufficient to reproduce the line ratios measured in this sample in the principal diagnostic diagram (light grey datapoints and model tracks in the right panel, same as in Figure \ref{fig: cool_excess}). A grid of radial gradient models with temperature slope and cutoff radius as labeled is shown for reference in the diagnostic diagram, their line ratios are reported in Appendix \ref{app: slab models params} in Table \ref{tab: radial gradients}. The temperature slopes inferred from this plot for CI~Tau ($\gtrsim 0.7$) and GQ~Lup ($\lesssim 0.5$) match well those measured from the line broadening in Figure \ref{fig: Radial_profiles}. A range in pebble mass fluxes is estimated using Equation 11 in \cite{munozromero24b} from two extreme cases of minimum/maximum cold water enrichment as shown by the shaded areas in the middle panel.}}
\label{fig: diagnostic_diagram_gradients}
\end{figure*}

\subsubsection{Combining line flux and broadening information}
The detection of Doppler broadening in MIRI lines provides additional physical information to improve the interpretation of the position of a given disk in Figure \ref{fig: cool_excess} in terms of the radial distribution of water. The line broadening in Figure \ref{fig: Radial_profiles} shows that a disk like CI~Tau that sits close to the 1,1 point in the diagnostic diagram has lines from all $E_u$ emitting from a compact inner disk annulus at $\sim$~0.1--0.3~au, with only a slightly larger emitting radius for the lower $E_u \sim 1500$~K. This explains why it has been found in previous work to be well reproduced by a single hot temperature component \citep{banz23b,munozromero24b}. Instead, a disk like GQ~Lup that sits along the W+H+C model track in Figure \ref{fig: cool_excess} shows a line broadening gradient in Figure \ref{fig: Radial_profiles} that corresponds to a much larger span of disk radii, with lower-$E_u$ lines emitted from up to $10\times$ larger radii than the higher-$E_u$ lines. This explains why this disk has such strong 1500~K lines indicating a strong cold component, as shown above in this work and in \citet{munozromero24b}. An extended radial emitting region for water was generally expected from disk models and velocity-resolved surveys \citep[Figure 13 in][]{banz23}, but has never been directly observed from the broadening of water lines at $> 13$~$\mu$m before this work.

We now compare the flux and broadening measurements from MIRI spectra in Figure \ref{fig: Kepl_ratios+cartoon}, using the sub-sample of disks in this work where Doppler broadening is detected across all energy levels (Appendix \ref{app: sample measurements}). The figure shows the ratio of Keplerian radii (each radius obtained from the deconvolved HWHM as in Figure \ref{fig: Radial_profiles}) as a function of the diagnostic line flux ratios used above in this work. To use a representative Keplerian radius for each bin in $E_u$, we take the median radius of lines included in Figure \ref{fig: Radial_profiles} for each of these ranges: 1400--1700~K, 3000--4000~K, and 6000--6500~K. Figure \ref{fig: Kepl_ratios+cartoon} shows the example of the 1500/6000~K line ratios (their flux ratio and Keplerian radius ratio), which maximizes the range of values measured in the spectra (as the cold component should be generally the most radially extended). With the exception of IQ~Tau (the same high-inclination disk where wind absorption is detected, see Section \ref{sec: absorption}), the general trend in this figure shows that a higher line flux ratio corresponds to emission in the cold component from larger disk radii.
Therefore, the line broadening measurements independently support the interpretation of Figure \ref{fig: cool_excess} as providing a quick reference for the radial distribution of water in the inner disk, which we approximated above with three temperature regions as illustrated in the right panel of Figure \ref{fig: Kepl_ratios+cartoon}, adapted from \cite{banz23b}.

\subsubsection{Radial gradients in the main diagnostic diagram}
\rev{After demonstrating that both the line flux ratios and the line broadening show the radial distribution of water emission in inner disks, we illustrate in Figure \ref{fig: diagnostic_diagram_gradients} the interpretation of the main diagnostic diagram introduced in this work in the context of radial gradients. We adopt the parametrization in temperature and column density recently applied to fit water spectra for part of this sample in \cite{munozromero24b}. For a general demonstration, we adopt the power-law profiles adopted in that work as $T = T_0 ( r / 0.5 \text{au})^{-\alpha}$ and $N = N_0 ( r / 0.5 \text{au})^{-\beta}$ with representative values from their best-fit results: $T_0 = 450 K$, $N_0 = 1 \times 10^{18}$~cm$^{-2}$, $\beta = 1$, and $\alpha$ between 0.4 and 0.7 (Figure \ref{fig: diagnostic_diagram_gradients}). We also consider a cutoff radius for the radial profiles between 0.75~au and 10~au, beyond which the water column density drops to zero \citep[as an approximation for the tapered profiles in][]{munozromero24b}. In the diagnostic diagram, the line ratios provided by this grid of models reproduce the entire range of those measured in this sample (right panel in Figure \ref{fig: diagnostic_diagram_gradients}) similarly to the discrete temperature components described in Section \ref{sec: cool excess}, again supporting that they can be used to approximate a power-law radial gradient \citep{temmink24b,munozromero24b}.}

\rev{Similarly to the discrete temperature components used in Figure \ref{fig: cool_excess}, the power-law models in Figure \ref{fig: diagnostic_diagram_gradients} show that a larger 3600/6000~K line ratio indicates a less steep temperature gradient where the inner disk has more emission from warm water at intermediate radii. A larger 1500/3600~K line ratio, instead, is indicative of colder water at larger distances out to the cutoff radius. Most of the sample included in this work can be well reproduced by gradients with temperature slopes between 0.7 (for the disks dominated by hotter emission, e.g. CI~Tau) and 0.5 (for the disks with significant emission from colder water, e.g. GK~Tau and GQ~Lup), and cutoff radii between $< 1$~au and 4~au, matching well the fit results found in part of the sample in \cite{munozromero24b}. This demonstrates that the main water diagnostic diagram introduced in this work (with the 1500/3600~K and 3600/6000~K line flux ratios) can be used as a simple proxy for temperature gradients too, and to provide input to more detailed fits. We note instead that the 3340~K line flux ratio introduced above as a diagnostic for the column density at $\sim$~400~K shows a more complex dependence in the grid of power-law radial profiles, suggesting that it may be more degenerate than what the discrete components show in Figure \ref{fig: cool_excess}. A more complete and detailed analysis of the diagnostic line ratios in terms of radial gradients is left to future work.}

\subsubsection{What regulates the water abundance in inner disks}
\rev{The correlations reported in Section \ref{sec: trends} are indicative of some of the major processes that determine the water distribution in inner disks. The emitting area of the hotter, optically thick inner water reservoir can be considered to be set by the disk irradiation and heating, based on the strong trend with accretion luminosity (Figure \ref{fig: Lum_corr}). Water lines at lower energy levels also correlate with accretion, but with a larger scatter that can be interpreted as due to other processes that become increasingly important for the colder water reservoir.
Of these processes, radial transport of water ice towards the snowline especially through pebble drift, followed by ice sublimation that enriches the observed water vapor at low temperatures, continues to be supported by the correlations between the diagnostic line ratios and the pebble disk size (Figure \ref{fig: cool_excess_ALMA}), building evidence on modeling predictions and observational correlations emerged in previous work \cite{cc06,najita13,banz20,banz23b,schneider21,kalyaan21,kalyaan23,mah24}, Houge et al. (submitted). By combining these correlations to the new analysis of radial gradients and Doppler broadening (Figures \ref{fig: Radial_profiles} and \ref{fig: diagnostic_diagram_gradients}, and \citet{munozromero24b}), this work now suggests that drift-dominated disks have shallower temperature gradients with an extended cold disk surface enriched by ice sublimation, while disks with strong pebble traps that reduce the influx of pebbles from the outer disk have steeper temperature profiles with much reduced emission from temperatures $< 300$~K, as summarized in Figure \ref{fig: Kepl_ratios+cartoon}. }

\rev{The scatter in the larger sample in Figure \ref{fig: cool_excess_ALMA} indicates that the outer millimeter disk radius may not be the best tracer of pebble drift through the snowline, as suggested in previous work. If multiple gaps are present in a disk, as commonly observed in ALMA images \citep[][]{andrews20,bae22}, the outer radius is set by the outer gap while ice delivery through the snowline is regulated by the innermost gap \citep[as long as it provides an effective trap to pebbles,][and Easterwood et al. in press]{kalyaan21,kalyaan23}. This highlights the importance for future work to investigate the dependence of the measured water line diagnostics on the observed inner gap properties from high-resolution ALMA images.
As the sample of disks observed with MIRI grows, other system parameters that are emerging as important for pebble drift efficiency, including stellar mass, age, and multiplicity \citep[e.g.][Long et al. in press]{cy23_Sz114,grant24} should become more clear.}

\rev{One important caveat of the main diagnostic diagram based on water lines at MIRI wavelengths is that it is only partly sensitive to the snowline region expected at temperatures of 120--180~K \citep{lodders03}. In fact, the power-law models in Figure \ref{fig: diagnostic_diagram_gradients} show to be sensitive to the colder region at $< 200$~K only in the case of shallow temperature slopes. We suggest, however, that a simple power-law (even with a tapered profile in column density) may not be the best approximation to the observed water spectra. In fact, the best-fit models in \citet{munozromero24b} still under-predict the flux of the 1500~K lines, and the maximum asymmetry in the 1448/1615~K line ratio from the model grid in Figure \ref{fig: diagnostic_diagram_gradients} is only 1.1 (for cutoff radii at 10~au), lower than what measured in half of the sample included in this work (up to 1.3, Figure \ref{fig: 23um_diagram}). A more complex radial profile that accounts for a surface layer beyond the midplane snowline may be necessary (Figure \ref{fig: Kepl_ratios+cartoon}).} To analyze in detail the water abundance near and across the snowline from midplane to surface, access to lower-energy levels at $> 30$~$\mu$m will be needed \citep[e.g.][]{kzhang13,blevins16,banz23b}, which would be provided by a future far-infrared observatory \citep{pont18,pontoppidan_prima,kamp21}.

Dynamic disk models including dust evolution and water processing are beginning to unfold how the observable water columns can evolve with time under the effect of pebble drift \citep[][Houge et al. submitted]{sellek24}. It would be very interesting in future work to generate evolutionary tracks of such models in the context of the diagnostic diagrams presented in this work, to add the time dimension to the interpretation of the diagnostic line ratios measured in this and future samples. \rev{Here we note that by integrating the water mass at temperatures of $< 300$~K from the radial gradients in Figure \ref{fig: diagnostic_diagram_gradients} and converting those into a pebble mass delivered to the snowline using Equation 11 from \cite{munozromero24b} gives pebble mass fluxes between a few $10^{-5}$ M$_{\oplus}$~yr$^{-1}$ (in the case of steep temperature profiles with a small cutoff radius, representative of disks with pebble traps) and a few $10^{-4}$ up to $10^{-3}$ M$_{\oplus}$~yr$^{-1}$ (in the case of shallow temperature profiles with a larger cutoff radius, representative of drift-dominated disks), consistent with typical predictions from dust evolution models \citep[e.g.][]{birnstiel12,Drazkowska21,mulders21}.}

\begin{figure*}
\centering
\includegraphics[width=1\textwidth]{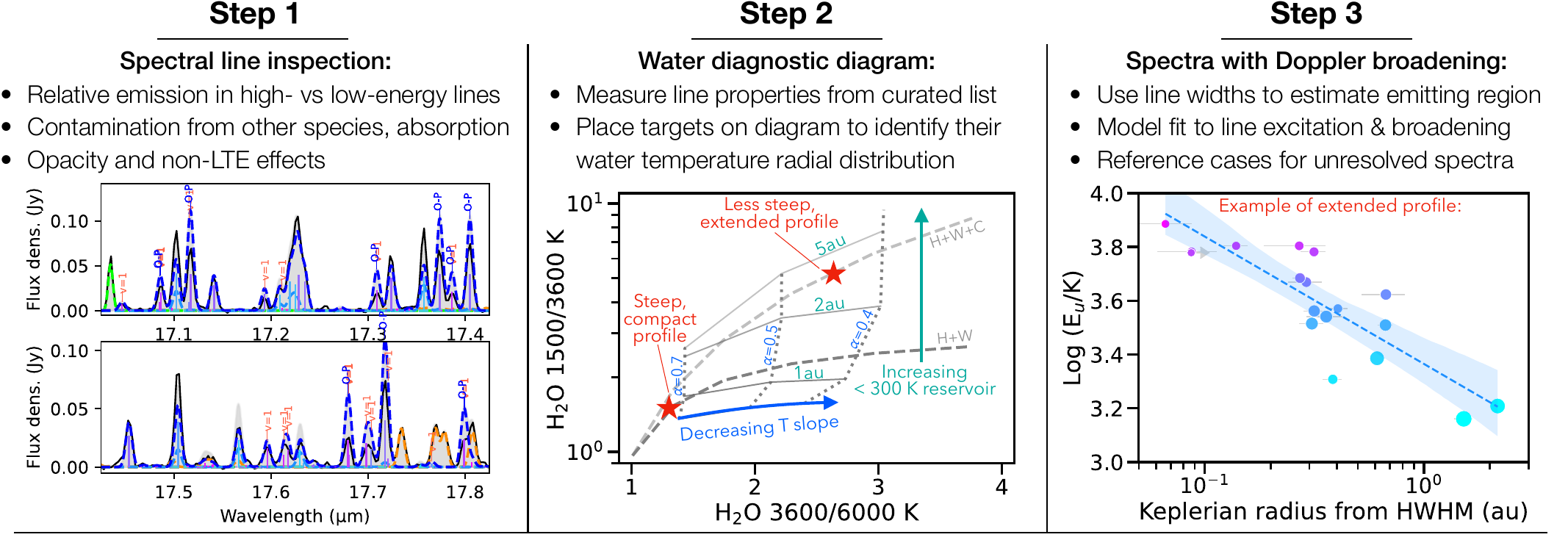}
\caption{General procedure for the analysis of water spectra based on the findings and analysis presented in this work. The plot in Step 1 shows only a narrow range extracted from Figures \ref{fig: WATER_atlas_ROVIB} to \ref{fig: WATER_atlas_ROTlong}, as an example. The plot in Step 2 shows the principal diagnostic diagram introduced and explained in Figures \ref{fig: cool_excess} and \ref{fig: diagnostic_diagram_gradients}. The plot in Step 3 gives one example of a radial excitation gradient as measured from water lines where Doppler broadening is detected (Figures \ref{fig: Eu_broadening} and \ref{fig: Radial_profiles}).}
\label{fig: roadmap}
\end{figure*}

\subsection{A procedure for the analysis of water spectra}
We propose now in Figure \ref{fig: roadmap} a simple general procedure for the analysis of water spectra, by combining the findings and tools presented above to the list of guidelines provided recently in \cite{banz23}, which was based on lessons learned from ground-based surveys of water emission from protoplanetary disks.

\subsubsection{Step 1 - Spectral line inspection}
The first step after reducing the spectra and subtracting the continuum (see Appendix \ref{app: cont_sub} for guidelines on that step) is to inspect them carefully across MIRI wavelengths for the general identification of a series of properties. These properties, related to line blending and excitation, will be useful and in some cases fundamental for a correct analysis of water emission to study physical and chemical processes in inner disks.

\paragraph{Relative emission in high- vs low-energy lines} 
With guidance from the atlas presented in this work, the identification of the general distribution of water temperature components can be visualized from the relative ratio of lines dominated by higher-$E_u$ versus lower-$E_u$. The 16--18~$\mu$m range is well suited for the hot and warm components, by including lines with $E_u$ from above 6000~K down to 2000~K (Figure 4 in \cite{banz23b} and Figure \ref{fig: Eu_broadening} in this work). The best lines at MIRI wavelengths where colder emission consistent with the snowline region emerges are two transitions around 23.85~$\mu$m \citep{kzhang13,banz23b,munozromero24b,temmink24b}, including their relative asymmetry which at lower temperatures is prominent in a stronger 1448~K line at 23.82~$\mu$m (Figure \ref{fig: Line_selection_demo}). This will give a first general impression of the global water distribution in the inner disk of a given target or sample, see e.g. Appendix \ref{app: atlas_compact} for the sample in this work.

\paragraph{Contamination from other species}
Next, before water emission lines shall be extracted and used for analysis, the atlas can be used for general reference of contamination from other species. To evaluate that carefully on individual targets, models that at least approximately reproduce the observed emission from other molecules should be used, as done in the example of CI~Tau in Figures \ref{fig: WATER_atlas_ROVIB} to \ref{fig: WATER_atlas_ROTlong}. With this process, it is possible to identify which water lines can be extracted and used for analysis without having to subtract other emission models. In this work we provide a curated list of single un-blended lines that should provide the most reliable measurements in most situations (Section \ref{sec: atlas}); in case of weak water relative to organic emission, lines at 12--$16 \mu$m should be checked for contamination and possibly removed from the list. 

\paragraph{Absorption}
If high-resolving-power infrared spectra are available from ground-based instruments, those can be used to check for the presence of absorption in the ro-vibrational lines. If not, the MIRI spectra at $<9$~$\mu$m should be inspected to identify potential blue-shifted absorption (e.g. Figure \ref{fig: IQTau_abs}), especially disks observed at inclinations $> 50$~deg (Figure \ref{fig: ROTAT_broadening}) that may generally have blue-shifted absorption by intercepting an inner disk wind \citep{pont11,banz22}.
Identifying absorption solely from MIRI data needs a strong absorption blue-shift and a broad emission component and will likely be possible only in a limited number of cases. Ground-based observations demonstrate that absorption is rather common in ro-vibrational CO spectra and may be blended into emission producing weaker line fluxes in MIRI spectra \citep[Figure 15 in][]{banz23}, making their interpretation more challenging. In those cases where the absorption blue-shift is large enough to enable detection in MIRI spectra, the ro-vibrational lines will provide a useful new probe of the molecular content of inner disk winds. To date, there is no evidence for absorption in the rotational lines \citep[except for tentative evidence in VW~Cha, see Figure 8 in][]{banz23}, which could therefore be unaffected. 

\paragraph{Opacity and non-LTE effects}
There are specific lines that should be handled carefully depending on the analysis and modeling tools that are going to be used. As shown above in Section \ref{sec: saturation}, there is a large number of ortho-para pairs of lines that overlap in wavelength, and where line opacity overlap is necessary to correctly model their combined flux. Not all modeling tools include line opacity overlap, including some of the thermo-chemical codes that are being used for the analysis of infrared spectra from disks, so these lines should in case be excluded from the modeling.

Another important case that we have illustrated in this work is the presence of several $v=1-1$ lines intermixed with $v=0-0$ lines (Section \ref{sec: v1-1lines}). These should generally be populated in non-LTE conditions and will bias the results from model fits that assume LTE, either from simple slab models or more sophisticated disk models. These lines should either be ignored when fitting spectra with LTE models, or accounted for by implementing non-LTE excitation. The measured flux in these lines could be used to characterize the density of the emitting gas in the inner hot region where they are excited \citep{meijerink09}, expanding what ground-based spectra have initially provided \citep[Section 5.1.2 in][]{banz23}. 

\subsubsection{Step 2 - Water diagnostic diagrams}
After the water spectrum has been inspected and any contaminated lines removed, the line list provided with this work can be used to measure line properties for a series of general analysis steps.
The fundamental lines listed in Table \ref{tab: cool excess line list} can be directly used to obtain the line flux ratios and place any target on Figure \ref{fig: cool_excess} for a general identification of its radial water distribution in the inner disk, whether approximated with discrete components or as a temperature gradient, and in Figure \ref{fig: 23um_diagram} for the coldest water detected at MIRI wavelengths. Figure \ref{fig: roadmap} visualizes the principal directions to interpret an object position on the diagram in terms of a decreasing temperature slope, which increases the emitting area of an intermediate 400~K component, and an increasing reservoir at lower temperatures down to ice sublimation at the snowline.
Objects that exceed the model series to the right of the diagram are increasingly dominated by a pure $\approx 400$~K component, those that exceed the diagram at the top are increasingly dominated by a colder component down to $\approx 150$~K. A comprehensive characterization of the region around the snowline will need the availability of high-resolution far-infrared spectra.

\subsubsection{Step 3 - Use Doppler line broadening, if detected}
While the water line-flux-ratio diagnostic diagrams have broad applicability to any observed MIRI spectrum, in the limited cases where Keplerian broadening by disk rotation is detected the measured line widths can be used to better characterize the radial distribution of water across the different temperatures (Figures \ref{fig: Radial_profiles} and \ref{fig: Kepl_ratios+cartoon}). Beyond the very simple Doppler mapping procedure presented above, which can be used for quick reference and comparison across objects, the advantage of using Doppler-broadened line widths will be in the simultaneous modeling of line excitation and broadening that can be done for specific cases in the future. These limited cases will also provide an important reference for comparison to the larger number of cases where lines are unresolved (or stellar mass or disk inclination may be unknown) and only the line fluxes can be used in modeling the spectra (e.g. Figure \ref{fig: Kepl_ratios+cartoon}).

From what observed in this sample, the ro-vibrational lines at $< 9$~$\mu$m are commonly Doppler-broadened due to their excitation in an innermost region within $\approx$~0.1--0.2~au, as shown by fully spectrally resolved data from the ground \citep{banz23}. In this case, future work should be able to use the observed flux and line broadening to study the density of a region near the inner disk wall, provided that non-LTE excitation is accounted for in the modeling.

\section{Summary and concluding remarks}
The study of water emission in protoplanetary disks is now reaching the end of its second decade, with a plethora of spectra obtained from ground and space observatories with a wide range of resolving powers (from 700 of Spitzer-IRS up to 90,000 of VLT-CRIRES and IRTF-iSHELL). By combining water spectra from multiple surveys, \cite{banz23} recently summarized a series of fundamental questions:

\begin{enumerate}
    \item which inner disk region(s) do the infrared water lines trace, and are there multiple water reservoirs (with different temperature and density);

    \item what is the water abundance in inner disks and what regulates it (chemistry vs dynamics);

    \item what is the relative role of different excitation processes, and is water emission in LTE;

    \item is water present in a molecular inner disk wind; and

    \item how can we correctly interpret the complex water spectra observed across infrared wavelengths within a unified picture of inner disks?
\end{enumerate}

After two years of JWST observations, steps forward have been made in most of these topics, especially the detection of multiple temperature components or a temperature gradient in water emission from inner disks (see Section \ref{sec: intro}). In this work, by analyzing MIRI-MRS high-quality spectra observed in 25 disks as part of the JDISC Survey, we have learned and demonstrated some general properties of water in inner disks, as summarized below (following the numbered questions above):

\begin{enumerate}
    \item[1.a] \rev{The line flux diagnostics introduced in this work demonstrate that water in inner disks is generally distributed with a radial temperature gradient that can be approximated with three discrete components in LTE ($\sim 850$~K, $\sim 400$~K, $\sim 170$--200~K) or a power-law profile with negative index of $\sim$~0.4--0.7 and cutoff radii of $\sim1$--10~au.} The diagnostic line ratios measured in this sample are consistent with a continuum of radial profiles, from disks with a steeper gradient dominated by a compact hot region (e.g. CI~Tau) to disks with a shallower gradient and a 2--10 times more extended region reaching ice sublimation temperatures (e.g. GK~Tau, GQ~Lup). The observed emission is typically optically thick at $\gtrsim 400$~K. 

    \item[1.b] \rev{The detection of Doppler line broadening from disk rotation demonstrates, for the first time at wavelengths $> 13 \mu$m directly from the line widths, that the observed water spectra emit from disk radii between the inner disk rim ($\lesssim 0.1$~au) and the water snowline (a few au). This discovery opens up the possibility to combine line flux and broadening measurements across MIRI-MRS spectra, from the ro-vibrational band at $< 9 \mu$m to the rotational lines at 10--28~$\mu$m, to estimate the water abundance and its evolution across the inner planet-forming region.}

    \item[2.] The line flux ratios of low (1500~K) and intermediate (3600~K) energy levels in comparison to the high levels (6000~K) anti-correlate with the ALMA dust disk radius in a selected sample of disks, supporting the role of icy pebble drift in regulating the water abundance within snowline as proposed before \citep{cc06,banz20,banz23b}. The broader sample in this work, which includes a range of dust structures (including multiple gaps and inner dust cavities) in disks of single and multiple-star systems, and slightly younger embedded objects, shows a larger spread likely due to multiple effects related to age, dust depletion, stellar multiplicity, and the depth of dust gaps that should be investigated in future work.
    
    \item[3.] Expanding what found from ground-based surveys \citep{banz23}, MIRI spectra show that the excitation of the entire ro-vibrational band at 5--8~$\mu$m is not in LTE and the broader line widths in this band demonstrate a smaller emitting radius than the $v=0-0$ rotational band. Evidence for non-LTE excitation is demonstrated for the first time also in rotational lines in the first vibrational level ($v=1-1$). The relative excitation of these different bands should enable obtaining estimates of the molecular gas density across inner disk radii.

    \item[4.] Water is indeed present in a dense molecular region of inner disk winds at $\approx 1$~Myr, as shown for the first time by detecting blue-shifted absorption in ro-vibrational lines observed from the high-inclination disk of IQ~Tau. Going forward, under specific geometric conditions MIRI spectra may provide a new probe of the physical and chemical composition of inner disk winds close to their launching radii at the disk surface.

    \item[5.] In this work, we have presented a number of guidelines and tools to identify, analyze, and interpret different excitation and broadening effects in water spectra as observed with MIRI-MRS, which should be beneficial to the community for producing reliable analyses and homogeneous comparisons across large datasets in the future. \rev{In particular, the diagnostic diagrams introduced in this work (Figures \ref{fig: cool_excess}, \ref{fig: 23um_diagram}, and \ref{fig: diagnostic_diagram_gradients}) will make comparative analyses across $>100$ disks (the total sample obtained by the end of Cycle 4) sustainable without extensive computing time and without depending on the different assumptions and limitations of different modeling tools.}

\end{enumerate}

\rev{This work demonstrates that protoplanetary disk spectra observed with MIRI-MRS provide a large content of information (the spectral line flux distribution and line broadening as a function of upper level energy) about the distribution in temperature and density of water in planet-forming regions from the inner disk rim out to the snowline, highlighting the role of JWST as a leading observatory that will continue to deliver new discoveries on the origins of planetary chemistry (including water) for many years.}
While water emission from temperatures down to ice sublimation is detected in some disks \citep[Figure \ref{fig: 23um_diagram}, and][]{munozromero24b}, with only a handful of water lines with $E_u \sim$~1000--1500~K observed with MIRI a detailed characterization of the water abundance across the snowline will necessitate a high-resolution far-infrared observatory.


\acknowledgments
We thank the referee for multiple suggestions that improved the clarity and usefulness of this work.
This work includes observations made with the NASA/ESA/CSA James Webb Space Telescope. The JWST data used in this paper can be found in MAST: \dataset[10.17909/7w5s-f430]{http://dx.doi.org/10.17909/7w5s-f430}. The data were obtained from the Mikulski Archive for Space Telescopes at the Space Telescope Science Institute, which is operated by the Association of Universities for Research in Astronomy, Inc., under NASA contract NAS 5-03127 for JWST. The observations are associated with JWST GO Cycle 1 programs 1549, 1584, and 1640.
Part of this research was carried out at the Jet Propulsion Laboratory, California Institute of Technology, under a contract with the National Aeronautics and Space Administration (80NM0018D0004).
The authors acknowledge support from NASA/Space Telescope Science Institute grants: JWST-GO-01640, JWST-GO-01584, and JWST-GO-01549.
G.A.B. gratefully acknowledges support from NASA grant 80NSSC24K0149.

\facilities{JWST}

\software{
Matplotlib \citep{matplotlib}, NumPy \citep{numpy}, SciPy \citep{scipy}, Seaborn \citep{seaborn}, Astropy \citep{astropy:2013, astropy:2018, astropy:2022}, LMFIT \citep{lmfit}, iSLAT \citep{iSLAT_code}, spectools\_ir \citep{spectools_ir}
}

\begin{deluxetable}{l c c c c c c}
\tabletypesize{\small}
\tablewidth{0pt}
\tablecaption{\label{tab: sample props} Sample properties used in this work as taken from the literature.}
\tablehead{Name & Dist. & $M_{\star}$ & log $L_{\rm{acc}}$ & Incl. & $R_{\rm{disk}}$ & Notes \\
 & (pc) & ($M_{\odot}$) & ($L_{\odot}$) & (deg) & (au) & }
\tablecolumns{7}
\startdata
AS205N & 127 & 0.87 & -0.07 & 20.0 & 50 & binary \\
AS209 & 121 & 0.96 & -1.12 & 35.0 & 138 & \nodata \\
CITau & 160 & 0.71 & -0.87 & 50.0 & 191 & \nodata \\
DoAr25 & 139 & 0.62 & $<$ -2.13 & 67.4 & 166 & cloud \\
DoAr33 & 143 & 0.69 & \nodata & 42.0 & 27 & \nodata \\
Elias20 & 138 & 0.48 & -0.09 & 49.0 & 65 & cloud \\
Elias24 & 143 & 0.78 & 0.44 & 29.0 & 135 & cloud \\
Elias27 & 118 & 0.49 & -0.57 & 56.0 & 257 & cloud \\
FZTau & 129 & 0.51 & 0.34 & 22.0 & 12 & \nodata \\
GKTau & 129 & 0.67 & -1.38 & 38.8 & 13 & \nodata \\
GOTau & 139 & 0.35 & -2.0 & 53.9 & 170 & cloud \\
GQLup & 151 & 0.78 & -0.36 & 60.5 & 56 & binary \\
HPTau & 177 & 1.20 & \nodata & 18.3 & 22 & \nodata \\
HTLup & 154 & 1.27 & -1.18 & 48.0 & 25 & binary \\
IQTau & 131 & 0.50 & -1.4 & 62.1 & 110 & \nodata \\
IRAS-04385 & 160 & 0.50 & \nodata & \nodata & 22 & cloud/HH \\
MYLup & 157 & 1.23 & $<$ -2.3 & 73.2 & 87 & high incl. \\
RULup & 158 & 0.55 & -0.01 & 19.0 & 63 & \nodata \\
RYLup & 158 & 1.27 & -1.4 & 68.0 & 135 & cavity \\
SR4 & 132 & 0.68 & -0.12 & 22.0 & 31 & cloud\\
Sz114 & 162 & 0.17 & -2.7 & 21.0 & 60 & cloud \\
Sz129 & 159 & 0.83 & -1.14 & 34.0 & 76 & cavity \\
TWCha & 183 & 0.70 & -1.54 & 31.0 & 53 & cavity \\
VZCha & 191 & 0.50 & -0.33 & 16.0 & 39 & \nodata \\
WSB52 & 142 & 0.48 & -1.11 & 54.0 & 32 & cloud\\
\enddata
\tablecomments{References -- distances are from GAIA \citep{gaia_mission,gaiaDR3}, stellar and accretion properties are from \cite{simon16,Fang18,McClure19,alcala17,alcala19,gangi22,Manara23}; $R_{\rm{disk}}$ (taken as the radius enclosing 90--95\% of emission depending on what reported in the original work) and disk inclinations are from \citet{long18,tazzari18,MacGregor17,Ansdell18,long19,huang18,kurtovic18,hendler20,Ribas20}. Cloud contamination as a sign of a more embedded younger object is reported when moderate to severe, as identified in \cite{dsharp,long22}. IRAS-04385 is IRAS~04385+2550 (Haro~6-33), a more embedded disk in Taurus associated with the Herbig-Haro object HH~408 \citep{stapelfeldt99,schaefer09,bally12}. }
\end{deluxetable}

\newpage
\appendix

\section{Sample properties and measurements} \label{app: sample measurements}
Tables \ref{tab: sample props}, \ref{tab: flux measurements}, and \ref{tab: width measurements} report the sample properties and some fundamental measurements extracted in this work. All line measurements in this work are extracted using the ``fit saved lines" function in iSLAT, which implements the least-square minimization code \texttt{lmfit} \citep{lmfit} to perform single-Gaussian fits and measure the line centroid, FWHM, and flux and their uncertainties. The pixel flux errors are adopted as estimated from the JDISCS pipeline \citep{pontoppidan24}. When a line is not detected, the integrated line flux is just the integral of the pixels in the line range (which could be negative, if there are more negative pixels) and its error is from the propagation of pixel flux errors and upper limits reported in figures in this work use the 2-$\sigma$ flux error.

\begin{deluxetable*}{l c c c c c c c c c c}
\rotate
\tabletypesize{\small}
\tablewidth{0pt}
\tablecaption{\label{tab: flux measurements} Line flux measurements from this work.}
\tablehead{Target & 1500 K & 3600 K & 6000 K & 3340 K (a) & 3340 K (b) & 1448 K & 1615 K & $v=0-0$ &  $v=1-0$ & $v=1-1$}
\tablecolumns{11}
\startdata
AS205N & 140.79(11.41) & 63.19(3.10) & 28.26(2.34) & 58.78(7.06) & 42.33(1.61) & 74.79(6.09) & 66.00(5.32) & 45.74(4.02) & 23.46(2.83) & 3.33(2.67) \\
AS209 & 12.47(3.79) & 6.34(0.76) & 4.09(0.41) & 2.75(0.97) & 3.23(1.44) & 6.74(1.94) & 5.73(1.86) & 5.19(1.21) & 4.34(4.13) & 1.92(1.06) \\
CITau & 10.07(0.35) & 6.51(0.12) & 4.83(0.22) & 4.77(0.23) & 5.57(0.16) & 4.75(0.16) & 5.32(0.19) & 6.24(0.12) & 2.74(0.30) & 1.25(0.01) \\
DoAr25 & 3.70(0.48) & 2.39(0.11) & 1.19(0.08) & 2.59(0.06) & 1.61(0.30) & 1.78(0.25) & 1.93(0.23) & 2.04(0.14) & 0.46(0.16) & 0.27(0.12) \\
DoAr33 & 2.60(0.17) & 1.58(0.06) & 0.71(0.05) & 1.95(0.36) & 1.23(0.10) & 1.21(0.15) & 1.39(0.02) & 1.22(0.04) & 0.67(0.26) & 0.24(0.08) \\
Elias20 & 45.83(0.67) & 15.48(0.38) & 6.34(0.23) & 11.75(2.12) & 13.63(0.37) & 23.57(0.55) & 22.26(0.12) & 10.60(0.51) & 2.32(0.65) & 1.90(0.19) \\
Elias24 & 104.82(1.36) & 32.91(0.80) & 13.37(0.45) & 24.73(3.18) & 31.39(0.45) & 54.70(0.04) & 50.12(1.32) & 24.61(0.84) & 4.62(4.41) & 5.39(0.68) \\
Elias27 & 21.92(0.39) & 10.77(0.17) & 5.22(0.17) & 9.12(0.29) & 7.99(0.51) & 11.36(0.09) & 10.56(0.30) & 8.38(0.22) & 3.06(0.37) & 1.46(0.13) \\
FZTau & 43.34(0.71) & 26.87(1.15) & 14.43(0.77) & 23.53(1.67) & 20.60(0.75) & 21.15(0.65) & 22.19(0.06) & 21.93(1.05) & 12.09(1.06) & 3.34(0.04) \\
GKTau & 22.27(0.63) & 6.44(0.29) & 3.32(0.24) & 4.22(0.52) & 6.72(0.20) & 12.19(0.40) & 10.08(0.22) & 4.80(0.19) & 1.45(0.41) & 0.64(0.14) \\
GOTau & 0.81(0.16) & 0.49(0.05) & 0.20(0.02) & 0.43(0.11) & 0.40(0.12) & 0.43(0.10) & 0.37(0.05) & 0.31(0.02) & 0.12(0.07) & -0.07(0.07) \\
GQLup & 30.80(0.42) & 7.12(0.33) & 3.00(0.17) & 2.34(0.28) & 8.60(0.45) & 16.50(0.30) & 14.30(0.12) & 4.28(0.13) & 1.89(0.74) & 0.17(0.33) \\
HPTau & 10.97(0.97) & 4.32(0.24) & 2.02(0.31) & 2.39(0.27) & 4.14(0.36) & 6.14(0.42) & 4.83(0.55) & 2.73(0.31) & 1.41(0.47) & -0.48(0.50) \\
HTLup & 8.01(1.84) & 6.26(0.40) & 4.02(0.73) & 4.78(0.51) & 2.80(0.52) & 3.79(0.86) & 4.22(0.98) & 5.86(0.29) & 5.83(1.34) & 0.31(1.01) \\
IQTau & 9.85(0.25) & 3.93(0.18) & 2.88(0.16) & 2.21(0.32) & 3.03(0.05) & 5.60(0.13) & 4.25(0.12) & 4.00(0.27) & 1.29(0.41) & 0.61(0.02) \\
IRAS-04385 & 21.85(1.15) & 3.98(0.18) & 1.16(0.07) & 3.94(0.38) & 3.24(0.89) & 12.47(0.76) & 9.38(0.40) & 2.42(0.05) & 0.63(0.14) & 0.06(0.38) \\
MYLup & 1.65(0.34) & 0.24(0.05) & 0.05(0.07) & 1.19(0.56) & 0.49(0.12) & 0.84(0.16) & 0.80(0.18) & -0.06(0.05) & -0.01(0.11) & 0.12(0.10) \\
RULup & 37.79(2.40) & 23.88(1.07) & 14.06(0.72) & 17.37(1.02) & 15.96(1.72) & 19.28(1.50) & 18.51(0.90) & 19.88(0.61) & 9.57(0.75) & 1.22(0.90) \\
RYLup & 5.55(0.73) & 1.39(0.45) & 0.42(0.27) & 0.19(0.23) & 0.26(0.34) & 2.98(0.42) & 2.56(0.30) & 0.23(0.21) & -0.05(0.60) & -0.24(0.35) \\
SR4 & 9.79(1.36) & 6.48(0.49) & 5.08(0.22) & 3.59(0.68) & 5.61(0.58) & 4.78(0.62) & 5.02(0.74) & 5.02(0.14) & 2.61(0.20) & -0.29(0.81) \\
Sz114 & 9.62(0.57) & 2.96(0.06) & 1.20(0.07) & 2.08(0.40) & 2.26(0.21) & 5.34(0.24) & 4.28(0.32) & 1.95(0.05) & 1.56(0.08) & 0.25(0.06) \\
Sz129 & 8.17(0.22) & 2.46(0.06) & 1.29(0.04) & 1.04(0.08) & 2.47(0.17) & 4.49(0.11) & 3.68(0.12) & 1.73(0.04) & 0.60(0.11) & 0.12(0.18) \\
TWCha & 16.62(0.29) & 5.61(0.11) & 3.02(0.09) & 2.44(0.06) & 5.86(0.14) & 8.86(0.10) & 7.76(0.19) & 4.16(0.15) & 1.82(0.19) & 0.65(0.07) \\
VZCha & 15.59(0.15) & 6.50(0.24) & 4.50(0.19) & 3.83(0.35) & 6.14(0.03) & 8.10(0.12) & 7.49(0.03) & 5.67(0.27) & 2.24(0.26) & 1.22(0.21) \\
WSB52 & 57.58(1.74) & 27.69(0.40) & 11.06(0.29) & 30.29(0.26) & 20.31(0.94) & 29.67(0.96) & 27.91(0.77) & 20.83(0.49) & 7.94(0.33) & 3.58(0.12) \\
\enddata
\tablecomments{Lines are labeled as defined in Table \ref{tab: cool excess line list}. Line fluxes are reported in units of $10^{-15}$ erg s$^{-1}$ cm$^{-2}$. 1-$\sigma$ uncertainties are shown in parentheses.}
\end{deluxetable*}

\begin{deluxetable*}{l c c c c c c c c c c c c c}
\rotate
\tabletypesize{\small}
\tablewidth{0pt}
\tablecaption{\label{tab: width measurements} Line FWHM measurements and Keplerian radii estimates from this work.}
\tablehead{Target & 1500 K & $R_{1500}$ & 3600 K & $R_{3600}$ & 6000 K & $R_{6000}$ & Ro-vib. & $< 18 \mu$m & $> 18 \mu$m & OH 11,000~K & OH 4000 K & CO P26 & \ce{H2}}
\tablecolumns{14}
\startdata
AS205N & 119(7) & \nodata & 104(6) & \nodata & 118(9) & 0.05(0.01) & 104(9) & 109(8) & 138(7) & 134(14) & 124(15) & 111(6) & 79(2) \\
AS209 & \nodata & \nodata & 151(19) & 0.09(0.03) & 160(24) & 0.07(0.03) & 126(18) & 147(26) & 167(49) & 101(15) & 276(70) & 171(6) & 87(4) \\
CITau & 131(3) & 0.42(0.07) & 134(4) & 0.21(0.02) & 125(3) & 0.22(0.02) & 128(8) & 131(6) & 160(12) & 108(8) & 157(1) & 148(3) & 95(2) \\
DoAr25 & 123(13) & \nodata & 156(6) & 0.14(0.01) & 140(12) & 0.13(0.02) & 167(33) & 152(12) & 172(26) & 107(23) & 304(27) & 189(20) & 99(11) \\
DoAr33 & 145(7) & 0.16(0.03) & 136(6) & 0.14(0.02) & 150(12) & 0.09(0.02) & 153(18) & 136(10) & 155(12) & 216(59) & 139(15) & 158(41) & 84(2) \\
Elias20 & 126(1) & 0.66(0.05) & 111(4) & 0.40(0.06) & 122(5) & 0.17(0.02) & 130(56) & 115(10) & 138(8) & 148(12) & 139(9) & 119(10) & 87(4) \\
Elias24 & 124(1) & 0.63(0.05) & 112(4) & 0.33(0.06) & 107(4) & 0.15(0.03) & 104(12) & 110(8) & 140(6) & 115(14) & 136(5) & 122(5) & 76(6) \\
Elias27 & 130(2) & 0.45(0.03) & 110(4) & 0.43(0.07) & 115(2) & 0.28(0.02) & 114(2) & 113(6) & 139(6) & 132(4) & 137(4) & 124(9) & 84(1) \\
FZTau & 121(1) & \nodata & 107(4) & \nodata & 104(4) & \nodata & 101(5) & 104(4) & 135(9) & 98(4) & 140(4) & 117(10) & 101(2) \\
GKTau & 126(3) & 0.57(0.07) & 122(4) & 0.17(0.02) & 116(7) & 0.15(0.03) & 122(6) & 121(8) & 158(15) & 104(7) & 143(4) & 139(4) & 106(16) \\
GOTau & 124(18) & \nodata & 143(9) & 0.08(0.01) & 138(15) & 0.08(0.03) & 122(43) & 142(12) & 152(26) & 164(25) & 170(3) & 167(19) & 85(4) \\
GQLup & 125(1) & 1.78(0.13) & 128(5) & 0.36(0.04) & 136(13) & 0.19(0.05) & 113(24) & 127(8) & 148(22) & 102(9) & 160(10) & 124(4) & 93(4) \\
HPTau & 119(8) & \nodata & 136(10) & 0.05(0.02) & 114(26) & \nodata & 118(17) & 127(9) & 142(31) & 94(8) & 185(14) & 111(14) & 96(21) \\
HTLup & 118(20) & \nodata & 150(7) & 0.21(0.03) & 146(18) & 0.20(0.07) & 175(22) & 149(8) & 167(41) & 138(38) & \nodata & 187(5) & 57(5) \\
IQTau & 136(3) & 0.34(0.03) & 179(7) & 0.07(0.01) & 162(11) & 0.09(0.01) & 94(12) & 172(10) & 180(31) & 121(7) & 185(6) & 103(18) & 99(4) \\
IRAS-04385 & 118(4) & \nodata & 120(8) & \nodata & 124(19) & \nodata & 138(7) & 122(13) & 148(14) & 126(25) & \nodata & 145(16) & 72(6) \\
MYLup & 122(22) & \nodata & 166(30) & \nodata & \nodata & \nodata & \nodata & 154(35) & 151(38) & 118(18) & \nodata & 304(130) & 96(2) \\
RULup & 118(6) & \nodata & 107(3) & \nodata & 110(5) & \nodata & 99(5) & 106(6) & 139(16) & 92(4) & 136(6) & 114(4) & 92(3) \\
RYLup & 110(10) & \nodata & 187(59) & \nodata & \nodata & \nodata & 110(10) & 178(47) & 114(17) & 77(19) & 141(5) & 136(14) & 79(9) \\
SR4 & 112(11) & \nodata & 124(8) & 0.07(0.01) & 140(15) & 0.04(0.01) & 108(8) & 122(4) & 144(38) & 106(10) & 142(3) & 127(7) & 111(21) \\
Sz114 & 117(5) & \nodata & 112(5) & \nodata & 99(6) & \nodata & 97(4) & 105(10) & 134(10) & 136(18) & 134(9) & 100(2) & 90(4) \\
Sz129 & 127(3) & 0.49(0.06) & 116(4) & 0.32(0.05) & 123(4) & 0.18(0.02) & 117(11) & 119(6) & 142(13) & 126(9) & 141(6) & 120(9) & 84(3) \\
TWCha & 127(2) & 0.35(0.03) & 111(2) & 0.33(0.03) & 106(2) & 0.28(0.02) & 103(4) & 106(5) & 143(12) & 109(3) & 134(2) & 118(8) & 95(6) \\
VZCha & 126(1) & 0.08(0.01) & 112(3) & 0.06(0.01) & 107(3) & 0.05(0.01) & 101(6) & 106(4) & 141(9) & 92(8) & 131(4) & 113(10) & 100(1) \\
WSB52 & 127(3) & 0.64(0.09) & 117(3) & 0.38(0.04) & 122(3) & 0.26(0.02) & 117(4) & 117(6) & 144(6) & 115(6) & 150(10) & 146(3) & 100(1) \\
\enddata
\tablecomments{Lines are labeled as defined in Table \ref{tab: cool excess line list} and shown in Figure \ref{fig: ROTAT_broadening}. Line FWHM are reported in units of km~s$^{-1}$. 1-$\sigma$ uncertainties are shown in parentheses. Keplerian radii (in units of au) are reported only where Doppler broadening is detected.}
\end{deluxetable*}

\section{Continuum subtraction} \label{app: cont_sub}
The continuum-subtraction algorithm presented in \cite{pontoppidan24}\footnote{The code is available at \url{https://github.com/pontoppi/ctool}.} is a very effective empirical procedure designed to remove broad dust features under narrow gas line emission. As noted in the original paper, the procedure may subtract a small fraction of the gas emission in some regions of dense clustering of lines. For this reason, in applying the method to this sample we excluded the following regions: 6.4--6.91~$\mu$m for the most densely clustered part of the ro-vibrational water bands, 7.45--7.515~$\mu$m in case of strong HI emission (which sits on a cluster of water lines, see Figure \ref{fig: WATER_atlas_ROVIB}), 13.4--14.1~$\mu$m for the broad Q-branches of HCN and \ce{C2H2}, and 14.9--15~$\mu$m for the Q-branch of \ce{CO2} when stronger than the nearby water emission. We also exclude the region at $> 28$~$\mu$m where the MRS sensitivity drops \citep{pontoppidan24}. A modification we make to the algorithm is to apply different smoothing pixel windows and number of iterations at short versus long wavelengths, with a separation at 8--10$\mu$m; this is found necessary in most spectra in this sample to account for the different clustering of lines, the different spectral resolution, and the different dust features in the two wavelength ranges, with the long wavelengths often including more dust features \citep[see e.g. the case of GK~Tau shown in][]{banz23b}.

\begin{figure*}
\centering
\includegraphics[width=0.87\textwidth]{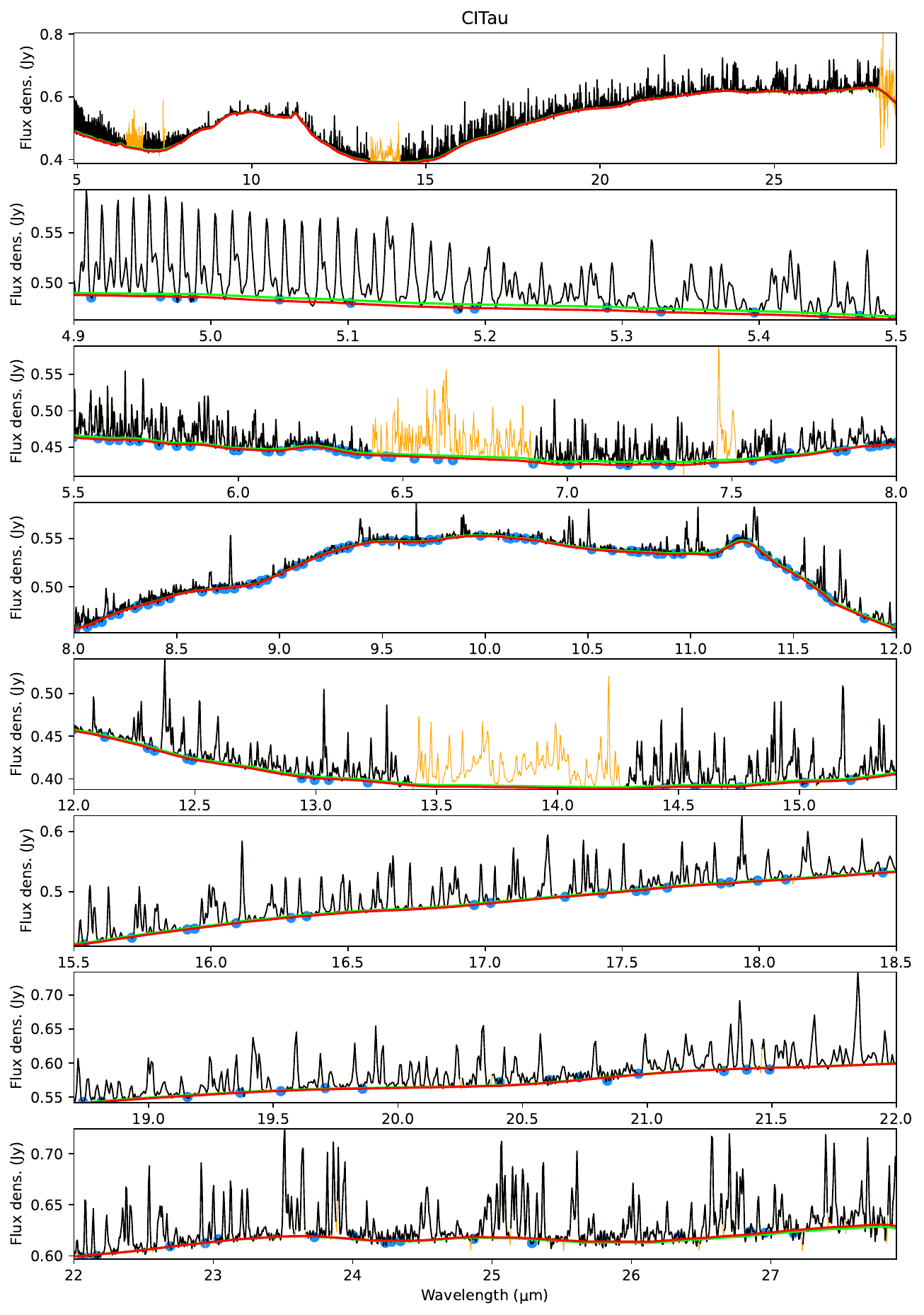}
\caption{Example of the continuum subtraction procedure as applied to CI~Tau. The entire original spectrum is shown at the top in black, and divided into smaller portions in the other panels. The excluded regions listed in Section \ref{app: cont_sub} are colored in orange. The continuum estimated with the procedure described in \cite{pontoppidan24} is shown in light green. The red line shows the continuum after applying a wavelength-dependent offset determined from the line-free regions marked with blue points (see Section \ref{app: cont_sub}).}
\label{fig: cont_sub}
\end{figure*}

Additionally, we use 200 line-free regions identified from the slab models in Figures \ref{fig: WATER_atlas_ROVIB} to \ref{fig: WATER_atlas_ROTlong} to apply a final small wavelength-dependent offset informed on where the flux is expected to be dominated by dust continuum. We find that this final step helps in getting closer to the underlying continuum especially at $<8$~$\mu$m, where the clustered ro-vibrational bands of CO and water produce a pseudo-continuum, and at $> 20$~$\mu$m, where fringe residuals would otherwise be interpreted as emission lines in the original code, pushing the continuum too low. This final offset turns out to also be important when molecular absorption is present, since the procedure from \cite{pontoppidan24} is built on the assumption that any gas feature in the spectrum is in emission (i.e. the algorithm assumes the continuum to be at the bottom of the spectrum, not at the top). Absorption spectra from a molecular inner disk wind (like the one identified in IQ~Tau in Figure \ref{fig: IQTau_abs}) or a stellar photosphere (e.g. in MY~Lup as identified in Salyk et al. submitted, and included in Figure \ref{fig: MIRI_spectra_atlas3r}) would instead have the continuum in between or at the top of any gas features, and the line-free regions are fundamental in identifying the level of the actual continuum especially in these cases (see the application of this procedure in Long et al. 2024, in press). An example of this procedure as applied to CI~Tau is shown in Figure \ref{fig: cont_sub}.

\section{Line list used for the analysis in this work} \label{app: line list}
Tables \ref{tab: total line list} and \ref{tab: total line list continued} report the list of single un-blended water transitions defined in Section \ref{sec: atlas} and used for the analysis in this work. The entire line list has been added to the GitHub repository of iSLAT and named ``MIRI\_general''. In addition to the water lines, we have included the un-blended CO transitions ($v=1-0$ P26 and P27, $v=2-1$ P23 and P27) and some coincident pairs of OH transitions that are used for reference to water in this work, in addition to the single transition at 14.62~$\mu$m (Figures \ref{fig: Res_Pow_MIRI} and \ref{fig: Radial_profiles}).

\begin{deluxetable}{l l c c}
\tabletypesize{\small}
\tablewidth{0pt}
\tablecaption{\label{tab: total line list} List of single un-blended water transitions defined in Section \ref{sec: atlas} and used for the analysis in this work. }
\tablehead{\colhead{Wavelength} & \colhead{Transitions (upper-lower)} & \colhead{$A_{ul}$} & \colhead{$E_{u}$} \\
\colhead{($\mu$m)} & \colhead{(level format: $v_1 v_2 v_3~~J_{\:K_a \:K_c}$)} & \colhead{(s$^{-1}$)} & \colhead{(K)} }
\tablecolumns{4}
\startdata
5.34529 & 010-000 \: $10_{\:3\:8} - 9_{\:2\:7}$ & 5.81 & 4420 \\
5.63179 & 010-000 \: $7_{\:2\:6} - 6_{\:1\:5}$ & 6.87 & 3335 \\
5.64107 & 010-000 \: $3_{\:3\:0} - 2_{\:2\:1}$ & 5.82 & 2744 \\
6.07545 & 010-000 \: $3_{\:2\:1} - 3_{\:1\:2}$ & 6.86 & 2617 \\
6.14316 & 010-000 \: $2_{\:0\:2} - 1_{\:1\:1}$ & 3.78 & 2395 \\
6.1854 & 010-000 \: $1_{\:1\:0} - 1_{\:0\:1}$ & 10.94 & 2360 \\
6.34443 & 010-000 \: $1_{\:0\:1} - 1_{\:1\:0}$ & 12.63 & 2328 \\
6.43355 & 010-000 \: $5_{\:1\:4} - 5_{\:2\:3}$ & 11.52 & 2878 \\
6.49224 & 010-000 \: $2_{\:1\:2} - 3_{\:0\:3}$ & 7.21 & 2412 \\
6.52896 & 010-000 \: $7_{\:3\:4} - 7_{\:4\:3}$ & 9.45 & 3543 \\
6.97738 & 010-000 \: $9_{\:0\:9} - 9_{\:1\:8}$ & 4.24 & 3614 \\
6.99328 & 010-000 \: $3_{\:2\:2} - 4_{\:3\:1}$ & 8.57 & 2609 \\
7.14692 & 010-000 \: $4_{\:2\:3} - 5_{\:3\:2}$ & 5.63 & 2745 \\
7.21253 & 010-000 \: $7_{\:2\:5} - 8_{\:3\:6}$ & 4.48 & 3442 \\
7.27924 & 010-000 \: $5_{\:3\:2} - 6_{\:4\:3}$ & 8.01 & 3065 \\
7.30659 & 010-000 \: $5_{\:3\:3} - 6_{\:4\:2}$ & 7.83 & 3059 \\
8.0696 & 010-000 \: $10_{\:5\:6} - 11_{\:6\:5}$ & 5.95 & 4867 \\
9.90602 & 000-000 \: $21_{\:4\:17} - 20_{\:3\:18}$ & 10.81 & 8270 \\
10.1132 & 000-000 \: $17_{\:7\:10} - 16_{\:4\:13}$ & 1.56 & 6371 \\
10.76435 & 000-000 \: $14_{\:9\:6} - 13_{\:6\:7}$ & 0.76 & 5302 \\
10.85307 & 000-000 \: $15_{\:6\:9} - 14_{\:3\:12}$ & 0.66 & 4996 \\
11.00168 & 000-000 \: $12_{\:6\:7} - 11_{\:1\:10}$ & 0.07 & 3501 \\
11.17771 & 000-000 \: $20_{\:8\:13} - 19_{\:5\:14}$ & 14.51 & 8556 \\
11.26877 & 000-000 \: $20_{\:7\:14} - 19_{\:4\:15}$ & 16.82 & 8257 \\
11.64764 & 000-000 \: $17_{\:3\:14} - 16_{\:2\:15}$ & 4.61 & 5483 \\
11.70161 & 000-000 \: $13_{\:5\:8} - 12_{\:2\:11}$ & 0.20 & 3783 \\
11.96812 & 000-000 \: $14_{\:8\:7} - 13_{\:5\:8}$ & 1.27 & 4985 \\
12.26544 & 000-000 \: $18_{\:7\:12} - 17_{\:4\:13}$ & 12.28 & 6953 \\
12.5645 & 000-000 \: $10_{\:6\:5} - 9_{\:1\:8}$ & 0.04 & 2697 \\
12.89409 & 000-000 \: $12_{\:5\:7} - 11_{\:2\:10}$ & 0.26 & 3310 \\
12.98575 & 000-000 \: $12_{\:7\:5} - 11_{\:4\:8}$ & 0.74 & 3759 \\
13.13243 & 000-000 \: $16_{\:7\:10} - 15_{\:4\:11}$ & 7.03 & 5763 \\
13.29319 & 000-000 \: $15_{\:3\:12} - 14_{\:2\:13}$ & 3.78 & 4431 \\
13.31231 & 000-000 \: $16_{\:4\:12} - 15_{\:3\:13}$ & 6.83 & 5213 \\
13.50312 & 000-000 \: $11_{\:7\:4} - 10_{\:4\:7}$ & 0.49 & 3340 \\
14.34608 & 000-000 \: $14_{\:3\:11} - 13_{\:2\:12}$ & 3.38 & 3941 \\
14.42757 & 000-000 \: $15_{\:4\:11} - 14_{\:3\:12}$ & 6.09 & 4668 \\
14.51301 & 000-000 \: $13_{\:2\:11} - 12_{\:1\:12}$ & 1.23 & 3232 \\
14.89513 & 000-000 \: $14_{\:5\:10} - 13_{\:2\:11}$ & 5.49 & 4198 \\
15.62568 & 000-000 \: $13_{\:3\:10} - 12_{\:2\:11}$ & 2.99 & 3474 \\
\hline
\enddata
\tablecomments{Line properties are from HITRAN \citep{hitran20}. The full line list, named ``MIRI\_general'', is included in iSLAT at \url{https://github.com/spexod/iSLAT}.}
\end{deluxetable}

\begin{deluxetable}{l l c c}
\tabletypesize{\small}
\tablewidth{0pt}
\tablecaption{\label{tab: total line list continued} List of single un-blended water transitions defined in Section \ref{sec: atlas} and used for the analysis in this work (continued).}
\tablehead{\colhead{Wavelength} & \colhead{Transitions (upper-lower)} & \colhead{$A_{ul}$} & \colhead{$E_{u}$} \\
\colhead{($\mu$m)} & \colhead{(level format: $v_1 v_2 v_3~~J_{\:K_a \:K_c}$)} & \colhead{(s$^{-1}$)} & \colhead{(K)} }
\tablecolumns{4}
\startdata
15.83495 & 000-000 \: $18_{\:8\:10} - 17_{\:7\:11}$ & 43.99 & 7252 \\
15.96622 & 000-000 \: $13_{\:5\:9} - 12_{\:2\:10}$ & 4.68 & 3721 \\
16.27136 & 000-000 \: $15_{\:5\:10} - 14_{\:4\:11}$ & 9.23 & 4835 \\
16.50525 & 000-000 \: $17_{\:7\:10} - 16_{\:6\:11}$ & 28.79 & 6371 \\
16.54402 & 000-000 \: $11_{\:6\:6} - 10_{\:3\:7}$ & 1.37 & 3082 \\
16.59123 & 000-000 \: $16_{\:9\:7} - 15_{\:8\:8}$ & 56.44 & 6369 \\
17.10254 & 000-000 \: $12_{\:5\:8} - 11_{\:2\:9}$ & 3.78 & 3273 \\
17.14148 & 000-000 \: $16_{\:8\:8} - 15_{\:7\:9}$ & 42.49 & 6053 \\
17.19352 & 010-010 \: $13_{\:4\:9} - 12_{\:3\:10}$ & 6.27 & 6005 \\
17.32395 & 000-000 \: $16_{\:8\:9} - 15_{\:7\:8}$ & 41.53 & 6051 \\
17.35766 & 000-000 \: $11_{\:2\:9} - 10_{\:1\:10}$ & 0.96 & 2432 \\
17.50436 & 000-000 \: $13_{\:4\:9} - 12_{\:3\:10}$ & 4.94 & 3645 \\
17.56683 & 000-000 \: $8_{\:6\:3} - 7_{\:3\:4}$ & 0.15 & 2030 \\
17.59626 & 010-010 \: $15_{\:7\:8} - 14_{\:6\:9}$ & 35.12 & 7686 \\
18.25429 & 000-000 \: $11_{\:5\:7} - 10_{\:2\:8}$ & 2.81 & 2857 \\
19.12996 & 000-000 \: $15_{\:7\:9} - 14_{\:6\:8}$ & 27.00 & 5214 \\
19.24597 & 000-000 \: $11_{\:3\:8} - 10_{\:2\:9}$ & 2.27 & 2608 \\
19.34995 & 000-000 \: $10_{\:5\:6} - 9_{\:2\:7}$ & 1.83 & 2472 \\
19.68805 & 000-000 \: $14_{\:7\:8} - 13_{\:6\:7}$ & 27.30 & 4696 \\
20.42595 & 000-000 \: $13_{\:7\:7} - 12_{\:6\:6}$ & 27.03 & 4211 \\
20.66181 & 000-000 \: $7_{\:4\:3} - 6_{\:1\:6}$ & 0.07 & 1339 \\
21.33317 & 000-000 \: $12_{\:7\:6} - 11_{\:6\:5}$ & 26.27 & 3759 \\
21.61495 & 000-000 \: $17_{\:6\:12} - 16_{\:5\:11}$ & 16.87 & 6073 \\
21.7488 & 000-000 \: $15_{\:6\:10} - 14_{\:5\:9}$ & 15.59 & 4953 \\
22.08091 & 000-000 \: $11_{\:4\:7} - 10_{\:3\:8}$ & 4.58 & 2732 \\
22.13775 & 000-000 \: $12_{\:6\:6} - 11_{\:5\:7}$ & 17.69 & 3507 \\
22.37473 & 000-000 \: $11_{\:7\:4} - 10_{\:6\:5}$ & 25.29 & 3340 \\
22.99881 & 000-000 \: $12_{\:6\:7} - 11_{\:5\:6}$ & 16.51 & 3501 \\
23.31846 & 000-000 \: $10_{\:6\:5} - 10_{\:3\:8}$ & 0.09 & 2697 \\
23.81676 & 000-000 \: $8_{\:3\:6} - 7_{\:0\:7}$ & 0.61 & 1447 \\
23.89518 & 000-000 \: $8_{\:4\:5} - 7_{\:1\:6}$ & 1.04 & 1615 \\
24.05845 & 010-010 \: $16_{\:5\:12} - 15_{\:4\:11}$ & 13.76 & 7640 \\
24.41975 & 010-010 \: $9_{\:3\:6} - 8_{\:2\:7}$ & 2.42 & 4179 \\
24.91403 & 010-010 \: $9_{\:6\:3} - 8_{\:5\:4}$ & 18.64 & 4778 \\
25.14613 & 000-000 \: $10_{\:6\:5} - 9_{\:5\:4}$ & 16.05 & 2697 \\
26.05384 & 000-000 \: $10_{\:5\:5} - 9_{\:4\:6}$ & 9.89 & 2481 \\
26.25519 & 000-000 \: $13_{\:5\:9} - 12_{\:4\:8}$ & 9.33 & 3721 \\
26.64294 & 000-000 \: $9_{\:6\:4} - 8_{\:5\:3}$ & 15.33 & 2346 \\
26.72651 & 000-000 \: $11_{\:4\:7} - 11_{\:1\:10}$ & 0.12 & 2732 \\
26.90847 & 000-000 \: $17_{\:3\:14} - 16_{\:4\:13}$ & 15.93 & 5483 \\
27.0272 & 000-000 \: $7_{\:3\:5} - 6_{\:0\:6}$ & 0.48 & 1175 \\
\hline
\enddata
\tablecomments{Line properties are from HITRAN \citep{hitran20}.}
\end{deluxetable}

\section{Reference slab models for the atlas and the water diagnostic diagrams} \label{app: slab models params}
Table \ref{tab: slab params} reports the reference models used for the general water atlas in Figures \ref{fig: WATER_atlas_ROVIB} to \ref{fig: WATER_atlas_ROTlong}. In the case of CO, we use HITEMP data as currently available from HITRAN.org \citep{HITEMP10,HITEMP_CO} and de-couple the excitation of different vibrational bands to approximately reproduce their excitation, which is known to not be in LTE \citep[e.g. Figure 11 in][]{banz22}. 

\begin{figure*}
\centering
\includegraphics[width=1\textwidth]{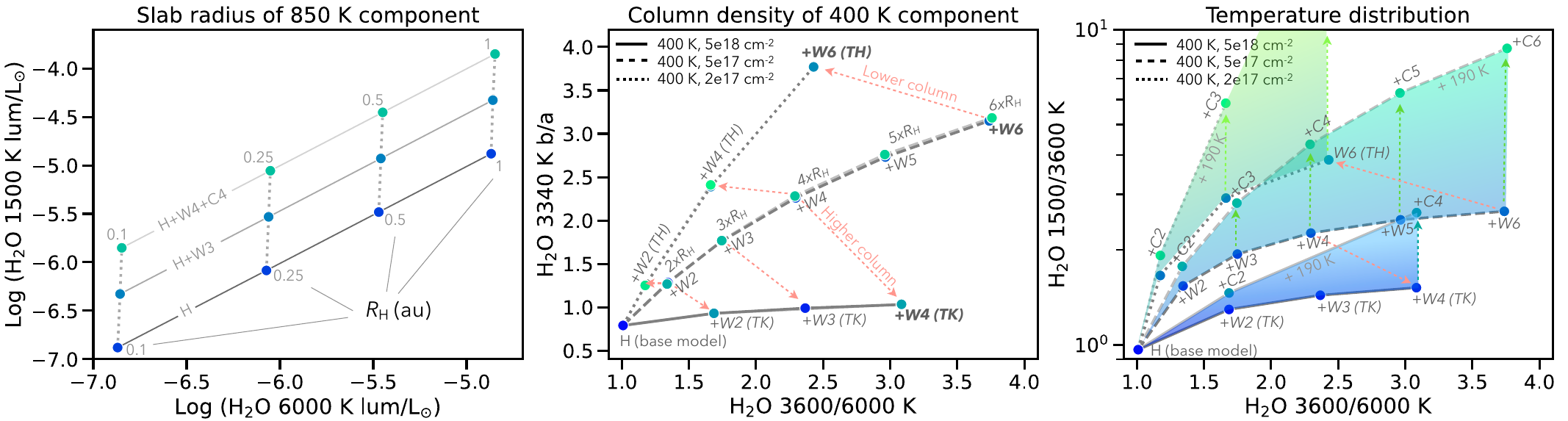}
\caption{Slab model series used as reference frame for the diagnostic line-ratio diagram in Figure \ref{fig: cool_excess}. ``H" stands for the hot (850~K) model, ``W" for the warm (400~K) model, and ``C" for the cold (190~K) model. These models are summed up into a ``H+W" and ``H+W+C" series producing the model tracks shown in this figure (see text for details), with specific model parameters reported in Table \ref{tab: slab params}.}
\label{fig: diagnostic_diagram_models}
\end{figure*}

Figure \ref{fig: diagnostic_diagram_models} and Table \ref{tab: cool excess models} illustrate the reference models used in Figure \ref{fig: cool_excess} for the water line-ratio diagnostic diagrams.
\rev{The discrete temperature component models described in the main text above are summed up into a ``H+W" and ``H+W+C" series producing the model tracks shown in Figure \ref{fig: diagnostic_diagram_models} as follows.
The H+W series takes the base hot model (850~K) and adds emission from a warm water component (400~K) by progressively increasing its emitting area to mimic a larger warm-water-rich disk region. The specific models are labeled by the size of the emitting area of the warm model in reference to the hot model, e.g. the W2 model has twice the emitting radius as the hot model, W3 three times the radius and so on. We explore models up to 6 times the hot model radius as they are sufficient to cover the range of line ratios observed in this sample. To illustrate the effects of having a more optically thick (``TK'') or thin (``TH'') emission, we reproduce the W models with a 10 times larger and 2.5 times lower column density, which produce the H+W(TK) and H+W(TH) model series. The 3340~K line ratio is most sensitive to the column density of the 400~K component, producing the spread of models in the middle panel.}

\rev{For the H+W+C series, we take each model in the H+W series and add a water emission component close to the snowline region (here assuming 190~K to represent the range estimated in Figure \ref{fig: 23um_diagram}) by progressively increasing its emitting area to mimic a larger cold-water-rich disk region. Similarly to how the H+W series is built, the specific models are labeled by the size of the emitting area of the cold model in reference now to the warm model, e.g. the C2 model has twice the emitting radius as the warm model, C3 three times the radius and so on. Also in this case, exploring models up to 6 times the radius of the warm region is enough to cover the observed line ratios. The 190~K component does not contribute to the flux of the 3340~K lines because it is too cold, therefore the ``H+W" and ``H+W+C" model series overlap perfectly in the middle panel of the figure.}

\rev{In Figure \ref{fig: diagnostic_diagram_models}, we only label some models to avoid over-crowding the plot. The arrows show the directions defined by changes in column density (red arrows in the middle and right panels) and by adding a cold component (green arrows in the right panel). The three shaded areas in the right panel identify the regions covered between a pure H+W and an H+W+C model in the three opacity regimes explored for the warm component. The overall information provided by this grid of models is summarized in Figure \ref{fig: cool_excess}: an increasing 3600/6000~K line ratio indicates a larger emitting area for the warm region, while an increasing 1500/3600~K ratio indicates increasing emitting area for the cold reservoir near the snowline. These effects are also consistent with what found by considering a continuous radial gradient rather than discrete temperature components, as shown and discussed in Section \ref{sec: discussion_radial}.}

\rev{Since line ratios are insensitive to specific value of the slab radius, the left panel in Figure \ref{fig: diagnostic_diagram_models} is provided to anchor the general models described here to the specific case of individual disks. The measured 6000~K line luminosity, being this line optically thick, mostly reflects the size of the emitting area once the temperature is assumed to be represented well by a 850~K component. Placing an object on the grid of models in the left panel will inform on the approximate emitting area of the hot model, which then can be used to estimate the specific areas for the warm and cold models as described above. Placing an object in the middle panel, instead, will help break degeneracies between different possible solutions in the plot to the right, where W models with different column density and the addition or not of a cold component may lie very close in the diagnostic diagram (e.g. the H+W5 and the H+W4(TK)+C models).}

\rev{The radial gradient models shown in Figure \ref{fig: diagnostic_diagram_gradients} are reported in Table \ref{tab: radial gradients}. The models use power-law profiles following work by \citet{munozromero24b} as $T = T_0 ( r / 0.5 \text{au})^{-\alpha}$ and $N = N_0 ( r / 0.5 \text{au})^{-\beta}$ with these fixed values: $T_0 = 450 K$, $N_0 = 1 \times 10^{18}$~cm$^{-2}$, $\beta = 1$, while $\alpha$ is varied between 0.4 and 0.7 and a cutoff radius $R_{\rm{cut}}$ between 0.75~au and 10~au, beyond which the water column density drops to zero. The cold water vapor mass  integrated over the temperature range $> 120$~K \citep[taken as the minimum sublimation temperature, from][]{lodders03} and $< 300$~K \citep[above which gas-phase formation becomes efficient, e.g.][]{glassgold09}, taken as having origin from ice sublimation at the snowline, is used to estimate an icy pebble mass flux through the snowline using Equation 11 and the same parameter values in \citet{munozromero24b}, with a water molecule mass of $3 \times 10^{-23}$ grams. This cold mass flux, called $\dot{M}_{\rm{<300K}}$, is reported for each model in Table \ref{tab: radial gradients}.}

\begin{deluxetable}{l c c c c}
\tabletypesize{\small}
\tablewidth{0pt}
\tablecaption{\label{tab: slab params} Slab model parameters adopted in the water atlas (Figures \ref{fig: WATER_atlas_ROVIB} to \ref{fig: WATER_atlas_ROTlong}).}
\tablehead{Species & $T$ & $N$ & $R_{\rm{slab}}$ & $A_{\rm{slab}}$ \\
 & (K) & (cm$^{-2}$) & (au) & (au$^2$)}
\tablecolumns{4}
\startdata
\ce{H2O} hot (ro-vibr.) & 850 & $1 \times 10^{18}$ & 0.25 & 0.2 \\
\ce{H2O} hot & 850 & $1 \times 10^{18}$ & 0.5 & 0.8 \\
\ce{H2O} warm & 400 & $5 \times 10^{17}$ & 0.9 & 2.5\\
\ce{H2O} cold & 170 & $5 \times 10^{16}$ & 4 & 50 \\
\ce{CO2} & 300 & $1 \times 10^{17}$ & 0.45 & 0.6\\
\ce{C2H2} & 800 & $1 \times 10^{16}$ & 0.32 & 0.32\\
\ce{HCN} & 950 & $1 \times 10^{16}$ & 0.55 & 0.95\\
\ce{H2} & 400 & $1 \times 10^{23}$ & 4 & 50 \\
\ce{OH} hot & 7000 & $1 \times 10^{16}$ & 0.1 & 0.03\\
\ce{OH} warm & 1000 & $1 \times 10^{16}$ & 0.7 & 1.5\\
\ce{CO} ($v=1-0$) & 1100 & $1 \times 10^{18}$ & 0.25 & 0.2\\
\ce{CO} ($v=1-0$) & 1500 & $1 \times 10^{19}$ & 0.1 & 0.03\\
\ce{CO} ($v=2-1$) & 1300 & $1 \times 10^{19}$ & 0.12 & 0.05\\
\ce{CO} ($v=3-2$) & 1500 & $1 \times 10^{19}$ & 0.07 & 0.02\\
\enddata
\tablecomments{Other parameters that are assumed in the models: a distance of 160 pc, a thermal line broadening of 1 km/s (FWHM), an instrumental+Doppler line broadening of 130 km/s at $< 18 \mu$m and 160 km/s at longer wavelengths (as measured in CI~Tau, see Section \ref{app: sample measurements}). The only exception is \ce{H2}, which is distinctly narrower and is simulated with 95 km/s.}
\end{deluxetable}

\begin{deluxetable*}{l c c c c | l c c c c}
\tabletypesize{\small}
\tablewidth{0pt}
\tablecaption{\label{tab: cool excess models} \rev{Series of slab models used in Figure \ref{fig: cool_excess}.}}
\tablehead{Model & $T$ & $N$ & $R_{\rm{slab}}$ & $M_{\rm{gas}}$ & Combined & 1500/ & 3600/ & 1500/ & 3400 K \\
 & (K) & (cm$^{-2}$) & (au) & ($\mu$M$_{\oplus}$) & models & 3600 K & 6000 K & 6000 K &  b/a}
\tablecolumns{10}
\startdata
H & 850 & $1 \times 10^{18}$ & $R_{\rm{H}}$ & $3.5 \times (R_{\rm{H}}/\rm{au})^2$ & H & 0.97 & 1.01 & 0.97 & 0.79  \\
W2 & 400 & $5 \times 10^{17}$ & $R_{\rm{W}}=2R_{\rm{H}}$ & $1.8 \times (R_{\rm{W}}/\rm{au})^2$ & H+W2 & 1.54 & 1.34 & 2.07 & 1.28 \\
W2(TK) & 400 & $5 \times 10^{18}$ & $R_{\rm{W}}=2R_{\rm{H}}$ & $18 \times (R_{\rm{W}}/\rm{au})^2$ & H+W2(TK) & 1.30 & 1.69 & 2.18 & 0.93 \\
W2(TH) & 400 & $2 \times 10^{17}$ & $R_{\rm{W}}=2R_{\rm{H}}$ & $0.7 \times (R_{\rm{W}}/\rm{au})^2$ & H+W2(TH) & 1.66 & 1.17 & 1.95 & 1.25 \\
W3 & 400 & $5 \times 10^{17}$ & $R_{\rm{W}}=3R_{\rm{H}}$ & $1.8 \times (R_{\rm{W}}/\rm{au})^2$ & H+W3 & 1.94 & 1.75 & 3.39 & 1.76 \\
W3(TK) & 400 & $5 \times 10^{18}$ & $R_{\rm{W}}=3R_{\rm{H}}$ & $18 \times (R_{\rm{W}}/\rm{au})^2$ & H+W3(TK) & 1.44 & 2.37 & 3.40 & 0.99 \\
W3(TH) & 400 & $2 \times 10^{17}$ & $R_{\rm{W}}=3R_{\rm{H}}$ & $0.7 \times (R_{\rm{W}}/\rm{au})^2$ & \nodata & \nodata & \nodata & \nodata & \nodata \\
W4 & 400 & $5 \times 10^{17}$ & $R_{\rm{W}}=4R_{\rm{H}}$ & $1.8 \times (R_{\rm{W}}/\rm{au})^2$ & H+W4 & 2.26 & 2.30 & 5.20 & 2.26 \\
W4(TK) & 400 & $5 \times 10^{18}$ & $R_{\rm{W}}=4R_{\rm{H}}$ & $18 \times (R_{\rm{W}}/\rm{au})^2$ & H+W4(TK) & 1.52 & 3.08 & 4.68 & 1.03 \\
W4(TH) & 400 & $2 \times 10^{17}$ & $R_{\rm{W}}=4R_{\rm{H}}$ & $0.7 \times (R_{\rm{W}}/\rm{au})^2$ & H+W4(TH) & 2.92 & 1.66 & 4.86 & 2.39 \\
W5 & 400 & $5 \times 10^{17}$ & $R_{\rm{W}}=5R_{\rm{H}}$ & $1.8 \times (R_{\rm{W}}/\rm{au})^2$ & H+W5 & 2.49 & 2.97 & 7.39 & 2.73 \\
W6 & 400 & $5 \times 10^{17}$ & $R_{\rm{W}}=6R_{\rm{H}}$ & $1.8 \times (R_{\rm{W}}/\rm{au})^2$ & H+W6 & 2.65 & 3.74 & 9.91 & 3.15 \\
C2 & 190 & $1 \times 10^{17}$ & $R_{\rm{C}}=2R_{\rm{W}}$ & $0.35 \times (R_{\rm{C}}/\rm{au})^2$ & H+W2+C2 & 1.77 & 1.34 & 2.37 & 1.27 \\
C3 & 190 & $1 \times 10^{17}$ & $R_{\rm{C}}=3R_{\rm{W}}$ & $0.35 \times (R_{\rm{C}}/\rm{au})^2$ & H+W3+C3 & 2.82 & 1.75 & 4.92 & 1.77 \\
C4 & 190 & $1 \times 10^{17}$ & $R_{\rm{C}}=4R_{\rm{W}}$ & $0.35 \times (R_{\rm{C}}/\rm{au})^2$ & H+W4+C4 & 4.33 & 2.29 & 9.91 & 2.28 \\
C5 & 190 & $1 \times 10^{17}$ & $R_{\rm{C}}=5R_{\rm{W}}$ & $0.35 \times (R_{\rm{C}}/\rm{au})^2$ & H+W5+C5 & 6.30 & 2.96 & 18.64 & 2.76 \\
C6 & 190 & $1 \times 10^{17}$ & $R_{\rm{C}}=6R_{\rm{W}}$ & $0.35 \times (R_{\rm{C}}/\rm{au})^2$ & H+W6+C6 & 8.72 & 3.76 & 32.78 & 3.18 \\
\enddata
\tablecomments{Other parameters that are assumed in the models: a thermal line broadening of 1 km/s (FWHM), an instrumental+Doppler line broadening of 130 km/s. $M_{\rm{gas}}$ is the observable gas mass (the product of area, column density, and molecular weight) in units of micro-Earth masses.}
\end{deluxetable*}

\begin{deluxetable}{l c c c c c}
\tabletypesize{\small}
\tablewidth{0pt}
\tablecaption{\label{tab: radial gradients} \rev{Series of slab models used in Figure \ref{fig: diagnostic_diagram_gradients}.}}
\tablehead{$\alpha$ & $R_{\rm{cut}}$ & $\dot{M}_{\rm{<300K}}$ & 1500/ & 3600/ & 1500/  \\
 & (au) & ($M_{\oplus}$~Myr$^{-1}$) & 3600~K & 6000~K & 6000~K }
\tablecolumns{6}
\startdata
0.4 & 0.75 & 0 & 1.57 & 2.40 & 3.76 \\
0.4 & 1 & 0 & 1.96 & 2.73 & 5.36 \\
0.4 & 2 & 133 & 3.86 & 3.02 & 11.7 \\
0.4 & 5 & 788 & 7.76 & 3.04 & 23.6 \\
0.4 & 10 & 1917 & 9.43 & 3.04 & 28.7 \\
0.5 & 0.75 & 0 & 1.51 & 1.96 & 2.95 \\
0.5 & 1 & 0 & 1.91 & 2.13 & 4.06 \\
0.5 & 2 & 199 & 3.46 & 2.22 & 7.66 \\
0.5 & 5 & 854 & 5.22 & 2.22 & 11.6 \\
0.5 & 10 & 1285 & 5.40 & 2.22 & 12.0 \\
0.7 & 0.75 & 0 & 1.36 & 1.37 & 1.88 \\
0.7 & 1 & 19 & 1.67 & 1.43 & 2.39 \\
0.7 & 2 & 239 & 2.42 & 1.43 & 3.47 \\
0.7 & 5 & 550 & 2.62 & 1.43 & 3.75 \\
0.7 & 10 & 550 & 2.62 & 1.43 & 3.75 \\
\enddata
\tablecomments{See text in Appendix \ref{app: slab models params} for details on model parameters. Other parameters that are assumed in the models: a thermal line broadening of 1 km/s (FWHM), an instrumental+Doppler line broadening of 130 km/s.}
\end{deluxetable}

\section{Linear regression parameters} \label{app: lin_reg_par}
Table \ref{tab: lin_fit_params} reports linear regression parameters for significant correlations detected in this work. The data to reproduce these correlations are in Appendix \ref{app: sample measurements}.

\begin{deluxetable*}{l l c c c}
\tablecaption{\label{tab: lin_fit_params} Linear regression parameters for correlations detected in this work.}
\tablehead{\colhead{$x$} & \colhead{$y$} & \colhead{$a(\sigma_a)$} & \colhead{$b(\sigma_b)$} & \colhead{Figure} }
\startdata
log L$_{1500K}$ & log L$_{acc}$ & -4.60(0.12) & 0.44(0.09) & \ref{fig: Lum_corr} \\ 
log L$_{3600K}$ & log L$_{acc}$ & -4.93(0.12) & 0.49(0.09) & \ref{fig: Lum_corr} \\ 
log L$_{6000K}$ & log L$_{acc}$ & -5.22(0.11) & 0.42(0.09) & \ref{fig: Lum_corr} \\ 
log L$_{3340K}$ & log L$_{acc}$ & -5.02(0.11) & 0.46(0.08) & \ref{fig: Lum_corr} \\ 
log L$_{1500K}$ & Incl & -4.45(0.24) & -0.014(0.005) & \ref{fig: Lum_corr} \\ 
log L$_{3600K}$ & Incl & -4.72(0.24) & -0.017(0.005) & \ref{fig: Lum_corr} \\ 
log L$_{6000K}$ & Incl & -5.14(0.25) & -0.012(0.006) & \ref{fig: Lum_corr} \\ 
log L$_{3340K}$ & Incl & -4.87(0.24) & -0.015(0.006) & \ref{fig: Lum_corr} \\ 
1500/6000~K & log R$_{disk}$ & 7.4(1.8) & -2.3(1.1) & \ref{fig: cool_excess_ALMA} \\ 
3600/6000~K & log R$_{disk}$ & 2.73(0.36) & -0.62(0.2) & \ref{fig: cool_excess_ALMA} \\ 
FWHM(ro-vib.) & Incl & 92(13) & 0.72(0.31) & \ref{fig: ROTAT_broadening} \\ 
FWHM(rot$<$18$\mu$m) & Incl & 96(8) & 0.70(0.20) & \ref{fig: ROTAT_broadening} \\ 
FWHM(rot$>$18$\mu$m) & Incl & 129(6) & 0.52(0.14) & \ref{fig: ROTAT_broadening} \\ 
FWHM(OH~4000K) & Incl & 109(21) & 1.21(0.48) & \ref{fig: ROTAT_broadening} \\ 
FWHM(CO~5000~K) & Incl & 100(12) & 0.76(0.27) & \ref{fig: ROTAT_broadening} \\ 
FWHM(rot$<$18$\mu$m) & FWHM(CO$_{ground}$) & 95(5) & 0.44(0.07) & \ref{fig: ROTAT_broadening} \\ 
FWHM(rot$>$18$\mu$m) & FWHM(CO$_{ground}$) & 129(3) & 0.32(0.04) & \ref{fig: ROTAT_broadening} \\ 
FWHM(OH~4000K) & FWHM(CO$_{ground}$) & 128(9) & 0.33(0.13) & \ref{fig: ROTAT_broadening} \\ 
\enddata
\tablecomments{Linear relations are in the form $y = a+b x$.}
\end{deluxetable*}

\section{Updates to the MIRI resolving power} \label{app: res power}
Table \ref{tab: resolving_power} reports updated fits to measured line FWHM to characterize the MIRI resolving power as presented in Section \ref{sec: broadening} and Figure \ref{fig: Res_Pow_MIRI}. We only update the intercept in sub-bands at $> 10\mu$m, since the slope was already well characterized in \cite{pontoppidan24}; the parameters for sub-bands not included in this table are unchanged and can be found in Table 3 in \cite{pontoppidan24}.

\begin{deluxetable}{l c c c}
\tablecaption{\label{tab: resolving_power} Updated MIRI resolving power.}
\tablehead{\colhead{Sub-band} & \colhead{$a$} & \colhead{$b$} & \colhead{Wavelength} \\
&  & & ($\mu$m) }
\startdata
1A & 2742  & 150 & 5.66--6.63  \\ 
2C & 430  & 264 & 10.02--11.70 \\ 
3C & -2240 & 312 & 15.41--17.98 \\ 
4A & -2066 & 225 & 17.70--20.95 \\ 
4B & -1076 & 150 & 20.69--24.48 \\ 
4C & -3451 & 216 & 24.19--28.10 \\
\enddata
\tablecomments{Updates to Table 3 in \cite{pontoppidan24}. The MIRI-MRS resolving power in each sub-band is reported as $\lambda/\Delta\lambda = R = a+b\lambda$.}
\end{deluxetable}

\section{Additional plots for the whole sample} \label{app: sample_plot_grids}
\begin{figure*}
\centering
\includegraphics[width=1\textwidth]{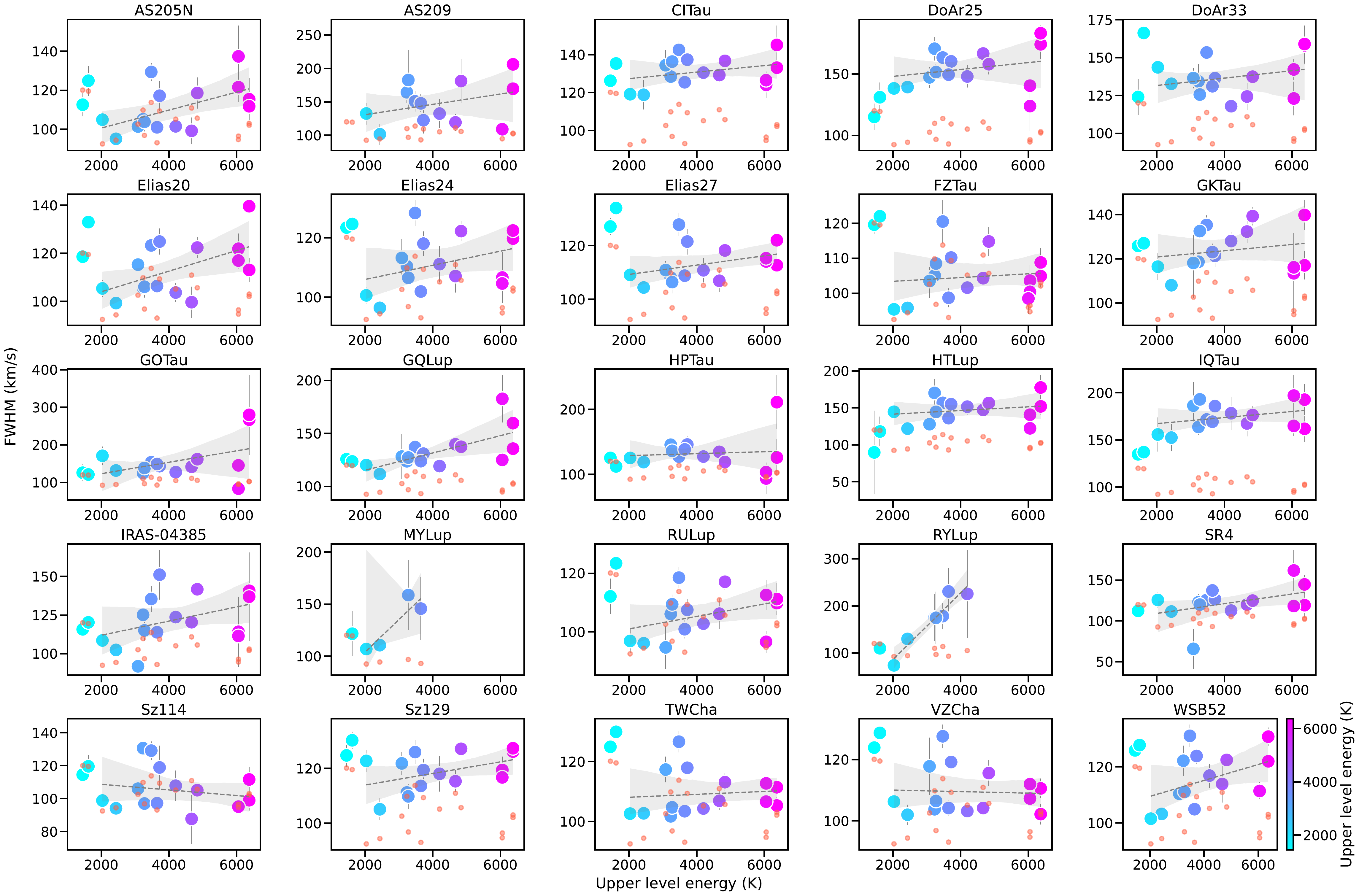}
\caption{Same as Figure \ref{fig: Eu_broadening}, but for the whole sample.}
\label{fig: Eu_broad_sample}
\end{figure*}

\begin{figure*}
\centering
\includegraphics[width=1\textwidth]{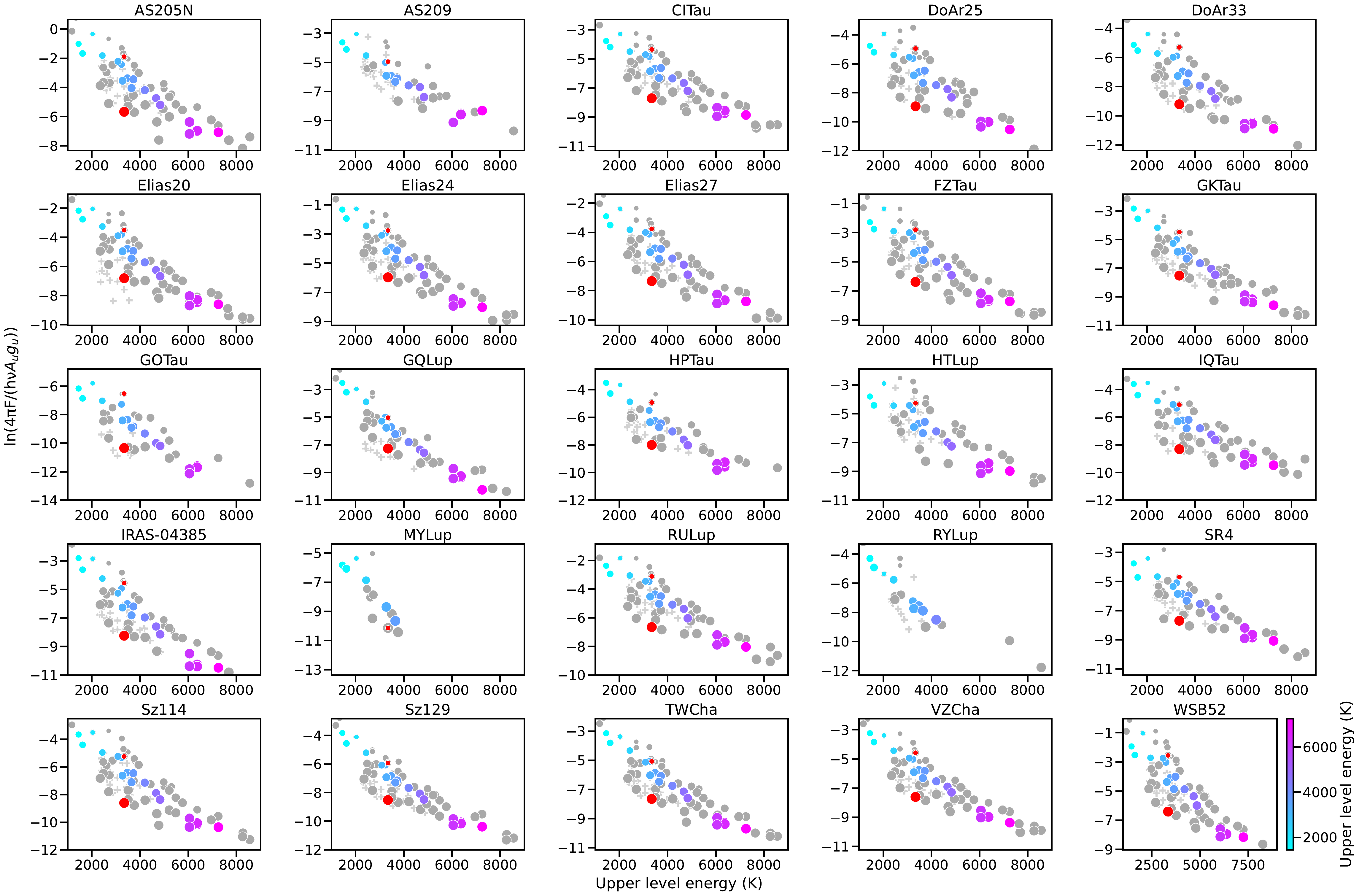}
\caption{Rotation diagrams for the whole sample, following the same style as in Figure \ref{fig: Line_selection_demo}. }
\label{fig: RotDiagr_sample}
\end{figure*}

Figures \ref{fig: Eu_broad_sample} and \ref{fig: RotDiagr_sample} report additional plots to complement those shown in the main text above.

\section{Compact water spectral atlas for JDISCS targets} \label{app: atlas_compact}
\begin{figure*}
\centering
\includegraphics[width=1\textwidth]{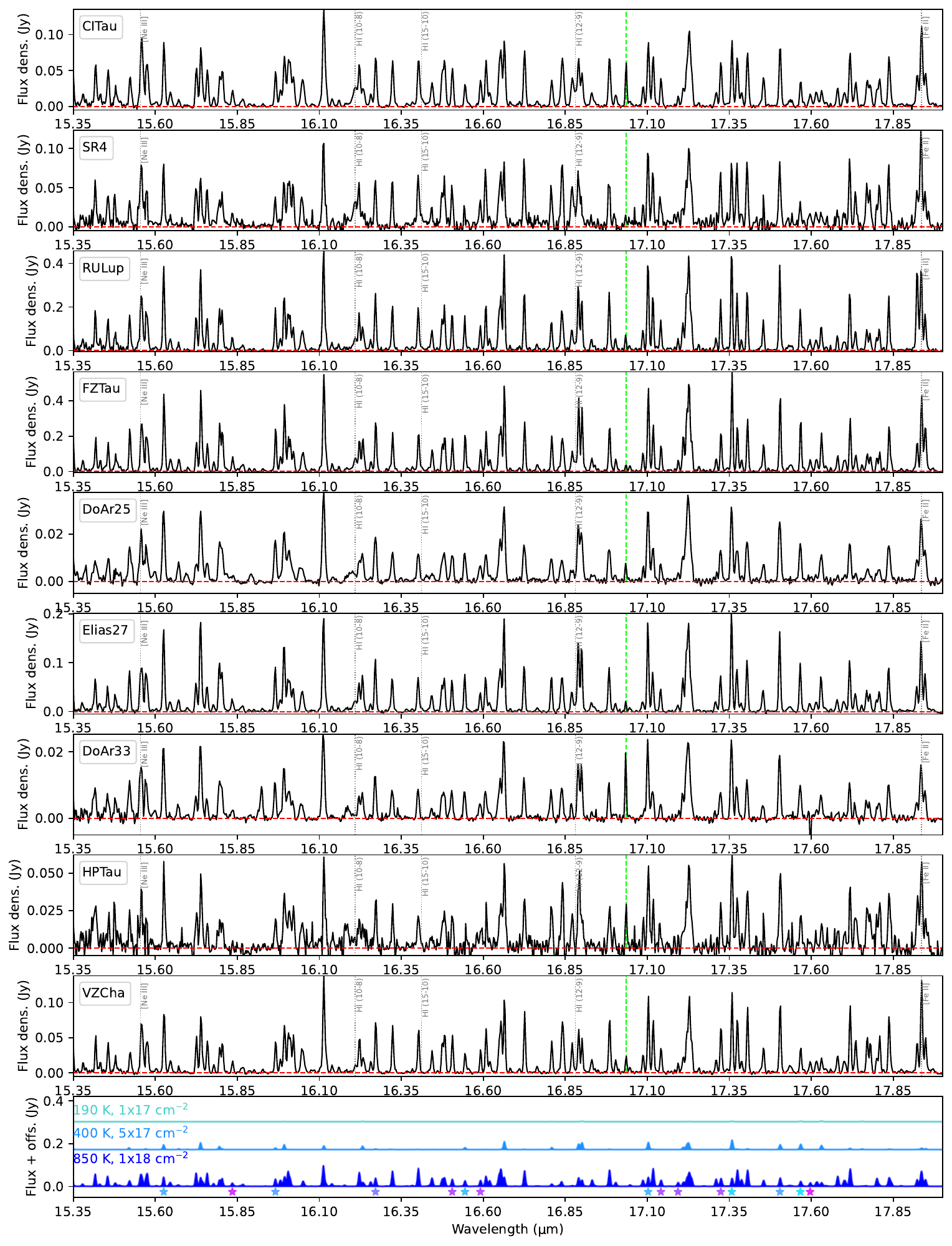}
\caption{Compact water spectral atlas for JDISCS targets, highlighting the three temperature components discussed in the text. This figure includes disks that are consistent with a small to moderate warm water component in Figure \ref{fig: cool_excess}.}
\label{fig: MIRI_spectra_atlas1}
\end{figure*}

\begin{figure*}
\centering
\includegraphics[width=1\textwidth]{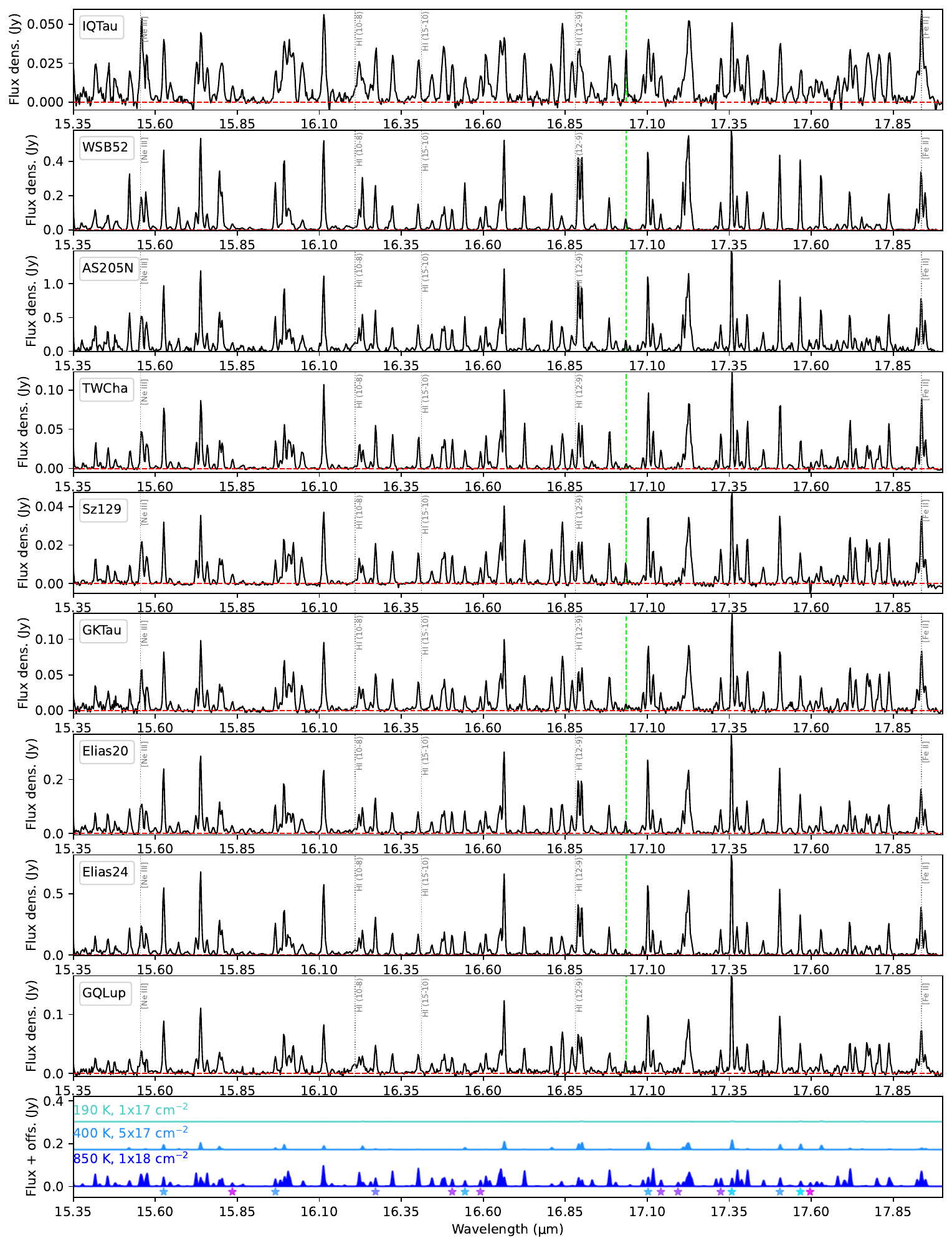}
\caption{Compact water spectral atlas for JDISCS targets (continued from Figure \ref{fig: MIRI_spectra_atlas1}). This figure includes disks that are consistent with a moderate to large cold water component in Figure \ref{fig: cool_excess}.}
\label{fig: MIRI_spectra_atlas2}
\end{figure*}

\begin{figure*}
\centering
\includegraphics[width=1\textwidth]{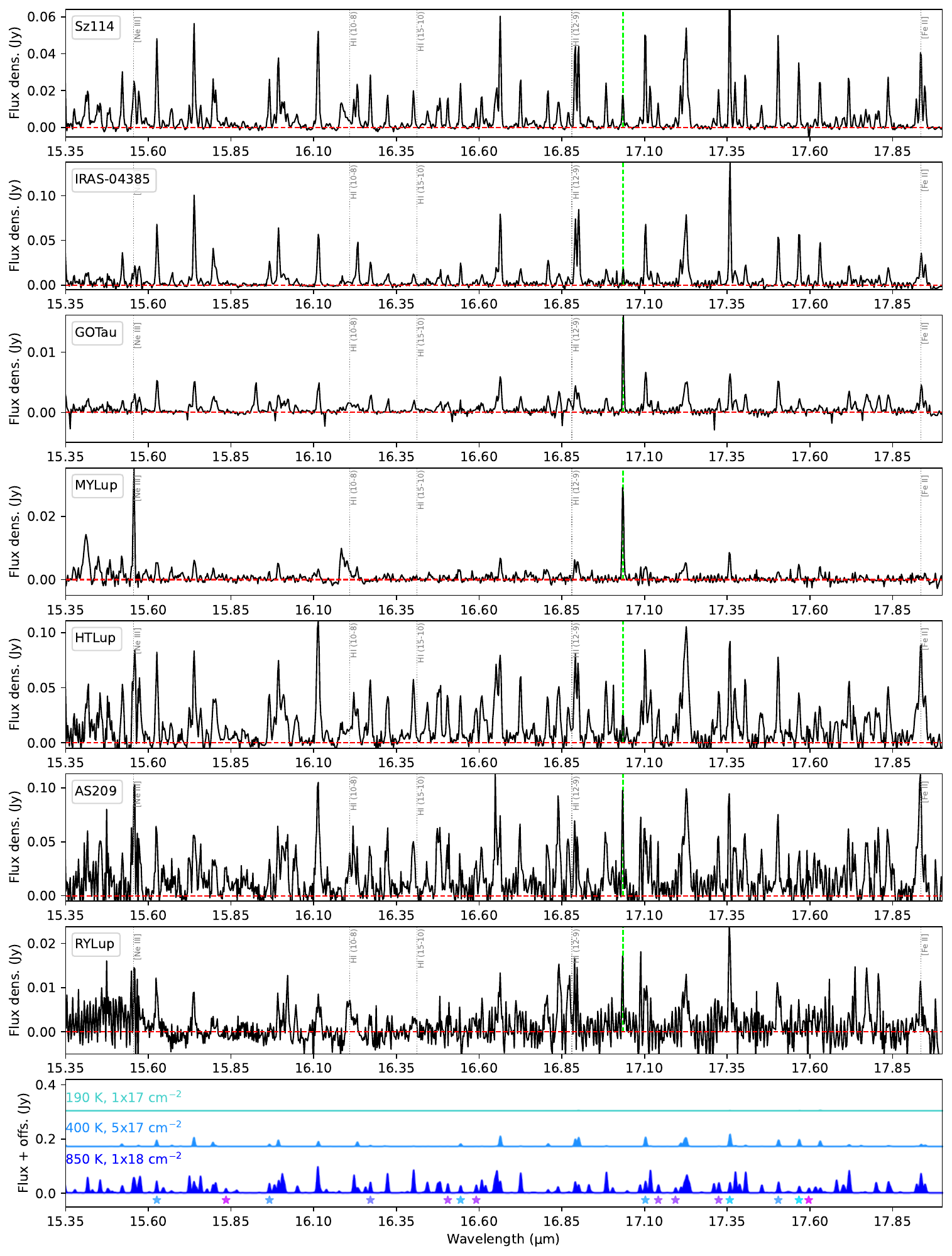}
\caption{Compact water spectral atlas for JDISCS targets (continued from Figure \ref{fig: MIRI_spectra_atlas1}). These targets have peculiarities discussed in the text  or lower S/N (bottom 2 targets).}
\label{fig: MIRI_spectra_atlas3}
\end{figure*}

\begin{figure*}
\centering
\includegraphics[width=1\textwidth]{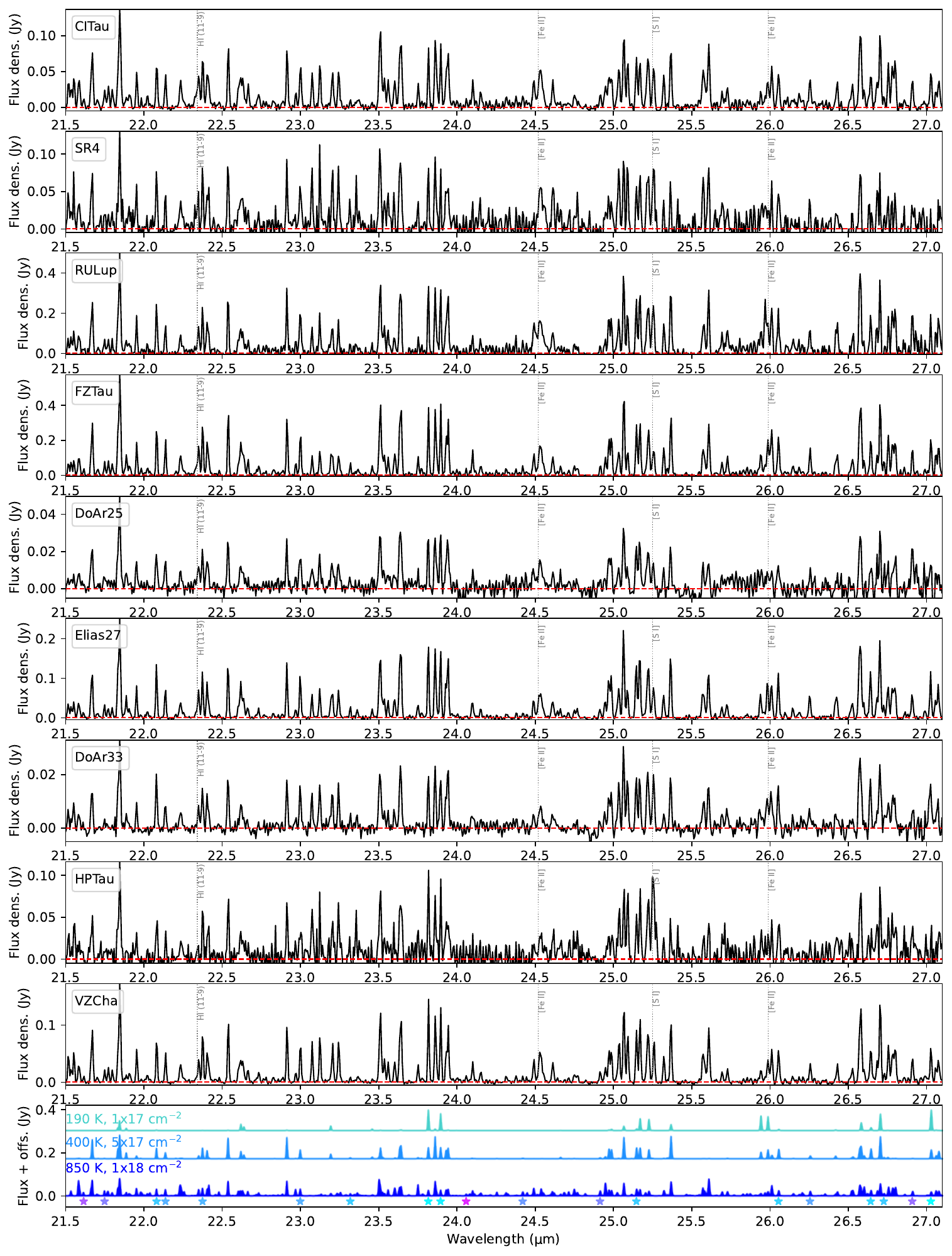}
\caption{Same as Figure \ref{fig: MIRI_spectra_atlas1}, but showing longer wavelengths.}
\label{fig: MIRI_spectra_atlas1l}
\end{figure*}

\begin{figure*}
\centering
\includegraphics[width=1\textwidth]{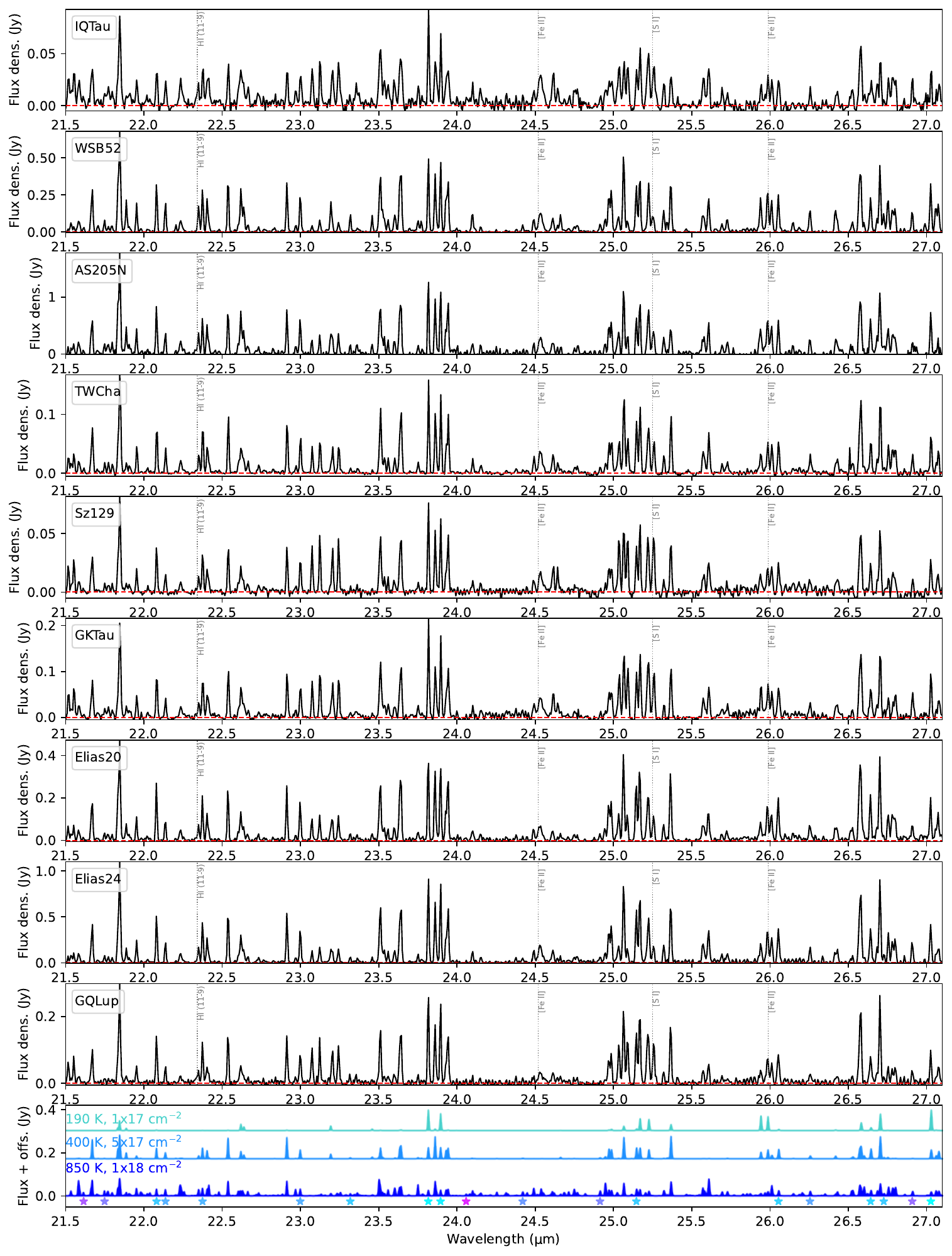}
\caption{Same as Figure \ref{fig: MIRI_spectra_atlas2}, but showing longer wavelengths.}
\label{fig: MIRI_spectra_atlas2l}
\end{figure*}

\begin{figure*}
\centering
\includegraphics[width=1\textwidth]{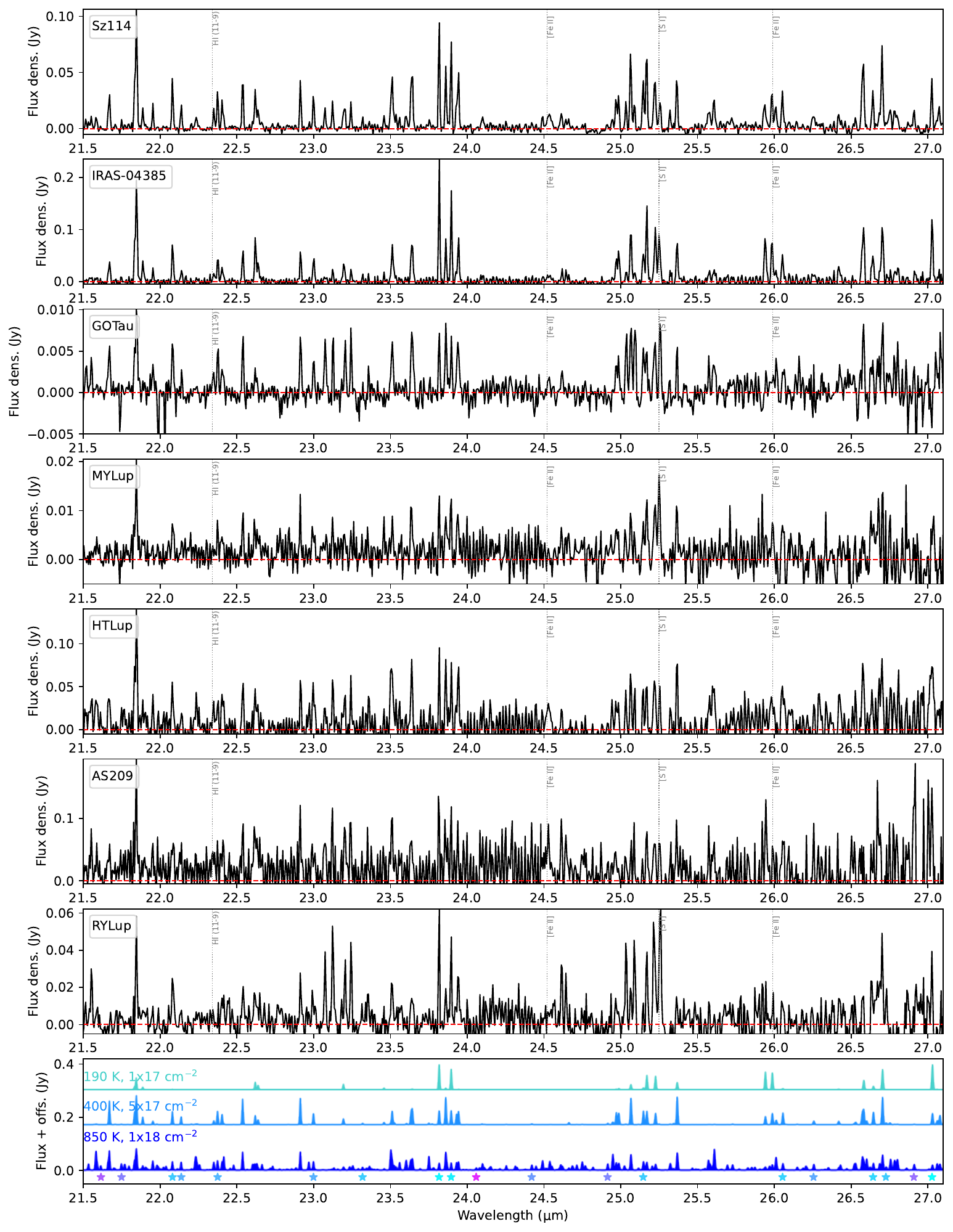}
\caption{Same as Figure \ref{fig: MIRI_spectra_atlas3}, but showing longer wavelengths.}
\label{fig: MIRI_spectra_atlas3l}
\end{figure*}

\begin{figure*}
\centering
\includegraphics[width=1\textwidth]{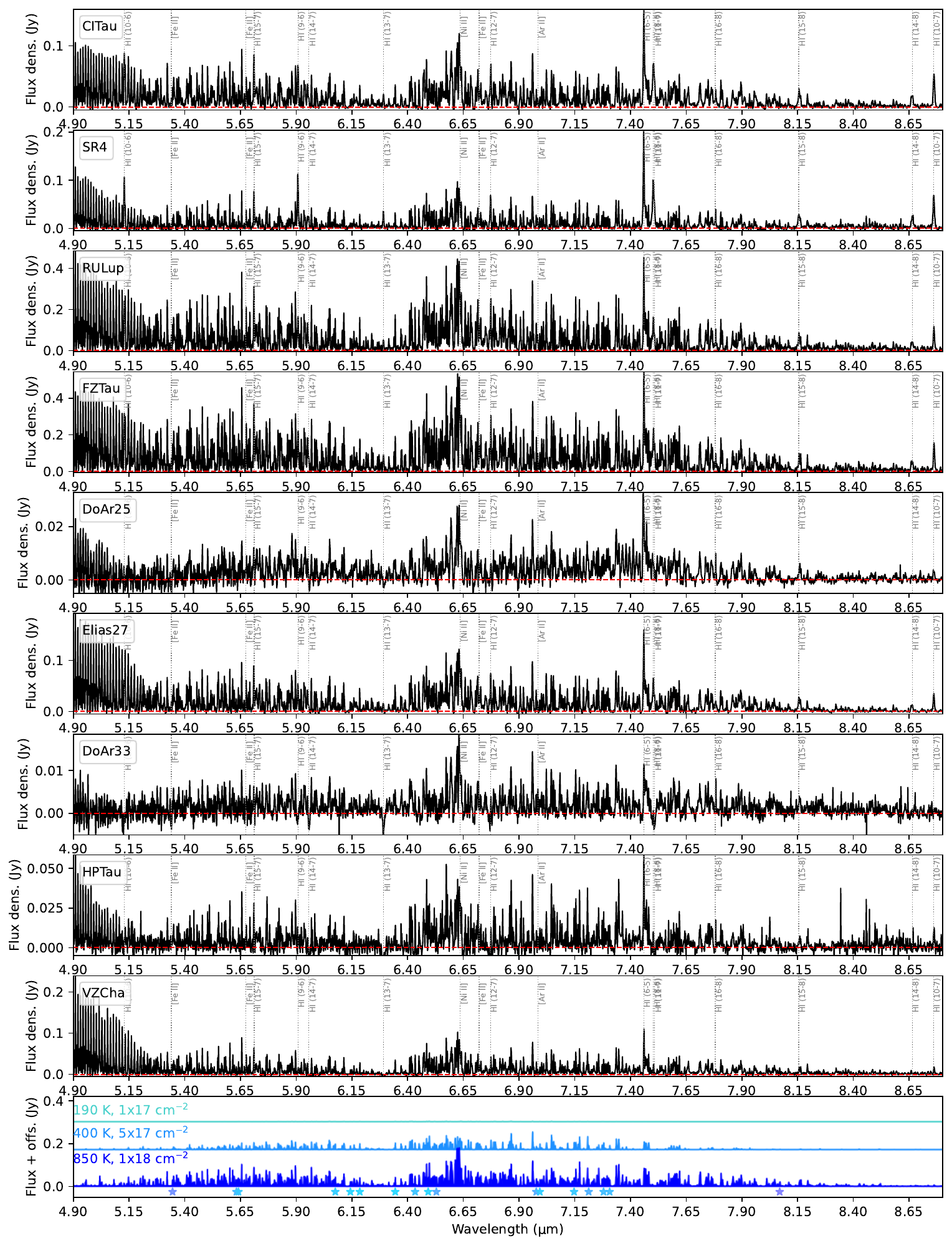}
\caption{Same as Figure \ref{fig: MIRI_spectra_atlas1}, but showing the ro-vibrational band.}
\label{fig: MIRI_spectra_atlas1r}
\end{figure*}

\begin{figure*}
\centering
\includegraphics[width=1\textwidth]{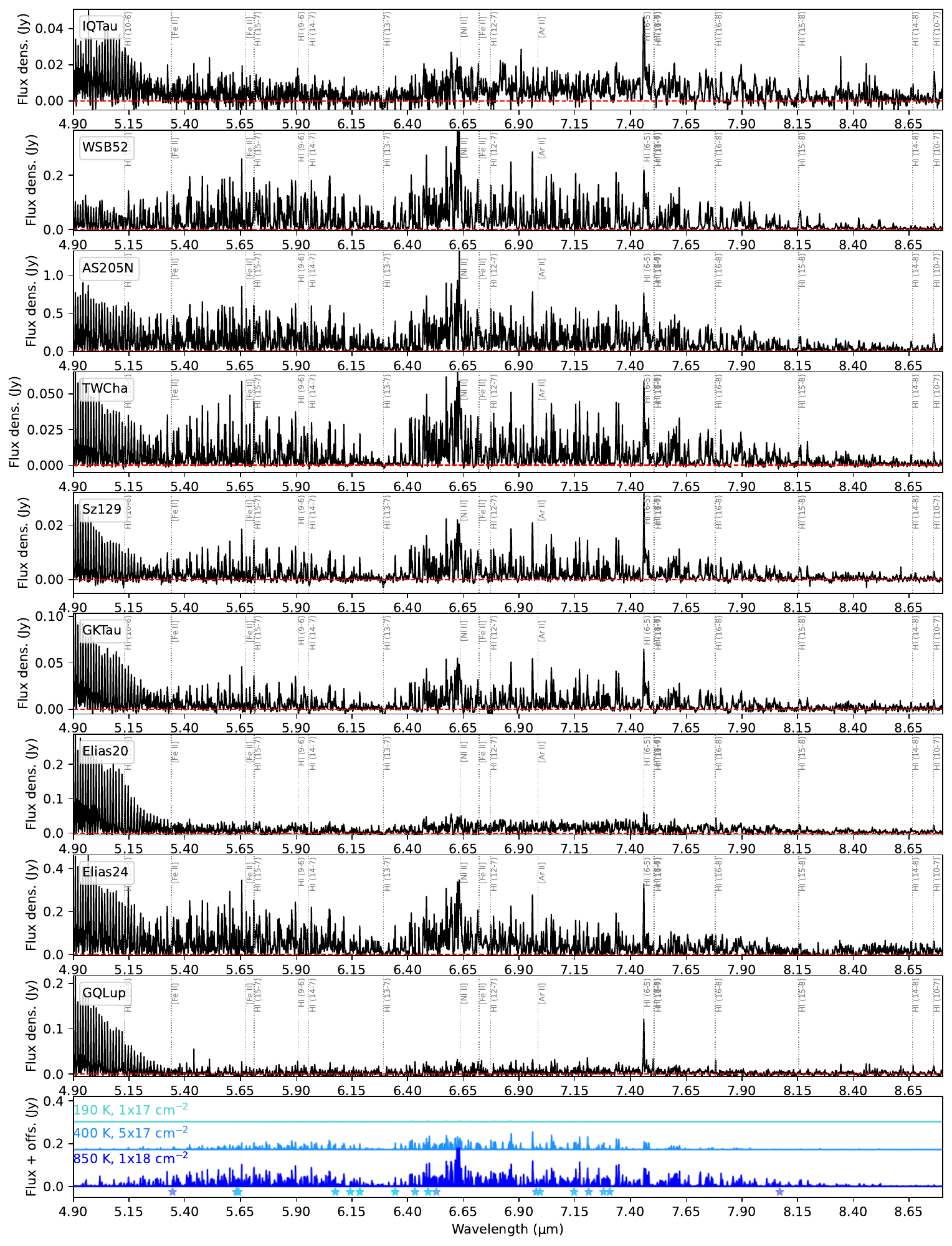}
\caption{Same as Figure \ref{fig: MIRI_spectra_atlas2}, but showing the ro-vibrational band.}
\label{fig: MIRI_spectra_atlas2r}
\end{figure*}

\begin{figure*}
\centering
\includegraphics[width=1\textwidth]{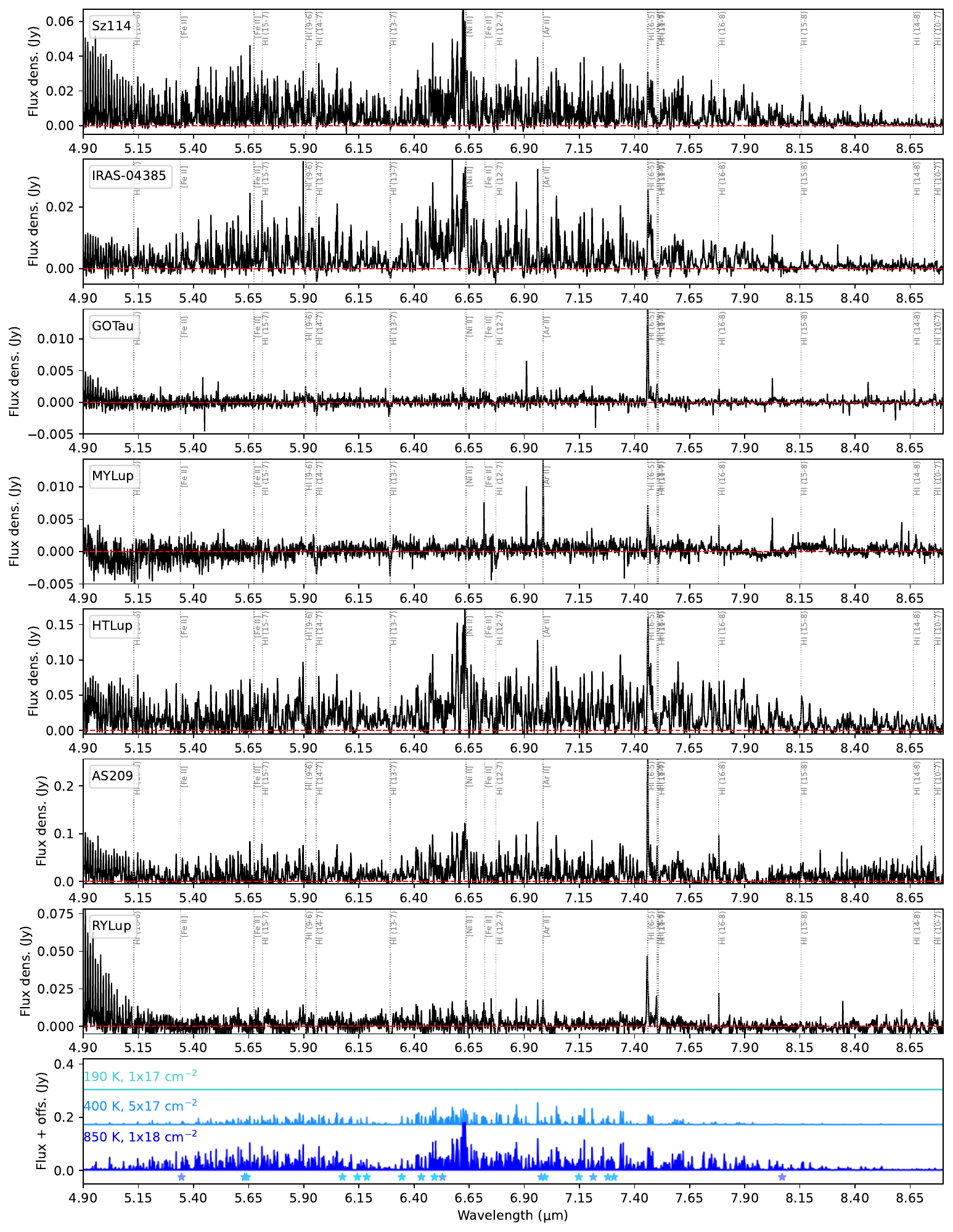}
\caption{Same as Figure \ref{fig: MIRI_spectra_atlas3}, but showing the ro-vibrational band.}
\label{fig: MIRI_spectra_atlas3r}
\end{figure*}

Figures \ref{fig: MIRI_spectra_atlas1} to \ref{fig: MIRI_spectra_atlas3r} show a compact version of a water spectral atlas for all disks included in this work in reference to the three temperature components as in Figure \ref{fig: MIRI_spectra_atlas_RED}. The spectra are split into three regions that are most important for the analysis presented in this work: the rotational lines at $\approx$~15--18~$\mu$m (typically dominated by water emission at higher temperatures), the rotational lines at $\approx$~21--27~$\mu$m (where cooler water emission, where present, becomes prominent), and, lastly, the ro-vibrational band from the bending mode at 5--8~$\mu$m. The region of organic emission at 12--15.5~$\mu$m is included in the JDISCS overview paper (Arulanantham et al. 2024, in prep.). The line list presented in Section \ref{app: line list} is identified with stars, color-coded as in Figure \ref{fig: WATER_atlas_ROVIB}. 
Targets in Figure \ref{fig: MIRI_spectra_atlas3} have some peculiarities: Sz~114 has the lowest stellar mass in this sample, IRAS 04385+2550 (labelled IRAS-04385 in the plots) is a younger more embedded disk \citep{schaefer09}, MY~Lup has an inclination $> 70$~deg and a cold water spectrum analyzed in Salyk et al. (submitted), HT~Lup is the spectrum of both A and B components of the triple system, AS~209 and RY~Lup have weaker molecular emission and larger fringe residuals than the rest of the sample.

\bibliography{water_atlas}{}
\bibliographystyle{aasjournal}
\pagebreak

\end{document}